\documentclass[preprintnumbers,nofootinbib,
superscriptaddress,amsmath,floatfix,prd]{revtex4}

\usepackage{amssymb}  
\usepackage{graphicx,subfigure}
\usepackage{exscale}
\usepackage{textcomp}
\usepackage{enumerate}
\usepackage [english]{babel}
\usepackage [autostyle, english = american]{csquotes}
\MakeOuterQuote{"}

\usepackage{color}

\usepackage[normalem]{ulem}  

\renewcommand\a{\alpha}

\renewcommand\k{\kappa}
\renewcommand\l{\lambda}

\newcommand\FE{U}

\newcommand\vf{\varphi}

\newcommand{\non}{\nonumber\\}

\newcommand\p{\partial}

\newcommand{\be}{\begin{equation}}
\newcommand{\ee}{\end{equation}}
\newcommand{\bea}{\begin{eqnarray}}
\newcommand{\eea}{\end{eqnarray}}
\newcommand{\ba}[1]{\begin{array}{#1}}
\newcommand{\ea}{\end{array}}

\begin{document}

\title{Critical magnetic fields in a superconductor coupled to a superfluid}

\author{Alexander Haber}
\email{ahaber@hep.itp.tuwien.ac.at}
\affiliation{Institut f\"{u}r Theoretische Physik, Technische Universit\"{a}t Wien, 1040 Vienna, Austria}
\affiliation{Mathematical Sciences and STAG Research Centre, University of Southampton, Southampton SO17 1BJ, United Kingdom}

\author{Andreas Schmitt}
\email{a.schmitt@soton.ac.uk}
\affiliation{Mathematical Sciences and STAG Research Centre, University of Southampton, Southampton SO17 1BJ, United Kingdom}

\date{19 June 2017}

\begin{abstract}

We study a superconductor that is coupled to a superfluid via density and derivative couplings. Starting from a Lagrangian for two 
complex scalar fields, we derive a temperature-dependent Ginzburg-Landau potential, which is then used to compute the phase diagram at nonzero temperature
and external magnetic field. 
This includes the calculation of the critical magnetic fields for the transition to an array of magnetic flux tubes, based on an 
approximation for the interaction between the flux tubes. We find that the transition region between type-I and type-II superconductivity changes qualitatively due to the presence of the superfluid: the phase transitions at the upper and lower critical fields in the type-II regime become first order, opening the possibility of clustered flux tube phases. These flux tube clusters may be realized in the core of neutron stars, where superconducting protons are expected to be coupled to superfluid neutrons. 
   
\end{abstract}

\maketitle


\section{Introduction}

\subsection{Goal}

An array of magnetic flux tubes is created in certain superconductors for intermediate strengths of an external magnetic field. Superconductors with this property are said to 
be of type II. This is in contrast to type-I superconductors, where the magnetic field is either completely expelled or completely destroys the superconducting state, but never 
penetrates partially through quantized flux tubes. The Ginzburg-Landau parameter $\kappa$ -- the ratio between the magnetic penetration depth and the coherence length of the superconducting condensate -- predicts whether a superconductor is of type I or of type II. 

The goal of this paper is to study the critical magnetic fields for the flux tube lattice in a two-component system, where
the superconductor is coupled to a superfluid. We consider a system of two complex scalar fields and an abelian gauge field, with the two scalar fields 
coupled to each other and one of them coupled to the gauge field  -- the neutral scalar field is then indirectly coupled to the gauge field through the 
charged scalar field. Various aspects of this system will be discussed, such as the effect of different forms of the coupling between the scalar fields (density coupling vs.~derivative
coupling), effects of nonzero temperature, and the interaction between magnetic flux tubes. Special emphasis will be put on the transition 
region between type-I and type-II behavior, because this region is changed qualitatively
by the presence of the superfluid, and one of the main results will be the topology of the phase diagram in this region. 

\subsection{Methods}

Our calculations are based on a Ginzburg-Landau free energy for two condensates. We start, however, from a field-theoretical Lagrangian from which 
we compute the thermal fluctuations of the system. This is necessary in order to generalize the 
standard temperature-dependent coefficients of the Ginzburg-Landau potential to the situation of two coupled fields. We shall work in a relativistic formalism, but the main results 
hold for non-relativistic systems as well because we only consider the static limit. The coupled equations of motion for the two 
condensates and the gauge field -- which yield the profile and energy of a single flux tube -- are computed numerically.
Nevertheless, where possible, we derive simple analytical results. For instance, when we compute the free energy of a flux tube array, we employ an approximation valid for 
sparse arrays, based on the numerical solution for a single flux tube, which is sufficient to derive certain aspects of the phase structure. 
For a complete study of the phase diagram a fully numerical calculation would be necessary. We believe that our results provide 
guidance and physical insights that can support and complement such a numerical calculation in future studies.

\subsection{Astrophysical context}

A superconductor that is coupled to a superfluid is expected to exist in the core of neutron stars in the form of superconducting protons which coexist with superfluid neutrons \cite{Bogolyubov1958,MIGDAL1959655,Page:2013hxa,Sedrakian:2006xm,Haber:2016ljn}.
Although we keep all our results as generic as possible, this is the application we have 
in mind when we make certain choices for the parameters of our model. It is also the main motivation for including a derivative coupling between the superconductor and the superfluid;
for a calculation of the strength of this coupling in dense nuclear matter see for instance Ref.\ \cite{Chamel:2006rc}. Microscopic 
calculations -- which have to be taken with care at these extreme baryon number densities -- suggest that the proton superconductor turns from type II to type I as the density increases, i.e., as we
move further towards the center of the star. In other words, a neutron star has a spatially varying $\kappa$, and the transition from type-II to type-I superconductivity might be 
realized as a function of the radius of the star \cite{Glampedakis:2010sk}. The resulting interface between the two superconducting phases might affect the evolution of the magnetic field in the star and is thus of potential relevance to observations. Even if this interface is not realized, be it because the central density is not sufficiently large or because 
quark matter is preferred before the necessary density is reached, it is important to understand the magnetic properties of the flux tube phase in the presence of the 
neutron superfluid.  

The energy gaps from nucleon Cooper pairing depend strongly on density, varying non-monotonically along the profile of the star, with a maximum of the order of $1\, {\rm MeV}$ 
at intermediate densities and being much smaller at higher densities deep in the core \cite{wambach1993quasiparticle}. 
Therefore, the critical temperatures, which can be as high as $T_c \sim 10^{10}\,{\rm K}$, are very small in certain regions of the star. And, the critical magnetic fields 
for proton superconductivity, at their maximum about $H_c \sim 10^{16}\, {\rm G}$ -- larger than the largest measured surface fields -- become very small as well. (A very feeble 
superconducting pairing gap is neither robust against temperature nor against a magnetic field.) This motivates us to study the behavior of the superconductor at magnetic fields
close to the critical fields, and it motivates us to include temperature. 
For predictions in the astrophysical context, the coefficients of our effective model should be made density-dependent, using results from more 
microscopic calculations (which, however, are prone to large uncertainties). In the present work we mainly focus on deriving general results and only mimic the situation 
of dense nuclear matter by varying our parameters in a way that is  reminiscent of the situation in a neutron star.

There are other possible two- or multi-fluid phases in the core of a neutron star, where at least one of the components is charged. For instance, hyperon condensation may yield further 
condensate species \cite{Gusakov:2009kc}, a charged hyperon condensate in coexistence with a proton superconductor possibly forming a two-superconductor system. 
Two-component systems are also possible in dense quark matter. In the color-flavor locked (CFL) phase \cite{Alford:1998mk}, the
pairing of all quarks is usually described by a single gap function. This is different in the presence of a magnetic field, and the study of
color-magnetic flux tubes \cite{Iida:2004if} or domain walls \cite{Giannakis:2003am} in a Ginzburg-Landau approach shows striking 
similarities with our two-component system. The color-magnetic flux tubes in CFL are not protected by topology \cite{Eto:2013hoa}, but if there is a mechanism to stabilize them, for instance an external magnetic field, they may have interesting implications for neutron star physics \cite{Glampedakis:2012qp}, like 
their analogues in 2SC quark matter \cite{Alford:2010qf}. In coexistence with 
a kaon condensate \cite{Bedaque:2001je,Alford:2007xm}, the CFL phase  couples a color superconductor with a superfluid and represents another interesting system to which our results can be potentially applied.

\subsection{Broader context}

A mixture of a superconductor with a superfluid is conceivable not only in neutron stars but also in the laboratory, for example in ultra-cold atomic 
systems, where Bose-Fermi mixtures have been produced \cite{2014Sci...345.1035F,2015PhRvL.115z5303D}. 
Atoms are, of course, neutral, and thus this is actually a mixture of two superfluids. 
However, at least for a single
atomic species, the coupling to a "synthetic magnetic field" has been realized, including the observation of analogues of magnetic flux 
tubes \cite{2009Natur.462..628L,2011RvMP...83.1523D,2014RPPh...77l6401G}. Therefore, 
future experiments may well allow for the creation of a laboratory version of a coupled superconductor/superfluid system. 

If we relax the condition of exactly one of the two components being charged, we find more realizations. Systems of two superconducting components have been discussed in the 
literature \cite{Carlstrom:2010wn,Brandt2011,2012arXiv1206.6786B,Wu:2015sqk} and can be realized in the form of two-band superconductors, or even in liquid metallic hydrogen \cite{babaev2004superconductor}. Two coexisting superfluids, besides atomic Bose-Fermi mixtures, are conceivable in $^3$He -- $^4$He 
mixtures \cite{2002JLTP..129..531T,PhysRevB.85.134529}, although in this case it is experimentally challenging to have both components in the superfluid state simultaneously. 

\subsection{Relation to previous work}
\label{sec:relation}

Our study makes use of and extends various results of the literature. The model we are using is a gauged version of the one of Ref.\ \cite{Haber:2015exa}, 
where two-stream instabilities 
in a system of coupled superfluids were discussed. Magnetic flux tubes from proton superconductivity in neutron stars have been studied extensively in the literature, usually 
with an emphasis on phenomenological consequences. 
More microscopic approaches often do not include a consistent treatment of both components and rather put together separate results
from the proton superconductor and the neutron superfluid (which may be a good approximation for certain quantities because of the small 
proton fraction in neutral, $\beta$-equilibrated nuclear matter). Studies relevant to our work that do include both components within a single model can be found in
Refs.\ \cite{Alpar:1984zz,Alford:2007np,2015arXiv150400570K,Sinha:2015bva}. In Ref.\ \cite{Alford:2007np}, flux tube profiles and energies are computed, results that we 
reproduce and utilize in the present paper.  Our calculation of the 
interaction between flux tubes is performed within an approximation valid for large flux tube separations, based on old literature for a single-component superconductor \cite{Kramer:1971zza}; for a different method leading to the same result see 
Ref.\ \cite{Speight:1996px}. 
Extensions to a system of a superconductor coupled to a superfluid can be found in Refs.\ \cite{Buckley:2003zf,Buckley:2004ca}, where the results were restricted to 
the symmetric situation of approximately equal self-coupling and cross-coupling strengths of the scalar fields
(which is unrealistic for neutron star matter \cite{Alford:2005ku}), and no derivative cross-coupling was taken into account. Interactions between flux tubes have also been 
computed, based on the same approximation, in the context of cosmic strings for one-component \cite{Bettencourt:1994kf} and two-component \cite{MacKenzie:2003jp} systems. 
Our study is also related to so-called type-1.5 superconductivity, predicted to occur in systems with two superconducting components 
\cite{PhysRevB.72.180502,PhysRevLett.105.067003,Carlstrom:2010wn}.  Although in our study only one component is 
charged, we shall find very similar effects, for instance the possibility of flux tube clusters. 

\subsection{Structure of the paper}

In Sec.\ \ref{sec:model}, we present the model, compute the free energy densities of the various phases at vanishing magnetic field, and introduce effects of nonzero temperature. 
In Sec.\ \ref{sec:critical_fields}, we derive the expressions for the critical magnetic fields $H_c$, $H_{c1}$, and $H_{c2}$ for our two-component system and use the 
flux tube - flux tube interaction to point out the possibilities of first-order phase transitions. Our numerical results, most of them in the form of phase diagrams, 
are presented in Sec.\ \ref{sec:phases}, together with a discussion of the type-I/type-II transition region. We give our conclusions and an outlook in Sec.\ \ref{sec:summary}. 
Throughout the paper, we use natural units $\hbar=c=k_B=1$ and Gaussian units for the electromagnetic fields, such that the elementary charge is $e = \sqrt{\alpha} \simeq  0.085$
with the fine structure constant $\alpha$.

\section{Model}
\label{sec:model}

\subsection{Lagrangian and basic phase structure}

Our calculation will essentially be a mean-field Ginzburg-Landau study, and we could thus, as a starting point, simply state the Ginzburg-Landau potential. 
We choose a slightly more 
general field-theoretical language, mainly because it provides us with the framework of thermal field theory to introduce temperature. Starting from a Ginzburg-Landau potential
directly, this would be less straightforward in our two-component system. In the following, we thus start with a Lagrangian for two complex scalar fields, 
and the zero-temperature Ginzburg-Landau 
potential simply is the tree-level potential of this Lagrangian. This is Eq.\ (\ref{Ux}). Temperature is then introduced in an approximation based on the thermal excitations of the system, providing a simple temperature dependence for the Ginzburg-Landau coefficients, given in Eqs.\ (\ref{UxT}) and (\ref{thermal_mass}).
    
The Lagrangian is 
\be \label{L}
{\cal L}={\cal L}_1 + {\cal L}_2 + {\cal L}_{\rm int} + {\cal L}_{\rm YM}\, , 
\ee
where  
\begin{subequations}
\bea 
{\cal L}_i&=&D_\mu\varphi_i(D^\mu \varphi_i)^*-m_i^2|\varphi_i|^2-\lambda_i|\varphi_i|^4  \, , \qquad i=1,2 \, , \\[2ex]
{\cal L}_{\rm int} &=& 2h|\varphi_1|^2|\varphi_2|^2-\frac{g_1}{2}\Big[\varphi_1\varphi_2(D_\mu\varphi_1)^*(D^\mu\varphi_2)^*+{\rm c.c.}\Big]-\frac{g_2}{2}\Big[\varphi_1\varphi_2^*(D_\mu\varphi_1)^*D^\mu\varphi_2+{\rm c.c.}\Big]  \, , \\[2ex]
 {\cal L}_{\rm YM}&=&-\frac{F_{\mu\nu}F^{\mu\nu}}{16\pi} \, ,
\eea
\end{subequations}
with the covariant derivative $D_\mu \varphi_i = (\partial_\mu+iq_iA_\mu)\varphi_i$, where $A_\mu$ is the gauge field and $q_1$, $q_2$ the electric charges, with 
the complex scalar fields $\varphi_1$, $\varphi_2$, the mass parameters  $m_i\ge 0$, the self-coupling constants $\lambda_i>0$, and the field strength tensor 
$F_{\mu\nu}=\p_\mu A_\nu-\p_\nu A_\mu$. We have included two types of cross-couplings between the fields: a density coupling with
dimensionless coupling constant $h$, and a derivative coupling which allows for two different structures with coupling constants $g_1$ and $g_2$ of mass dimension $-2$.
Due to this derivative coupling, the model is non-renormalizable and an ultra-violet cutoff is required in general. However, in our Ginzburg-Landau-like study 
we are only interested in an effective potential for which the only occurring momentum integral is made finite by nonzero temperature. Therefore, the non-renormalizability 
will not play any role in the following.    
The chemical potentials $\mu_1$ and $\mu_2$ are introduced in the usual way, they can be formally included in the Lagrangian as temporal components of the gauge fields
in the covariant derivatives, $q_iA_0\to -\mu_i$, including the covariant derivatives in the coupling terms \cite{Haber:2015exa}. In isolation, each of the fields would 
form a Bose-Einstein condensate if $\mu_i>m_i$. We parametrize the condensates by their moduli $\rho_i$ and their phases
$\psi_i$,
\be
\langle \vf_i \rangle = \frac{\rho_i}{\sqrt{2}}e^{-i\psi_i} \, .
\ee
Since we are interested in a superconductor coupled to a superfluid, we assume only one of the fields to be charged, say field 1, and the second to be neutral, 
\be
q\equiv q_1 \, , \qquad q_2=0 \, .
\ee
Moreover, we are only interested in static solutions and thus drop all time derivatives. Then, the zero-temperature tree-level potential 
$U=-{\cal L}_{\varphi_i\to\langle \varphi_i \rangle}$ is
\bea \label{Ux}
U(\vec{r}) &=& \frac{(\nabla\rho_1)^2}{2}+\frac{(\nabla\rho_2)^2}{2}- \frac{\mu_1^2-(\nabla\psi_1-q\vec{A})^2-m_{1}^2}{2}\rho_1^2-
\frac{\mu_2^2-(\nabla\psi_2)^2-m_{2}^2}{2}\rho_2^2+\frac{\lambda_1}{4}\rho_1^4 +\frac{\lambda_2}{4}\rho_2^4 \non[2ex]
&&-\frac{h+g\mu_1\mu_2}{2}\rho_1^2\rho_2^2 -\frac{G}{2}\rho_1\rho_2\nabla\rho_1\cdot\nabla\rho_2+\frac{g}{2}\rho_1^2\rho_2^2(\nabla\psi_1-q\vec{A})\cdot\nabla\psi_2 +\frac{B^2}{8\pi} \, ,
\eea
where we have reduced the Yang-Mills contribution to a purely magnetic term, $\vec{B} = \nabla\times\vec{A}$, and where we have introduced the abbreviations 
\be \label{Gg}
G\equiv \frac{g_1+g_2}{2} \, ,\qquad g\equiv \frac{g_1-g_2}{2} \,.
\ee
Boundedness of the tree-level potential requires $h+g\mu_1\mu_2<\sqrt{\lambda_1\lambda_2}$. In the remainder of the paper, we shall set 
$g=0$, mainly for the sake of simplicity\footnote{In Ref.\ \cite{Alford:2007np} the terms proportional to $g$ were not included from the beginning. In 
Ref.\ \cite{Haber:2015exa}, which did not discuss vortex solutions,  only the tree-level potential with $\nabla\rho_i=0$ was used, such that $G$ dropped out and $g$ was the only relevant derivative coupling. }. 
Some of our results would become more complicated with a nonzero $g$,
for example the large-temperature expansion in Sec.\ \ref{sec:temperature}. Also, by reducing the number of parameters, the parameter space of our model becomes 
a little less unwieldy. On the other hand, at least for zero temperature, $g$ does not play an important role for the magnetic flux tube
profiles because our solutions will not include any circulation of the neutral condensate, $\nabla\psi_2=0$, in which case we see from Eq.\ (\ref{Ux}) that $g$ appears merely as a 
modification of the density coupling $h$.   

\begin{center}
\begin{figure} [t]
\begin{center}
\includegraphics[width=0.35\textwidth]{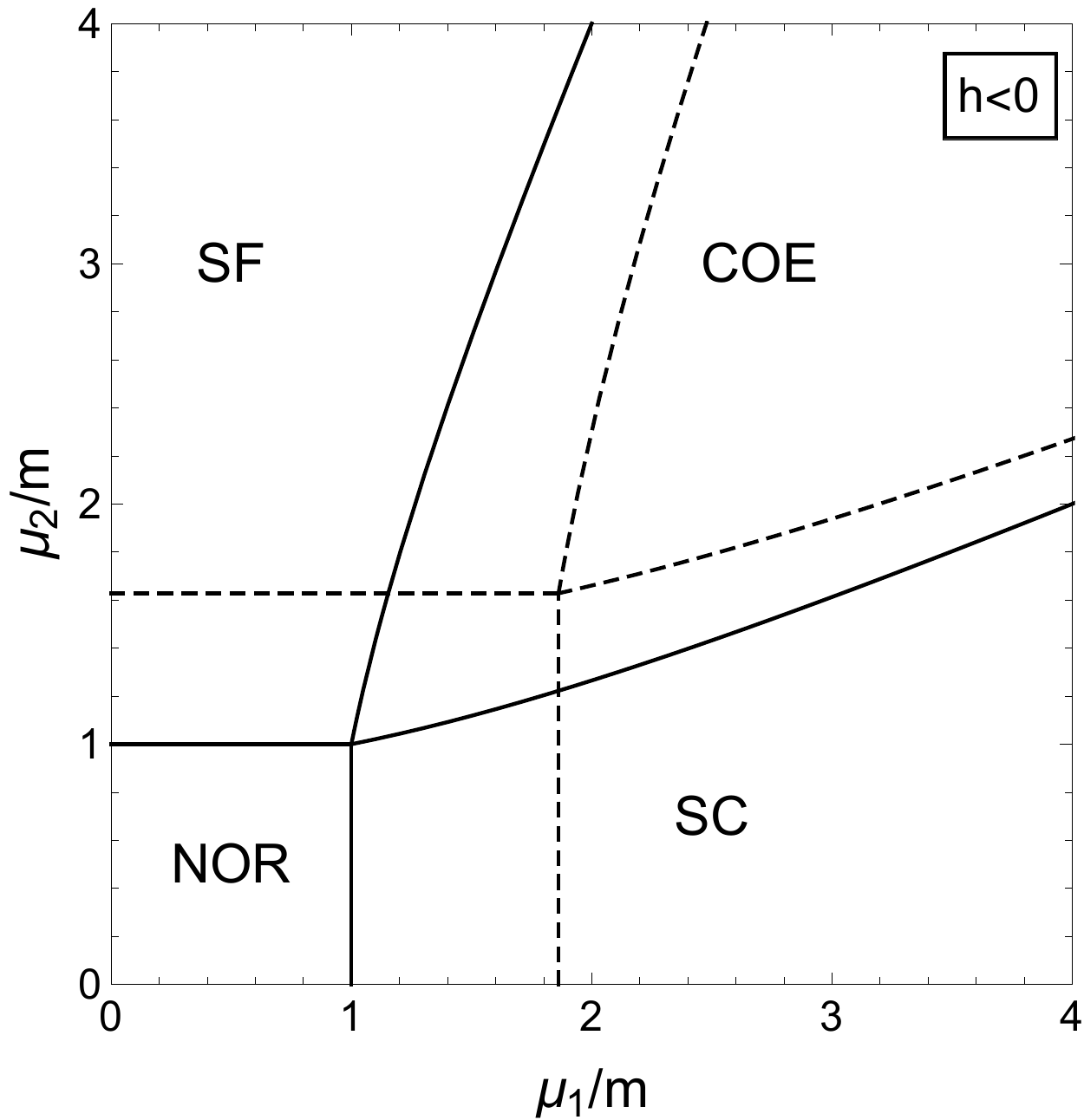}\hspace{2cm}\includegraphics[width=0.35\textwidth]{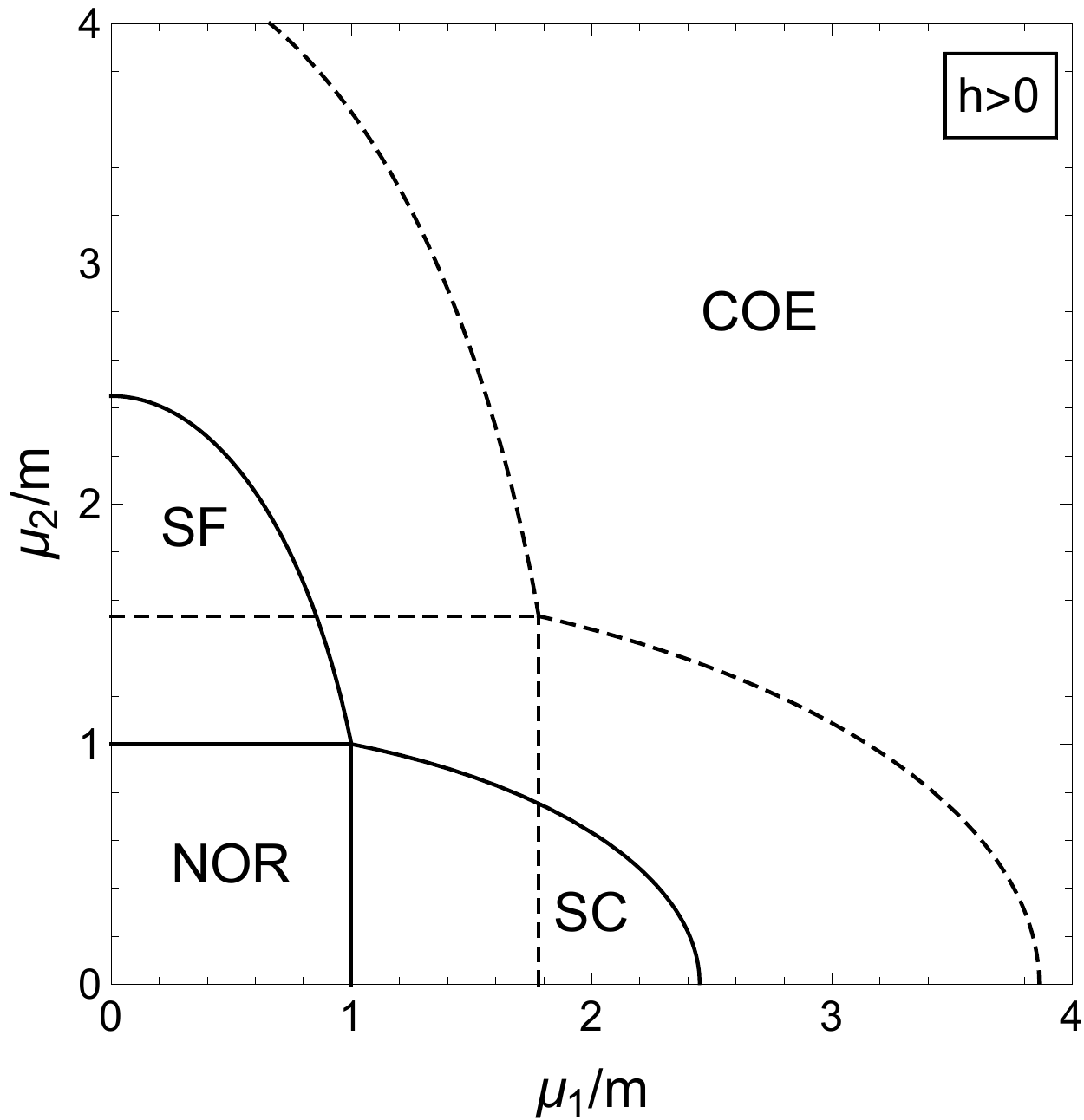}
\caption{Phases in the $\mu_1$-$\mu_2$-plane at zero temperature (solid curves) and nonzero temperature (dashed curves). All lines are second-order phase transitions. 
The density coupling disfavors ($h<0$, left panel) or favors ($h>0$, right panel) the COE phase. The effect of temperature on the SC and SF phases is 
asymmetric, even for identical self-coupling constants, because the thermal mass of the charged field depends on $q$. 
The values of the parameters are $q=2e$, $m_1=m_2\equiv m$, $\lambda_1 = \lambda_2 = 0.5$, $T=3m$, $h =\pm 0.1$, $G=0$.
}
\label{fig:phases}
\end{center}
\end{figure}
\end{center}

To define the possible phases of the system and establish the notation for their condensates, we start with the simplest case of spatially uniform condensates in the absence of a magnetic field, $\nabla\rho_1=\nabla\rho_2=\nabla\psi_1=\nabla\psi_2=\vec{A}=0$. As a consequence of these assumptions, the potential becomes independent of $G$. 
The local minima of the potential yield the possible phases, i.e., we need to solve the algebraic equations 
\be
\frac{\p U}{\p\rho_1}=0\, ,\qquad \frac{\p U}{\p\rho_2}=0\, ,
\ee 
which allow for the following solutions.
\begin{itemize}
\item In the normal phase ("NOR"), neither the charged nor the neutral field condenses,
\be \label{NOR}
\rho_1=\rho_2=0 \, , \qquad U_{\rm NOR}=0 \, .
\ee
\item In the (pure) superconductor ("SC"), only the charged field forms a condensate, whereas the condensate of the other field is zero,
\bea 
\rho_1^2 &=& \rho_{\rm SC}^2 \equiv\frac{\mu_1^2-m_{1}^2}{\lambda_1}  \, , \qquad  \rho_2=0\, , \qquad U_{\rm SC}=-\frac{\l_1\rho_{\rm SC}^4}{4} \, .\label{eq:rhoSC}
\eea
\item In the (pure) superfluid ("SF"), only the neutral field forms a condensate, while the charged fields remains uncondensed, 
\bea 
\rho_2^2 &=&  \rho_{\rm SF}^2 \equiv \frac{\mu_2^2-m_{2}^2}{\lambda_2} \, , \qquad  \rho_1=0\, , \qquad U_{\rm SF}=-\frac{\l_2\rho_{\rm SF}^4}{4} \, .\label{eq:rhoSF} 
\eea
\item In the coexistence phase ("COE"), both condensates exist simultaneously. Without coupling, the coexistence phase is realized if and only if 
both chemical potentials are larger than the corresponding masses. The coupling favors ($h>0$) or disfavors ($h<0$) the COE phase. The condensates and the free energy density are 
\begin{subequations}\label{COE}
\bea 
\rho_{1}^2 &=&   \rho_{01}^2\equiv\frac{\lambda_2(\lambda_1\rho_{\rm SC}^2+h\rho_{\rm SF}^2)}{\lambda_1\lambda_2-h^2} \, , \qquad 
\rho_{2}^2 = \rho_{02}^2 \equiv \frac{\lambda_1(\lambda_2\rho_{\rm SF}^2+h\rho_{\rm SC}^2)}{\lambda_1\lambda_2-h^2} \, ,\\[2ex]
U_{\rm COE} &=& -\frac{\lambda_1\lambda_2(\lambda_1\rho_{\rm SC}^4+\lambda_2\rho_{\rm SF}^4+2h\rho_{\rm SC}^2\rho_{\rm SF}^2)}{4(\lambda_1\lambda_2-h^2)} \, .
\label{UCOE}
\eea
\end{subequations}
\end{itemize}
The ground state is then found by determining the global minimum of $U$. The resulting phase diagram in the $\mu_1$-$\mu_2$ plane is shown in Fig.\ \ref{fig:phases},
for both signs of the coupling $h$. The figure also contains the phase transitions at nonzero temperature, which we discuss now. 

\subsection{Introducing temperature}
\label{sec:temperature}

We intend to include temperature $T$ into the potential (\ref{Ux}) in an effective way. In Ginzburg-Landau models this is usually done by 
introducing $T$-dependent coefficients, with a $T$-dependence that is strictly valid only close to the critical temperature. In our system, the form of these coefficients 
is not obvious because we have two fields and hence (at least) two critical temperatures. We thus proceed by introducing temperature in our underlying field theory 
and derive an effective potential. This will be done in a high-temperature approximation, assuming the condensates to be uniform, and without background magnetic field. 
Once we have derived the $T$-dependent Ginzburg-Landau potential, we shall reinstate the magnetic field for our discussion of the phase diagram and allow for
spatially varying condensates and gauge fields in a flux tube. Neglecting zero-temperature quantum 
corrections, the one-loop potential is 
\be
\label{eq:LOfiniteT}
\Omega(\mu_1,\mu_2,T) = U+T\sum_{i=1}^{6}\int\frac{d^3k}{(2\pi)^3}\ln\big(1-e^{-\epsilon_{ki}/T}\big) \, ,
\ee
where the sum is taken over all 6 quasiparticle excitations $\epsilon_{ki}$.  Without condensation, each of the complex scalar fields yields 2 excitations 
(both massive if $m_i>0$), corresponding to particle and anti-particle excitations, while the gauge field has two massless excitations, corresponding to the two possible polarizations of massless photons. These are 6 modes in total. 
In the coexistence phase both scalar fields condense. As a consequence, there is one Goldstone mode from the 
neutral field and one would-be Goldstone boson from the charged field, which becomes a third mode of the now massive gauge field. Together with the two massive modes from the scalar fields and the two original modes of the gauge field -- which are now massive as well -- these are again 6 modes. 
The excitations $\epsilon_{ki}$ are computed from the tree-level propagator. Their expressions are very complicated, but for the high-$T$ approximation we only need their behavior 
at large momenta. All details of this calculation are deferred to appendix \ref{app:prop}. From a field-theoretical perspective our high-$T$ approximation is very crude, and for a
quantitative evaluation of the model for all temperatures more sophisticated methods are needed, such as the two-particle irreducible formalism \cite{Alford:2013koa} or functional 
renormalization group techniques \cite{Fejos:2016wza}. These methods are beyond the scope of the present work because, firstly, if applied to our present context of magnetic flux tubes and their interactions, they would render the calculation much more complicated and purely numerical methods would be required. Secondly, having in mind the application of our model to nuclear matter, the next step towards a more sophisticated description should probably be to employ a fermionic model, rather than improving the bosonic one (note for instance that our bosonic system has well-defined quasiparticle excitations for all energies, while a fermionic one has 
a continuous spectral density for energies larger than twice the energy gap from Cooper pairing). 

We also simplify the result by only keeping the leading order contribution from the derivative coupling $G$.  As a result, all temperature corrections can be absorbed into thermal masses and a thermal density coupling, and we can work with the effective potential
\bea \label{UxT}
U(\vec{r}) &\simeq& \frac{(\nabla\rho_1)^2}{2}+\frac{(\nabla\rho_2)^2}{2}- \frac{\mu_1^2-(\nabla\psi_1-q\vec{A})^2-m_{1,T}^2}{2}\rho_1^2-
\frac{\mu_2^2-(\nabla\psi_2)^2-m_{2,T}^2}{2}\rho_2^2+\frac{\lambda_1}{4}\rho_1^4 +\frac{\lambda_2}{4}\rho_2^4 \non[2ex]
&&-\frac{h_T}{2}\rho_1^2\rho_2^2 -\frac{G}{2}\rho_1\rho_2\nabla\rho_1\cdot\nabla\rho_2 +\frac{B^2}{8\pi} \, ,
\eea
where 
 \begin{subequations}\label{thermal_mass}
\bea 
m_{1,T}^2 &=& m_1^2+\frac{2\l_1-h+6\pi q^2}{6}T^2\, ,\\[2ex]
m_{2,T}^2 &=& m_2^2+\frac{2\l_2-h}{6}T^2\, , \\[2ex]
h_T&=& h\left(1+\frac{GT^2}{6}\right) \, .
 \eea
 \end{subequations}
For the following, we can thus simply take Eqs.\ (\ref{NOR}) -- (\ref{COE}) and replace the masses and the density coupling by their thermal generalizations. 
The effect of nonzero temperature on the phase structure is shown in Fig.\ \ref{fig:phases}. 
Before we use the potential to compute the critical magnetic fields, we briefly comment on the critical temperatures of the coexistence phase without external magnetic field. 
In the presence of a derivative coupling $G$ the 
resulting expressions are very lengthy and not very insightful. Therefore, we set $G=0$ for the moment, such that the only effect of temperature is a modification of the 
masses $m_1$ and $m_2$. Inserting the thermal masses into Eqs.\ (\ref{COE}), we compute the
$T$-dependent condensates
\be
\rho_{0i}^2(T)=\rho_{0i}^2(T=0)\left(1-\frac{T^2}{T_{ci}^2}\right) \, , 
\ee
where the critical temperatures $T_{c1}$ and $T_{c2}$ indicate the phase transitions to the SF and SC phases, 
 \begin{subequations} \label{TcCOE}
 \bea 
T_{c1}^2&=&\frac{6(\lambda_1\lambda_2-h^2)}{\lambda_2(2\lambda_1+h+6\pi q^2)-h^2} \rho_{01}^2(T=0) \, , \label{Tc1} \\[2ex]
T_{c2}^2&=&\frac{6(\lambda_1\lambda_2-h^2)}{\lambda_1(2\lambda_2+h)-h(h-6\pi q^2)} \rho_{02}^2(T=0) \, .\label{Tc2}
\eea
\end{subequations}
In the limit $h=0$,  Eq.\ (\ref{Tc1}) reduces to the well-known result for a single charged field, see for instance Eq.\ (4.24) in Ref.\ \cite{Kapusta:1981aa} (in this 
reference Heaviside-Lorentz units are used, i.e., our charge $q$ has to be divided by $\sqrt{4\pi}$ to match that result exactly). If we set $h=0$ in Eq.\ (\ref{Tc2}) 
the result becomes independent of the charge $q$, as it should be because field 2 is neutral and couples to the gauge field only indirectly through field 1.
 
The critical temperatures (\ref{TcCOE}) and their more complicated versions with nonzero $G$ are interesting in themselves. For instance, they can be used 
to analyze systematically in which regions of parameter space the COE phase is superseded by the SF phase at high temperature (i.e., the charged condensate 
melts first, $T_{c1}<T_{c2}$) or by the SC phase (i.e., the neutral condensate melts first, $T_{c2}<T_{c1}$). Or, they can be used to 
identify regions in the parameter space where one or both critical temperatures squared become negative, indicating that one or both condensates "refuse" to melt.
This interesting observation -- although it may 
be an artifact of our approximation -- has been pointed out previously in the literature, see for instance appendix C in Ref.\ \cite{Alford:2007qa} and references therein. 
Here we shall not further analyze the critical temperatures and proceed with our main concern, phases at nonzero external magnetic field. 
None of the parameter sets we shall use in the following 
show this unusual behavior, i.e., we choose parameters such that $T_{c1}$ and $T_{c2}$ exist.

\section{Critical magnetic fields}
\label{sec:critical_fields}

The free energy can be computed from the potential (\ref{UxT}),
\be
F = \int d^3r \,U(\vec{r}) \, .
\ee
Since we are interested in the phase structure at fixed external (and homogeneous) magnetic field $\vec{H} = H\vec{e}_z$, we need to consider the Gibbs free energy
\be \label{Gibbsdef}
{\cal G} = F - \frac{\vec{H}}{4\pi}\cdot\int d^3r \, \vec{B} \, .
\ee
To determine the complete phase diagram, we would 
have to compute the Gibbs free energy for all possible phases at each point in the phase space given by the thermodynamic variables
$(\mu_1,\mu_2,T,H)$. The possible phases are the NOR, SF, SC, and COE phases listed above, and for the phases that are superconducting (SC and COE) we have 
to distinguish the Meissner phase, in which the magnetic field is completely expelled, $\vec{B}=0$, from the flux tube phase, where a lattice of magnetic flux tubes is formed,
admitting part of the applied magnetic field in the superconductor. We shall simplify this problem by not computing the Gibbs free energy for the flux tube phase in full generality, which 
would require us to determine the spatial profile of the condensate and the magnetic field, including the preferred lattice structure, fully dynamically. 
Instead -- following the usual textbook treatment \cite{tinkham2004introduction} -- we shall compute the critical magnetic fields $H_{c1}$, $H_{c2}$, and $H_c$, 
although they do not provide complete information 
of the phase diagram, not even for a single-component superconductor. To interpret their meaning for the phase diagram (in particular in our two-component system)
it is important to precisely recall how they are computed, and thus we start each of the following three subsections with the definition of the corresponding critical magnetic field
before we compute them for our system. 
In general, when we speak of the superconducting phase, this can be either the COE or the SC phase, while the normal-conducting phase can either be NOR or SF.
The concrete calculations will always be done for the most interesting case, where both charged and neutral condensates exist in the superconducting phase (COE)
and the normal conductor is the pure superfluid (SF).  The critical magnetic fields for the transition between the COE and NOR and 
between the SC and NOR phases are not needed for our main results, but can be computed analogously. The latter appears 
to be the standard textbook scenario. However, in our two-component system it is conceivable that in the SC phase a neutral condensate is induced in the center of a flux tube \cite{Forgacs:2016ndn,Forgacs:2016iva}. Therefore,  
the pure superconductor SC might acquire a superfluid admixture, which can affect  the critical magnetic fields for the transition to the 
completely uncondensed phase (NOR). In the present paper, we shall only consider flux tube
solutions that approach the COE phase, not the SC phase, far away from the center of the flux tube.

\subsection{Critical magnetic field $H_c$}
\label{sec:Hc}

{\it Definition.} The critical magnetic field $H_{c}$ is the magnetic field at which the Gibbs free energies of the superconducting phase in the Meissner 
state and the normal-conducting phase  are identical, resulting in a first-order phase transition between them.

\bigskip
The Gibbs free energy of the 
COE phase with complete expulsion of the magnetic field is
\be \label{GibbsCOE}
{\cal G}_{\rm COE} = VU_{\rm COE} \, ,
\ee
where $V$ is the total volume of the system and $U_{\rm COE}$ is the free energy density from Eq.\ (\ref{UCOE}), with the masses $m_1$, $m_2$ and the coupling $h$
replaced by their thermal generalizations $m_{1,T}$, $m_{2,T}$, $h_T$. We neglect any magnetization in the normal-conducting phases, 
and thus $\vec{B}=\vec{H}$ in the SF phase, which yields the Gibbs free energy 
\be \label{GibbsSF}
{\cal G}_{\rm SF} = V\left(U_{\rm SF} -\frac{H^2}{8\pi}\right) \, ,
\ee
with $U_{\rm SF}$ from Eq.\ (\ref{eq:rhoSF}). Note that the $H^2$ term is a sum of the magnetic energy $\propto B^2$ and the term $\propto HB$ in the Legendre transformation 
from the free energy $F$ to the Gibbs free energy ${\cal G}$. Therefore, the critical magnetic field, defined by ${\cal G}_{\rm COE} = {\cal G}_{\rm SF}$, becomes
\be \label{Hc}
H_c = \sqrt{8\pi(U_{\rm SF}-U_{\rm COE})} =  2\pi q\sqrt{2}\kappa\sqrt{1-\frac{h_T^2}{\l_1\l_2}} \rho^2_{01}\, .
\ee
Here we have introduced the Ginzburg-Landau parameter 
\be 
\kappa=\frac{\ell}{\xi}=\sqrt{\frac{\l_1}{4\pi q^2}} \, ,
\ee
with the magnetic penetration depth $\ell$  and the coherence length $\xi$, 
\be\label{ellxi}
\ell=\frac{1}{\sqrt{4\pi q^2}\rho_{01}} \, , \qquad \xi=\frac{1}{\sqrt{\l_1}\rho_{01}} \, .
\ee

\subsection{Critical magnetic field $H_{c2}$}
\label{sec:Hc2}

{\it Definition.} Suppose there is a second-order phase transition between the superconductor in the flux tube phase and the normal-conducting phase, such that 
the equations of motion can be linearized in the charged condensate.
Then, the critical magnetic field $H_{c2}$ is the maximal magnetic field allowed by the equations of motion. $H_{c2}$ is a {\it lower} bound for the actual transition from the flux tube phase to the normal-conducting phase because it does not exclude a first-order transition at some larger $H$. 
We call the critical field for such a first-order transition $H_{c2}'$. 

\bigskip

By definition, as we approach $H_{c2}$, the charged condensate approaches zero and the neutral condensate approaches the condensate of the SF phase.
For magnetic fields $H$ close to and smaller than $H_{c2}$, we can write the condensates and the gauge field as their values at $H_{c2}$ plus small perturbations.
Then, for the calculation of $H_{c2}$ itself the equations of motion linear in the charged condensate are sufficient. 
We are also interested in checking whether and in which parameter regime 
the flux tube phase is energetically preferred just below $H_{c2}$. This is done within the same calculation, but taking 
into account higher order terms in the equations of motion and the free energy. This calculation is somewhat lengthy 
and is explained in appendix \ref{app:Hc2}. Here we  summarize the results.
The critical magnetic field becomes
\be\label{HcHc2}
H_{c2} = \frac{1}{q\xi^2}\left(1-\frac{h_T^2}{\lambda_1\lambda_2}\right)=\sqrt{2}\kappa\sqrt{1-\frac{h_T^2}{\l_1\l_2}}H_c\, ,
\ee
where the second expression relates $H_{c2}$ to $H_c$ by using Eq.\ (\ref{Hc}). At zero temperature, $H_{c2}$ 
does not depend on the gradient coupling $G$. However, the difference in Gibbs free energies between the superconducting and the normal-conducting phases does depend on $G$, see Eq.\ (\ref{DeltaGfull}). For $G=0$ we have
\be\label{DeltaG}
\frac{{\cal G}_{\rm COE}}{V} = \frac{{\cal G}_{\rm SF}}{V} + \lambda_1\langle\bar{\varphi}_1^4\rangle\left[\frac{1}{2\kappa^2}-1+\frac{h^2}{\lambda_1\lambda_2}
{\cal I}_1(p)\right] \, ,
\ee
where $\langle\bar{\varphi}_1^4\rangle $ is the spatial average of the charged condensate (\ref{eq:SEsol}), where 
\be
p^2 = \frac{2\lambda_2\rho_{\rm SF}^2}{qH_{c2}} \, ,
\ee 
and where 
\be
{\cal I}_1(p) \equiv \frac{pe^{p^2/4}}{2\sqrt{2}} \int_{-\infty}^\infty dt\,e^{-t^2}\left\{
e^{pt}\left[1-{\rm erf}\left(\frac{p}{2}+t\right)\right]+e^{-pt}\left[1-{\rm erf}\left(\frac{p}{2}-t\right)\right]\right\} \, ,
\ee
with the error function erf.

In the limit of a single superconductor, $h = 0$, we recover the standard result: in that case, Eq.\ (\ref{HcHc2}) shows that the critical fields $H_c$ and $H_{c2}$ coincide 
at $\kappa^2=1/2$, and Eq.\ (\ref{DeltaG}) shows that the flux tube phase is preferred, ${\cal G}_{\rm COE}<{\cal G}_{\rm SF}$, if and only if $\kappa^2>1/2$. 
In the coupled system the situation is more complicated. Now, from Eq.\ (\ref{HcHc2}) we see that $H_c$ and $H_{c2}$ coincide at a larger value of $\kappa$ 
(since $h^2<\lambda_1\lambda_2$ to ensure the boundedness  of the  potential for $h>0$ and to ensure the existence of the COE phase for $h<0$, the square root is always real and smaller than 1). This appears to take away phase space from the flux tube phase. However, from Eq.\ (\ref{DeltaG}) we see that the difference in Gibbs free energies between the COE and the SF phases changes sign at a different point,
and this point is given not just by the coupling constant $h$, but also depends on $p$, i.e., on the magnitude of the neutral condensate $\rho_{\rm SF}$ compared to the 
square root of the critical magnetic field $H_{c2}$. 
Despite this dependence we can make a general statement: we find $0\le {\cal I}_1(p) <1$, and thus the factor ${\cal I}_1(p)$ weakens the 
effect of the term $h^2/(\lambda_1\lambda_2)$. At the value of $\kappa$ where $H_c$ and $H_{c2}$ are equal, the superconducting phase is preferred and -- for all $p$ --
remains preferred along $H_{c2}$, until the smaller $\kappa$ defined through Eq.\ (\ref{DeltaG}) is reached. This observation is indicative of 
the complications at the transition between type-I and type-II superconductivity in the two-component system, and we shall find further discrepancies to the standard scenario 
when we compute the critical field $H_{c1}$.

Anticipating the numerical results in Sec.\ \ref{sec:phases}, let us comment on a possible first-order phase transition at $H'_{c2}$, as mentioned in the definition at the 
beginning of this section. Suppose we are in a parameter region where the flux tube phase is favored just below $H_{c2}$, i.e., let $\kappa$ be larger than the critical $\kappa$ 
defined through Eq.\ (\ref{DeltaG}). Then, any phase transition from the flux tube phase to the normal phase at a critical field smaller than $H_{c2}$ is excluded because we know that the 
system prefers to be in the flux tube phase just below $H_{c2}$ (here we ignore the very exotic possibility that the system quits the flux tube phase and then re-enters it below $H_{c2}$). A phase transition at a critical magnetic field larger than $H_{c2}$ -- instead of the one at $H_{c2}$ -- is however possible. This phase transition must be of first order 
because by definition $H_{c2}$ is the largest magnetic field at which a second-order transition may occur. Putting these arguments together leads to the conclusion that $H_{c2}$ is a lower bound for the transition from the flux tube phase to the normal phase, possibly replaced by a first order transition at $H'_{c2}>H_{c2}$. Our numerical results
will indeed suggest such a first-order phase transition. However, we shall find $H'_{c2}<H_{c2}$, which, as we will explain, is an artifact of the approximation we apply 
for the interaction between flux tubes.  Nevertheless, our result will allow us to speculate about the correct critical field $H'_{c2}$, obtained in a more complete calculation 
that goes beyond our approximation.

\subsection{Critical magnetic field $H_{c1}$}
\label{sec:Hc1}

{\it Definition.} The critical magnetic field $H_{c1}$ is the magnetic field at which it becomes energetically favorable to 
put a single flux tube into the superconductor in the Meissner phase, resulting in a second-order phase transition from the Meissner phase into the 
flux tube phase. $H_{c1}$ is an {\it upper} bound for this transition because there can be a first-order transition at some smaller $H$, i.e., it can be 
favorable to directly form a flux tube lattice with a finite, not infinite, distance between the flux tubes. We call this first-order critical field $H_{c1}'$. 

\bigskip
According to the definition (\ref{Gibbsdef}), the Gibbs free energy for the COE phase with a single magnetic flux tube is 
\be \label{Gibbsfl}
\mathcal{G}_{\rm COE}^{\circlearrowleft}=VU_{\rm COE}+F_{\circlearrowleft}-\frac{Hn\Phi_0}{4\pi}  L \, ,
\ee
where $F_{\circlearrowleft}$ is the free energy of the flux tube, and where we have used 
\be \label{drB}
\int d^3r \, B = n\Phi_0 L \, ,
\ee
with the winding number $n$ of the flux tube, the length of the flux tube $L$, and the fundamental flux quantum $\Phi_0 = 2\pi/q$. Placing a single flux tube into the system results in a loss in 
(negative) condensation energy, and thus the free energy increases. However, at fixed magnetic field $H$, there is an energy gain from allowing magnetic flux into the system. 
As a consequence, there is a competition between these two contributions of opposite sign in Eq.\ (\ref{Gibbsfl}). 
At the critical point, the two contributions exactly cancel each other, 
\be 	\label{Hc1}
H_{c1}=\frac{2q}{n}\frac{F_{\circlearrowleft}}{L} \,.
\ee
The calculation of $H_{c1}$ thus amounts to the calculation of the free energy of a single flux tube $F_{\circlearrowleft}$, for which we can largely follow Ref.\ \citep{Alford:2007np}. 
We work in cylindrical coordinates, $\vec{r}=(r,z,\theta)$, and make the following, radially symmetric, ansatz for the condensates, 
\be 
\rho_i(r)=\rho_{0i} f_i(r)\, , \qquad \psi_1(\theta) = n\theta \, , \qquad \psi_2=0 \, ,
 \ee
and the gauge field
\be
\vec{A}(r)=\frac{na(r)}{qr}\vec{e}_{\theta}\quad \Rightarrow \qquad \vec{B}(r)=\frac{n}{qr}\frac{\partial a}{\partial r}\vec{e}_z \, .
\ee 
The profile functions $f_i$ and $a$ have to be computed numerically. Their boundary conditions are $f_{i}(\infty)=a(\infty)=1$, $f_1(0)=0$, 
and $\partial_rf_2(\infty)=\partial_r a(\infty)=0$, such that the condensates approach their homogeneous values $\rho_{0i}$ and the magnetic field vanishes far 
away from the center of the flux tube. The values of the neutral condensate and the gauge field at the center of the flux tube are determined dynamically.
We have set the winding number of the neutral condensate to zero because the flux tube does not induce a superfluid 
vortex \citep{Alford:2007np}. 

We insert our ansatz into the potential (\ref{UxT}) and separate the potential of the homogeneous COE phase,
\be
U(\vec{r}) =  U_{\circlearrowleft}(\vec{r}) + U_{\rm COE} \, ,
\ee
with
\be
U_{\rm COE} = -\frac{\mu_1^2-m_{1,T}^2}{2}\rho_{01}^2 -\frac{\mu_2^2-m_{2,T}^2}{2}\rho_{02}^2 +\frac{\lambda_1}{4}\rho_{01}^4 +\frac{\lambda_2}{4}\rho_{02}^4 
-\frac{h_T}{2}\rho_{01}^2\rho_{02}^2 \, .
\ee
To write the free energy of the flux tube in a convenient form, we introduce the dimensionless variable
\be
R=\frac{r}{\xi} \, ,
\ee
abbreviate the dimensionless gradient coupling by 
\be
\Gamma\equiv G\rho_{01}\rho_{02} \, ,
\ee
and the ratio of neutral over charged condensate by 
\be
x\equiv\frac{\rho_{02}}{\rho_{01}} \, .
\ee
It is also useful to write $\mu_1^2-m_{1,T}^2 = \lambda_1\rho_{\rm SC}^2 = \lambda_1\rho_{01}^2-h_T\rho_{02}^2$ and $\mu_2^2-m_{2,T}^2 = \lambda_2\rho_{\rm SF}^2 = \lambda_2\rho_{02}^2-h_T\rho_{01}^2$, which follows from Eq.\ (\ref{COE}). Then, we obtain the free energy per unit length
\bea \label{Efl}
\frac{F_{\circlearrowleft}}{L} &=& \frac{1}{L}\int d^3r\, U_{\circlearrowleft}(\vec{r}) \non[2ex]
&=&  \pi \rho_{01}^2 \int_0^\infty dR\,R\left\{\frac{n^2\kappa^2 a'^2}{R^2}+f_1'^2+f_1^2\frac{n^2(1-a)^2}{R^2}+\frac{(1-f_1^2)^2}{2} +x^2
\left[f_2'^2 + \frac{\lambda_2}{\lambda_1}x^2\frac{(1-f_2^2)^2}{2}\right]\right. \non[2ex]
&& \left.-\frac{h_T}{\lambda_1} x^2 (1-f_1^2)(1-f_2^2)-\Gamma x f_1f_2f_1'f_2'
 \right\} \, , 
\eea
where prime denotes derivative with respect to $R$.
This yields the equations of motion for $a$, $f_1$, $f_2$, 
\begin{subequations} \label{eomA}
\bea
a''-\frac{a'}{ R}&=&-\frac{f_1^2}{\kappa^2}\left(1-a\right)  \, , \label{eoma}\\[2ex]
0&=&f_1''+\frac{f_1'}{ R}+ f_1\left[1-f_1^2-\frac{n^2(1-a)^2}{R^2}\right] -\frac{h_T}{\lambda_1} x^2 f_1(1-f_2^2)
 - \frac{\Gamma x}{2}f_1\left[f_2'^2+f_2\left(f_2''+\frac{f_2'}{ R}\right)\right]  \, , \\[2ex]
0&=&f_2''+\frac{f_2'}{R}+ f_2\frac{\lambda_2}{\lambda_1}x^2\left(1-f_2^2\right)- \frac{h_T}{\lambda_1} f_2\left(1-f_1^2\right)-\frac{\Gamma}{2x}
f_2\left[f_1'^2+f_1\left(f_1''+\frac{f_1'}{ R}\right)\right] \, . 
\eea
\end{subequations}
We solve these equations numerically with a successive over-relaxation method. The profiles themselves have been discussed in detail in 
Ref.\ \cite{Alford:2007np}\footnote{Eqs.\ (\ref{eomA}) are identical to Eqs.\ (16) in Ref.\ \cite{Alford:2007np}  if we identify
\be
\frac{\Gamma}{2} \leftrightarrow \sigma \, , \qquad x \leftrightarrow \frac{\langle\phi_n\rangle}{\langle\phi_p\rangle} \, , \qquad \frac{h_T}{\lambda_1} 
\leftrightarrow -\frac{a_{pn}}{a_{pp}}  \, , \qquad \frac{\lambda_2}{\lambda_1} \leftrightarrow \frac{a_{nn}}{a_{pp}} \, . \nonumber
\ee}, and we do not further comment on them. Instead we continue with the asymptotic solution, which will be needed later.   

Far away from the center of the flux tube, all profile functions are close to one. Therefore, we write 
\bea \label{aff}
a(R) &=& 1+ Rv(R) \, , \qquad 
f_1(R) = 1+ u_1(R) \, , \qquad 
f_2(R) = 1+u_2(R) \, ,
\eea
and linearize the profile equations (\ref{eomA}) in $v$, $u_1$, and $u_2$, 
\begin{subequations} \label{vuM}
\bea
0&\simeq& R^2v''+Rv'-\left(1+\frac{R^2}{\kappa^2}\right) v    \, , \label{vuM1}\\[2ex]
\Delta  u  &\simeq& M u  \, ,
\eea
\end{subequations}
where 
\be
u \equiv \left(\begin{array}{c} u_1 \\[2ex] u_2 \end{array}\right) \, , \qquad 
M \equiv 2\left(\begin{array}{cc} 1 & -\frac{\Gamma x}{2} \\[2ex] -\frac{\Gamma}{2x} & 1 \end{array}\right)^{-1} \left(\begin{array}{cc} 1 & -\frac{h_T}{\lambda_1}x^2  \\[2ex] -\frac{h_T}{\lambda_1} & \frac{\lambda_2}{\lambda_1}x^2 \end{array}\right) \, .
\ee
We can decouple the equations for $u_1$ and $u_2$ by diagonalizing $M$, 
\be
{\rm diag}\,(\nu_+,\nu_-) = U^{-1}MU \, , \qquad U =  \left(\begin{array}{cc} \gamma_+ & \gamma_- \\[2ex] 1 & 1 \end{array}\right) \, , 
\ee
where $\nu_\pm$ are the eigenvalues of $M$ and $(\gamma_\pm,1)$ its eigenvectors, given by
\bea \label{nugam}
\nu_\pm = \frac{\lambda_1+\lambda_2 x^2-h_T\Gamma x\pm{\cal Q}}{\lambda_1(1-\Gamma^2/4)} \, , \qquad \gamma_\pm = \frac{x(\lambda_1-\lambda_2 x^2\pm {\cal Q})}{\lambda_1\Gamma -2h_T x} \, ,
\eea
where ${\cal Q} \equiv[(\lambda_1-\lambda_2 x^2)^2-2h_T\Gamma x(\lambda_1+\lambda_2 x^2)+x^2(4h_T^2+\Gamma^2\lambda_1\lambda_2)]^{1/2}$. 
This yields two uncoupled equations for $\tilde{u}_1$ and $\tilde{u}_2$, where $\tilde{u}=U^{-1}u$, which we solve with the boundary condition 
$\tilde{u}_1(\infty)=\tilde{u}_2(\infty)=0$ (which leaves one integration constant from each equation undetermined). We undo the rotation with $u=U\tilde{u}$, and, together with the
solution to Eq.\ (\ref{vuM1}), insert the 
result into Eq.\ (\ref{aff}) to obtain the asymptotic solutions
\begin{subequations}
\label{asympsol}
\bea
a(R) &\simeq& 1+C R K_1(R/\kappa) \, , \label{Qasym} \\[2ex]
f_1(R) &\simeq& 1+ D_+ \gamma_+K_0(\sqrt{\nu_+}R)+ D_- \gamma_-K_0(\sqrt{\nu_-}R)   \, , \label{f1asym} \\[2ex]
f_2(R) &\simeq& 1+ D_+ K_0(\sqrt{\nu_+}R) + D_-K_0(\sqrt{\nu_-}R)  \, , \label{f2asym}
\eea
\end{subequations}
where $K_0$ and $K_1$ are the modified Bessel functions of the second kind, and the constants $C$, $D_+$, $D_-$ can only be determined numerically by solving the full equations of motion, including the boundary conditions at $R=0$. In deriving the linearized equations (\ref{vuM}), we have not only used $u_1, u_2, v\ll 1$, but also $v^2\ll u_1, u_2$, which implies $e^{-2R/\kappa} \ll e^{-\sqrt{\nu_\pm}R}$. This assumption is violated if $\kappa$ is sufficiently large compared to $1/\sqrt{\nu_{\pm}}$ (compared
to $1/\sqrt{2}$ in a single superconductor), i.e., deep in the type-II regime. Later, when we use the solutions of  
the linearized equations for the interactions between flux tubes, we are only interested in the transition region between type-I and type-II behavior, where $1/\kappa \simeq \sqrt{\nu_\pm}$, i.e., for our purpose the linearization is a valid approximation.

\subsection{Interaction between flux tubes and first-order phase transitions}
\label{sec:inter}

If the phase transitions from the Meissner phase to the flux tube phase and from the flux tube phase to the normal-conducting phase were of second order we would be done. 
The critical magnetic fields of the previous sections would be sufficient to determine the phase structure. We shall see, however, that, due to the presence of the superfluid, 
first-order phase transitions become possible. To this end, we compute the Gibbs free energy of the entire flux tube lattice, rather than only of a single flux tube.
We shall do so in an approximation of flux tube distances much larger than the width of a flux tube. 

We generalize the Gibbs free energy (\ref{Gibbsfl}) to a system with flux tube area density $\nu$ and add a term that takes into account the interaction between the flux tubes
\be \label{Gibbs_lattice}
\frac{\mathcal{G}_{\rm COE}^{\circlearrowleft\circlearrowleft}}{V}\simeq U_{\rm COE}+\frac{n\nu}{2q}(H_{c1}-H)+\frac{t\nu}{2}\frac{F_{\rm int}^{\circlearrowleft}(R_0)}{L}\, ,
\ee
where we have eliminated $F_{\circlearrowleft}$ in favor of $H_{c1}$ with the help of Eq.\ (\ref{Hc1}), and where we have employed the nearest-neighbor approximation for the interaction 
term with the number of nearest neighbors $t$, and the dimensionless lattice constant $R_0$. For a hexagonal lattice, which we shall use in our explicit calculation, 
$t=6$ and $\nu=2/(\sqrt{3}R_0^2)$. The interaction energy $F_{\rm int}^{\circlearrowleft}(R_0)$ is defined by writing the 
total free energy of two flux tubes with distance $R_0$, say 
flux tubes $(a)$ and $(b)$, in terms of the free energy of the flux tubes in isolation plus the interaction energy, 
\be \label{Fintdef}
F_{\circlearrowleft}^{(a)+(b)}=F_{\circlearrowleft}^{(a)}+F_{\circlearrowleft}^{(b)}+F_{\rm int}^{\circlearrowleft}(R_0) \, .
\ee 
We calculate $F_{\rm int}^{\circlearrowleft}(R_0)$ in appendix \ref{app:fl_int} in an approximation that is valid for large $R_0$. This calculation makes use of the method 
first employed in Ref.\ \cite{Kramer:1971zza},  adapted to our two-component system with gradient coupling. All related references mentioned in Sec.\ \ref{sec:relation}
are based on this method or an equivalent one, and our results reproduce the ones of those references in various limits. The result is 
\bea \label{Fint}
\frac{F_{\rm int}^{\circlearrowleft}(R_0)}{L} &\simeq& 2\rho_{01}^2 R_0 \int_{R_0/2}^\infty \frac{dR}{\sqrt{R^2-(R_0/2)^2}}\bigg\{\frac{\kappa^2 n^2 a'(1-a)}{R^2}-(1-f_1)f_1'-x^2(1-f_2)f_2'\non[2ex]
&&\hspace{4cm}+\frac{\Gamma x}{4}(f_1+f_2+f_1f_2-1)[(1-f_1)f_2'+(1-f_2)f_1']\bigg\} \, . 
\eea
As explained in the appendix in more detail, the integration can be reduced to an integral over the plane  that separates the two Wigner-Seitz cells, which, in this simple 
setup, are two half-spaces. Since the integration along the direction of the flux tubes is trivial, we are left with a one-dimensional integral. As a consequence of the
approximation, only the profile functions of a single flux tube appear in the integrand. In the derivation we have also assumed the asymptotic values of the condensates to be identical to the homogeneous values in the Meissner phase, $\rho_{01}$ and $\rho_{02}$.  
We shall later insert our numerical solutions $f_1$, $f_2$, and $a$ into Eq.\ (\ref{Fint}) 
to compute the Gibbs free energy  numerically. Before we do so we extract some simple
analytical results with the help of the asymptotic solutions (\ref{asympsol}). Inserting them into Eq.\ (\ref{Fint}) yields a lengthy expression which is not 
very instructive, especially due to the terms proportional to the gradient coupling. In appendix \ref{app:asymp} we show that a simple expression can be extracted, even including
the gradient coupling, if we restrict ourselves to the leading order contribution at large distances. Here we proceed with the simpler case of vanishing gradient 
coupling, $\Gamma=0$, to obtain straightforwardly 
\be \label{Fintasymp}
\frac{F_{\rm int}^{\circlearrowleft}(R_0)}{L} \simeq2\pi\rho_{01}^2\Big[\kappa^2 n^2 C^2 K_0(R_0/\kappa) - D_+^2(\gamma_+^2+x^2) K_0(R_0\sqrt{\nu_+})- D_-^2(\gamma_-^2+x^2) K_0(R_0\sqrt{\nu_-})\Big] \, ,
\ee
where we have used $\gamma_+\gamma_-+x^2=0$ for $\Gamma=0$, which follows from Eqs.\ (\ref{nugam}), the derivatives $K_1'(x) = -K_0(x) - K_1(x)/x$, $K_0'(x) = -K_1(x)$, and 
the integral 
\be \label{Kint}
\int_{R_0/2}^\infty\frac{dR\, K_0(\alpha R)K_1(\alpha R)}{\sqrt{R^2-(R_0/2)^2}} = \frac{\pi  K_0(\alpha R_0)}{\alpha R_0} \, .
\ee
The result (\ref{Fintasymp}) shows that there is a positive contribution, which makes the flux tubes repel each other due to their magnetic fields, and there is a negative contribution, 
which makes the flux tubes attract each other due to the lower loss of (negative) condensation energy if the flux tubes overlap. Let us first see how the case of a single 
superfluid is recovered by switching off the coupling $h$.  (Since we have set $\Gamma=0$, there is no temperature dependence left in $h_T$ and we drop the subscript $T$ in 
this discussion.) As $h\to 0$, the quantities $\nu_\pm$ and $\gamma_\pm$ go to different limits, depending on the sign of $\lambda_1-\lambda_2x^2$. If $\lambda_2 x^2>\lambda_1$, we have $\gamma_+\sim h$ and $\gamma_-\sim h^{-1}$. Numerically, we find that while $\gamma_-$ diverges, 
the product $D_-\gamma_-$ goes to a finite value. Moreover, $D_+$ goes to zero, such that the attractive terms reduce to $- D_-^2\gamma_-^2 K_0(R_0\sqrt{2})$ since $\nu_-\to 2$ for $h\to 0$. In particular, all dependence on $x$, which contains the neutral condensate, has disappeared, as it should be. If, on the other hand, 
$\lambda_2 x^2<\lambda_1$, we see from Eqs.\ (\ref{nugam}) that now $\gamma_+\sim h^{-1}$ and $\gamma_-\sim h$, and it is the other term, $- D_+^2\gamma_+^2 K_0(R_0\sqrt{2})$, which survives, again reproducing the correct result of a single superconductor. The result can be used to find the sign of the interaction at $R_0\to \infty$, 
i.e., to determine whether the flux tubes repel or attract each other at large distances.
Since the Bessel functions fall off exponentially for large $R_0$, we simply compare the arguments of the Bessel functions of the negative and positive contributions. 
For the single superconductor, the long-distance flux tube interaction is thus attractive for $\kappa^2<1/2$ and repulsive for $\kappa^2>1/2$, i.e., the sign change 
appears exactly at the point where $H_c=H_{c2}$. 

Going back to the full expression (\ref{Fintasymp}) for the two-component system, we compare $\nu_-$ with $1/\kappa^2$, because
$\nu_-<\nu_+$, i.e., the term proportional to $K_0(R_0\sqrt{\nu_-})$ is less suppressed for $R_0\to\infty$. Therefore, the point at which the long-range interaction changes from repulsive to attractive is given by 
\bea\label{sign}
\frac{1}{\kappa^2} &=& 1+\frac{\lambda_2}{\lambda_1}x^2 
-\sqrt{\left(1-\frac{\lambda_2}{\lambda_1} x^2\right)^2 + \frac{4h^2 x^2}{\lambda_1^2}} \non[2ex]
&=& \frac{H_{c2}^2}{\kappa^2 H_c^2} \left[1-\frac{h^2}{\lambda_2^2 x^2}+{\cal O}\left(\frac{1}{x^4}\right)\right] \, .
\eea
By comparing Eq.\ (\ref{sign}) with Eq.\ (\ref{HcHc2}), we see that in the two-component system the long-distance interaction changes its sign at a point different from $H_c=H_{c2}$. 
This is made particularly obvious in the second line of Eq.\ (\ref{sign}), where we have expanded the result for large values of $x$, i.e., for large values of the 
neutral condensate compared to the charged one,  $\rho_{02}/\rho_{01} \gg 1$. This limit is interesting for the interior of neutron stars, where protons are expected to 
contribute only about 10\% to the total baryon number 
density\footnote{In Ref.\ \cite{Buckley:2004ca}, the limit $x \gg1 $ was considered ($n_1/n_2 \ll 1$ in the notation of that reference), and it was argued that the critical 
$\kappa$'s for $H_c=H_{c2}$ and the sign change of the long-range interaction are identical, in agreement with the leading-order contribution of our Eq.\ (\ref{sign}). Ref.\ \cite{Buckley:2004ca}
only considered the near-symmetric situation $\lambda_1=\lambda_2\equiv\lambda$, $h=-\lambda+\delta\lambda$ with $0<\delta\lambda\ll\lambda$ (notice that $h<0$ here). 
In this case, our results show that $H_c=H_{c2}$ occurs at $\kappa^2 \simeq \frac{\lambda}{4\delta\lambda}$ and the sign change in the long-range interaction energy at $\kappa^2\simeq\frac{\lambda}{4\delta\lambda}\frac{1+x^2}{x^2}$. Consequently, even in the near-symmetric situation the two 
critical $\kappa$'s are different and only become identical in the limit $x\gg 1$.
}. 
From Eq.\ (\ref{sign}) we recover $\kappa^2=1/2$ for $h=0$, but only if $\lambda_2 x^2>\lambda_1$. 
The reason is that the limits $R_0\to \infty$ and $h\to 0$ do not commute in general: 
in deriving Eq.\ (\ref{sign}) we have fixed $h$ at a nonzero value and let $R_0\to\infty$, while in our above discussion of the single superconductor, we fixed $R_0$ while first letting 
$h\to 0$.

An attractive long-distance interaction between the flux tubes can have very interesting consequences. Recall that $H_{c1}$ is the magnetic field at which the phase with 
a single flux tube is preferred over the phase with complete field expulsion. 
In other words, at $H_{c1}$ the flux tube density is zero and increases continuously, while the flux tube distance decreases continuously from 
infinity at $H_{c1}$. If the interaction at infinite distances is attractive, the flux tubes do not "want" to form an array with arbitrarily small density. Assuming that the interaction 
always becomes repulsive at short range [which our numerical results confirm if we extrapolate Eq.\ (\ref{Fint}) down to lower distances], there is a minimum in the flux tube - flux tube potential, which corresponds to a favored distance between the flux tubes. As a consequence, the transition from the Meissner phase to 
the flux tube phase occurs at a critical field lower than $H_{c1}$, which we call $H_{c1}'$, at which the flux tube density jumps from 
zero to a nonzero value. An instructive analogy is the onset of nuclear matter as a function of the baryon chemical potential $\mu_B$. If the nucleon - nucleon potential was purely repulsive, there would be a second-order onset at the baryon mass, $\mu_c=m_B$. In reality, there is a binding energy $E_b$, and the baryon onset is a first-order transition at a lower chemical potential $\mu_c'=m_B-E_b$. Here, the role of the chemical potential is played by the external field $H$, the role of the nucleons is played by the flux tubes with mass per unit length  $H_{c1}=2qF_{\circlearrowleft}/(nL)$, and the binding energy is generated by the attractive interaction between the flux tubes.

In the single-component system, this first-order phase transition is not realized because it occurs in the type-I regime.
More precisely, if we were to continue $H_{c1}$ into the type-I regime, then, at $H_{c1}$, it does not matter that the flux tube phase is made more favorable by an attractive interaction 
because the normal-conducting phase is the ground state (under the assumption that the gain in Gibbs free energy is not sufficient to overcome the 
difference to the normal phase). In the two-component system, however, the attractive interaction may exist in the regime where, at $H_{c1}$, the Meissner phase (and the phase with a single flux tube) is already preferred over the normal phase. Hence, any {\it arbitrarily small} binding energy will lead to a first-order phase transition at $H_{c1}'<H_{c1}$. 
As we move along $H_{c1}$ towards smaller values of $\kappa$, i.e., towards the type-I regime, we hit the critical point given by Eq.\ (\ref{sign}), where  
the second-order transition turns into a first-order transition. Since our approximation is accurate for infinitesimally small flux tube densities, our prediction for this point is exact. If we 
then keep moving along $H_{c1}'$, the flux tube density at the transition increases and our results have to be taken with care.

We can directly compute $H_{c1}'$ by equating the Gibbs free energy of the flux tube phase (\ref{Gibbs_lattice}) 
to the Gibbs free energy of the Meissner phase (\ref{GibbsCOE}). In the flux tube phase we have to find the preferred flux tube distance $R_0$ (or, equivalently, the
preferred flux tube density $\nu$), which is given by minimizing the Gibbs free energy. Hence, we compute $H_{c1}'$ by solving the coupled equations
\be \label{Hcprime}
{\cal G}^{\circlearrowleft\circlearrowleft}_{\rm COE} = {\cal G}_{\rm COE} \, , \qquad \frac{\partial {\cal G}^{\circlearrowleft\circlearrowleft}_{\rm COE}}{\partial R_0} = 0 
\ee
for $H$ and $R_0$.
We may use the same method to compute a potential first-order phase transition from the flux tube phase to the normal-conducting phase, i.e., in the free energy comparison
we replace ${\cal G}_{\rm COE}$ with ${\cal G}_{\rm SF}$ from Eq.\ (\ref{GibbsSF}) and compute the resulting critical field $H_{c2}'$. 

\section{Phase diagrams}
\label{sec:phases}

\subsection{Taming the parameter space}

\begin{figure} [t]
\begin{center}
\hbox{\includegraphics[width=0.5\textwidth]{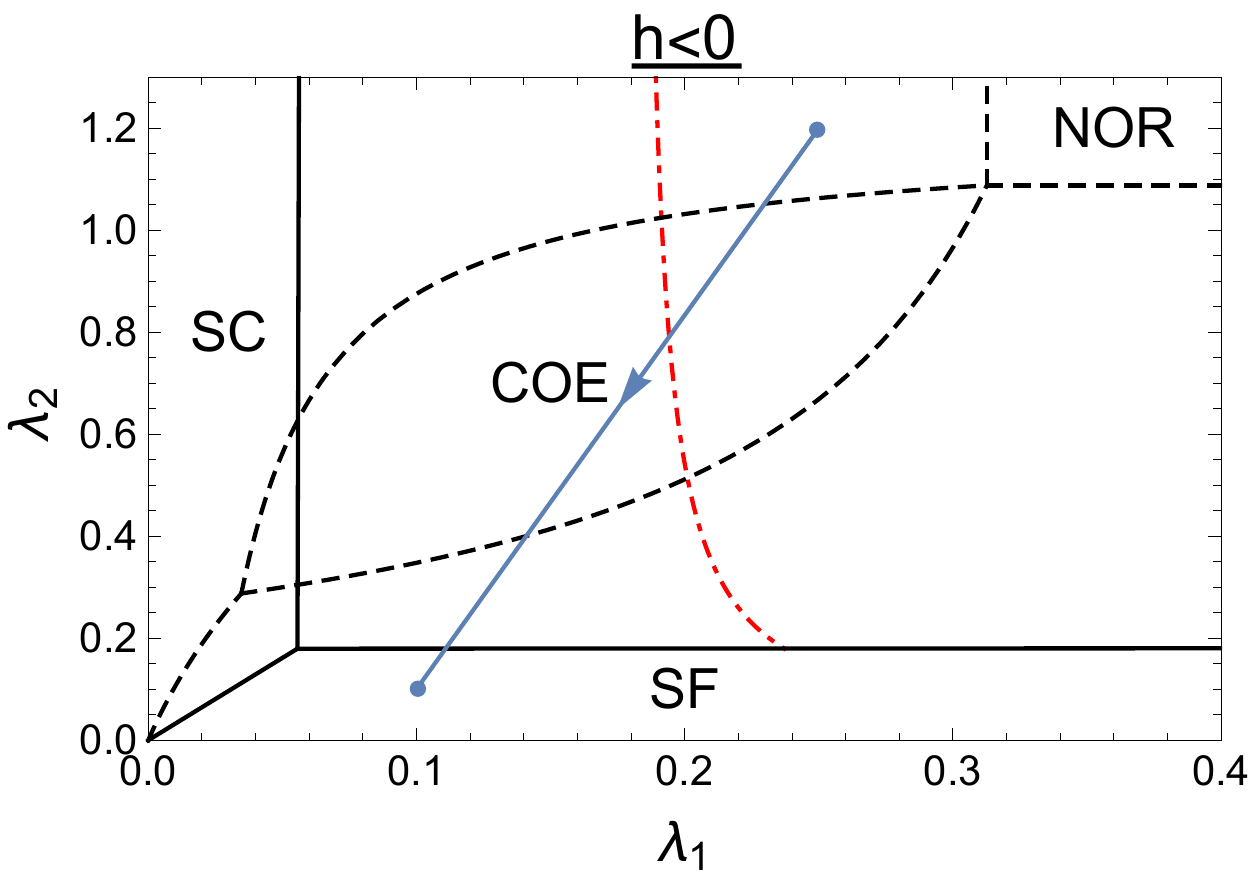}\includegraphics[width=0.5\textwidth]{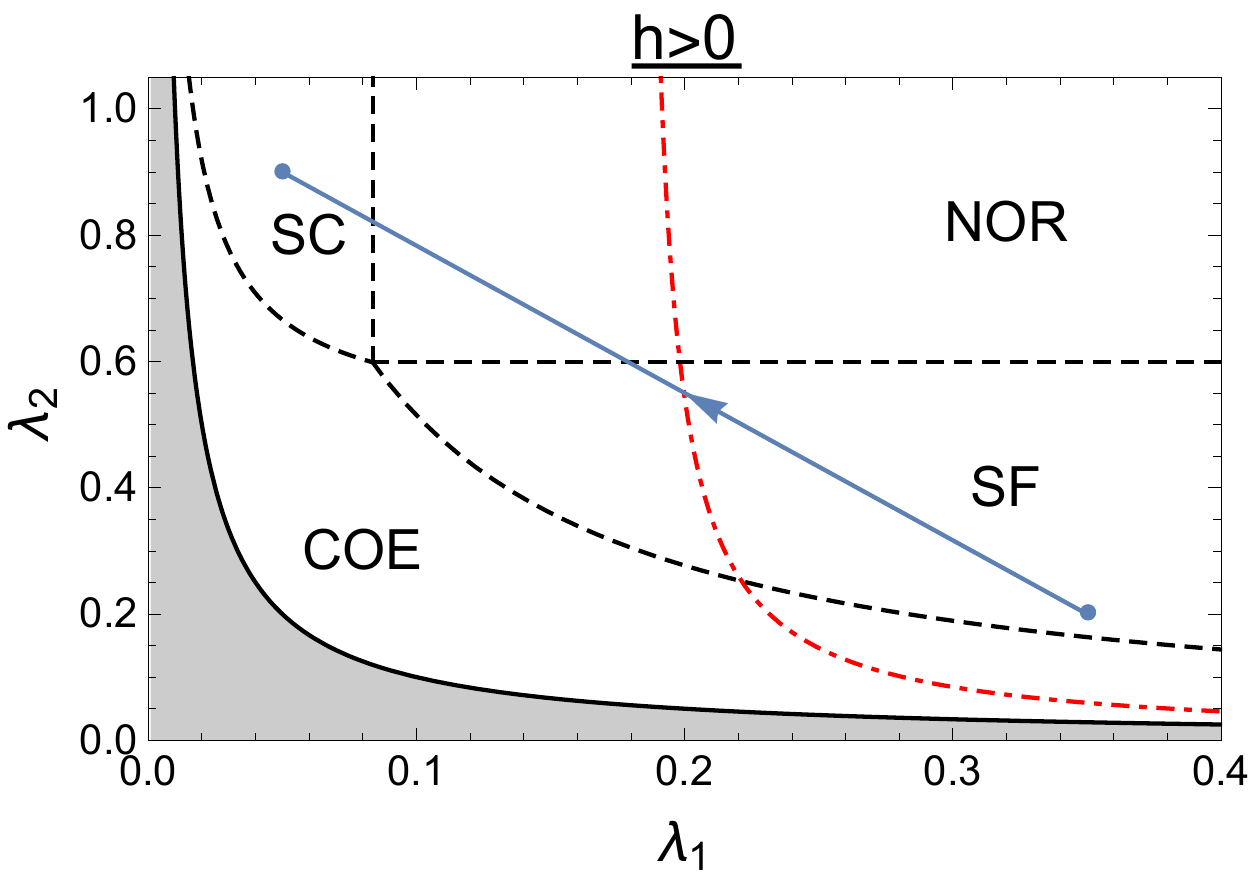}}

\vspace{0.5cm}
\hbox{\includegraphics[width=0.5\textwidth]{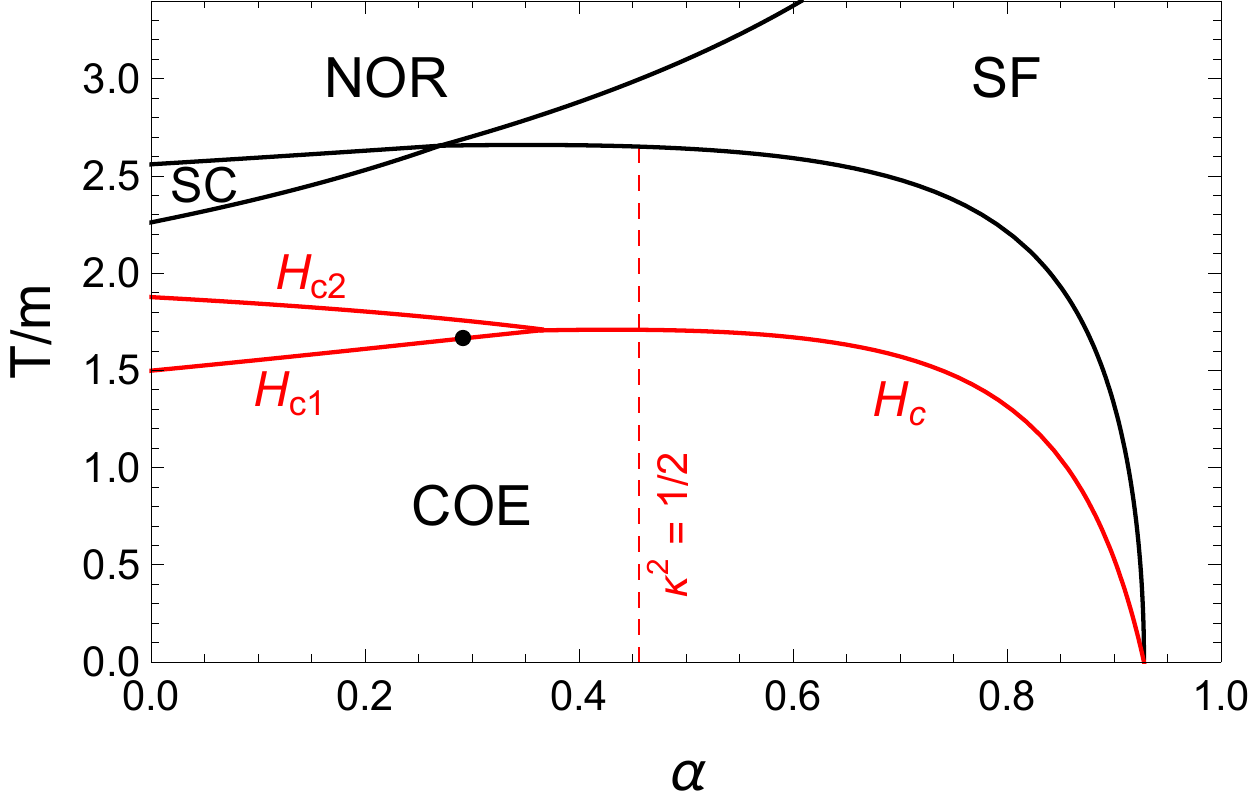}\includegraphics[width=0.49\textwidth]{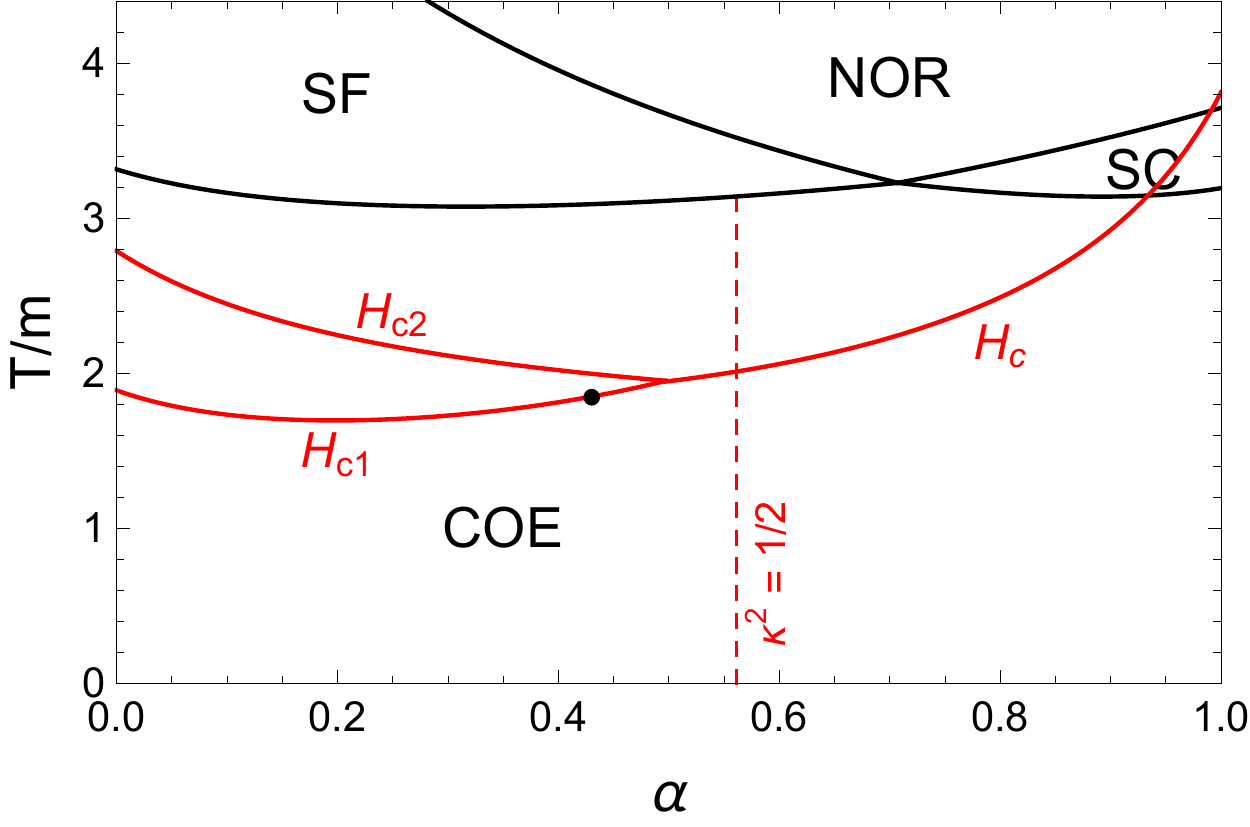}}
\caption{{\it Upper panels}: phases in the $\l_1$-$\l_2$-plane at $T=0$ [solid (black) curves] and $T>0$ [dashed (black) curves], at vanishing magnetic
 field, $H=0$. The shaded region in the upper right panel has to be excluded because there the potential is unbounded from below, $h>\sqrt{\lambda_1\lambda_2}$.  
 The dash-dotted (red) lines indicate $H_c=H_{c2}$ in the COE phase (this curve does not depend on temperature). The specific parameters are $m_1=m_2\equiv m$, 
 $\mu_1=1.5m$, $\mu_2=1.8m$ for all panels, and $h=-0.1$ and $T=2.43m$ (upper left), $h=0.1$, $T=3.5m$ (upper right). The (blue) paths in both upper panels 
 are used for the lower panels and following figures and are parametrized by $\alpha$, see Eq.\ (\ref{alpha}),  with
  $\vec{\lambda}_{\rm start}=(0.25,1.2)$, $\vec{\lambda}_{\rm end}=(0.1,0.1)$ for $h<0$ and 
 $\vec{\lambda}_{\rm start}=(0.35,0.2)$, $\vec{\lambda}_{\rm end}=(0.05,0.9)$ for $h>0$.
{\it Lower panels}: critical temperatures [upper (black) curves] and zero-temperature critical magnetic fields [lower (red) curves] along the paths from the upper panels. The magnetic fields are given in units of $m^2$ and are scaled down by 0.3 (left) and 
0.2 (right) to fit into the plot. The black dots on the $H_{c1}$-curves represent the onset of the first order phase transition.  The three critical magnetic fields do not intersect in a single point although they appear to do so in 
these plots, see Fig.\ \ref{fig:zoom} for a zoom-in. 
 }
\label{fig:phasesl1l2}
\end{center}
\end{figure}

The results in the previous sections have shown that the presence of the superfluid 
affects the transition from type-I to type-II superconductivity in a qualitative way, and we will make these results now more concrete by discussing the phase diagram of our model. 
To this end, we need to locate this transition in the parameter space. A priori, we have to deal with a large number of parameters, 
$m_1$, $m_2$, $\l_1$, $\lambda_2$, $q$, $h$, $G$, and the external thermodynamic parameters $T$, $H$, $\mu_1$, $\mu_2$. Having in mind a system of neutron and proton Cooper pairs we set $m_1=m_2\equiv m$ and $q=2e$, and express all dimensionful quantities in units of $m$. Many interesting 
results can already be obtained with a density coupling alone, and we shall therefore set the gradient coupling to zero, $G=0$, which implies $h_T=h$, for all numerical results. 
This leaves us with the 3 coupling constants $\l_1$, $\lambda_2$, $h$, plus 4 thermodynamic parameters. 
If we take the condition $H_{c2}=H_c$ as an indication for the location of the type-I/type-II transition, then Eq.\ (\ref{HcHc2}) shows that the transition is, for $G=0$ and fixed $q$, given by a surface in the $\lambda_1$-$\lambda_2$-$h$-space. (This surface is independent of $\mu_1$, $\mu_2$, and $T$, but these parameters of course determine the favored phase, and thus, if embedded in the larger parameter space, not everywhere on that surface the COE phase is the preferred phase at $H=0$.) 
Therefore, the phase diagrams in Fig.\ \ref{fig:phases}, 
where $\lambda_1$, $\lambda_2$, and $h$ are fixed, are not very useful for our present purpose, and it is more suitable to start from the $\lambda_1$-$\lambda_2$ plane, 
where, for a given cross-coupling $h$, we obtain a nontrivial curve $H=H_{c2}$. Two phase diagrams in the $\lambda_1$-$\lambda_2$ plane at vanishing magnetic field 
are shown in the upper panels of Fig.\ \ref{fig:phasesl1l2}, one for positive and one for negative cross-coupling $h$. We have chosen the chemical potentials to be larger than the common 
mass parameter, $\mu_i>m$, in which case it is always possible to find negative and positive values of $h$ such that at sufficiently low 
$T$ and $H$ there is a region in the phase diagram where the COE phase is preferred, cf.\ Fig.\ \ref{fig:phases}.

In the interior of a neutron star, as we move towards the center and thus increase the total baryon number, the system will take some complicated path in our multi-dimensional 
parameter space, under the assumption that the model describes dense nuclear matter reasonably well. Here we do not attempt to construct this path. But, we keep in mind
that nuclear matter is expected to cross the critical surface $H=H_{c2}$ if we move to sufficiently large densities. Therefore, we now choose a path with this property. 
Starting from the diagrams in Fig.\ \ref{fig:phasesl1l2}, the simplest way to do this is to choose a path in the $\lambda_1$-$\lambda_2$ plane with all other parameters held fixed. 
We parametrize the path by $\alpha\in [0,1]$, which is defined by 
\bea \label{alpha}
\vec{\lambda}  = \vec{\lambda}_{\rm start}+\alpha(\vec{\lambda}_{\rm end}-\vec{\lambda}_{\rm start})  \, ,
\eea
with $\vec{\lambda}=(\lambda_1,\lambda_2)$. In Fig.\  \ref{fig:phasesl1l2} we show the paths for positive and negative $h$ that we shall use in the following. Both paths cross from a type-II region for small $\alpha$ into a type-I region for large $\alpha$. In a very crude way, $\alpha$ plays the role of the baryon density in a neutron star. 
Since our paths are chosen such that $\lambda_1$ decreases along them and the charge $q$ is fixed, 
the Ginzburg-Landau parameter $\kappa$ decreases as $\alpha$ increases.

\subsection{Phases at nonzero temperatures and magnetic fields}

In the lower panels of Fig.\ \ref{fig:phasesl1l2} we show the zero-temperature critical magnetic fields $H_c$, $H_{c2}$, and $H_{c1}$, computed as explained in Secs.\ \ref{sec:Hc} -- \ref{sec:Hc1}, and the 
critical temperatures at zero magnetic field, computed from Eqs.\ (\ref{TcCOE}) for the transition between the COE phase and a single-condensate phase, and with the help of the condensates (\ref{eq:rhoSC}) and (\ref{eq:rhoSF})  together with the thermal masses 
(\ref{thermal_mass}) for the transitions from a single-condensate phase to the NOR phase.  
The horizontal axis is given by $\alpha$, i.e., we move through the 
$\lambda_1$-$\lambda_2$ plane along the paths shown in the upper panels of the figure. 
In principle, we can use the model straightforwardly to determine the phases in the entire $\alpha$-$H$-$T$-space. As a rough guide to this three-dimensional space 
notice that increasing the magnetic 
field at fixed $T$ will eventually destroy the charged condensate, i.e., if $H$ is sufficiently large only the SF and NOR phases 
survive, while increasing the temperature at fixed $H$ will eventually destroy all condensates, i.e., at sufficiently large $T$ only the NOR phase survives.  
Working out the details of the entire phase space might be interesting, but it is tedious and not necessary for the main purpose of this paper. 
Nevertheless, we emphasize that this possibility makes our model very useful for 
nuclear matter inside a neutron star. For instance, comparing our Fig.\ \ref{fig:phasesl1l2} with Fig.\ 1 in Ref.\ \cite{Glampedakis:2010sk}, we see that our results are -- on the one hand -- 
a toy version of more concrete calculations of dense nuclear matter, but -- on the other hand -- more sophisticated because they include all possible phases in a consistent way, 
not relying on any result within a single-fluid system.

Here we proceed with the discussion of the critical magnetic fields, and for the remainder of the paper we shall restrict ourselves to zero temperature.

\subsection{Type-I/type-II transition region}

At first sight, the phase structure in Fig.\ \ref{fig:phasesl1l2} regarding the critical magnetic fields looks as expected from a single superconductor, only with a critical $\kappa$ that is shifted from the standard value. But, we already know from Sec.\ \ref{sec:inter} that the point at which the second-order onset of flux tubes turns into a first-order transition is different from the point where $H_{c}$ and $H_{c2}$ intersect. We have marked this point in both lower panels of Fig.\ \ref{fig:phasesl1l2}. Moreover, 
in the presence of the superfluid, the three critical magnetic fields do not intersect in a single point. This is only visible 
on a smaller scale, and we discuss this transition region in detail now. With respect to that region, there is no qualitative difference between the two parameter sets chosen in 
Fig.\ \ref{fig:phasesl1l2}, and therefore we will restrict ourselves to the set with $h<0$.

\begin{figure} [t]
\begin{center}
\hbox{\includegraphics[width=0.5\textwidth]{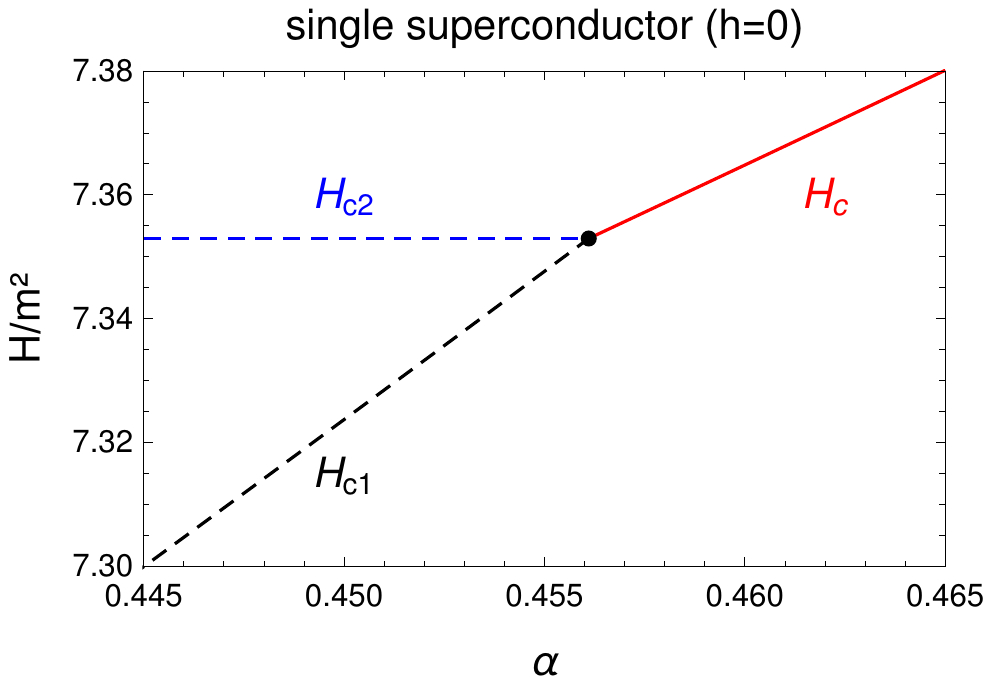}\includegraphics[width=0.5\textwidth]{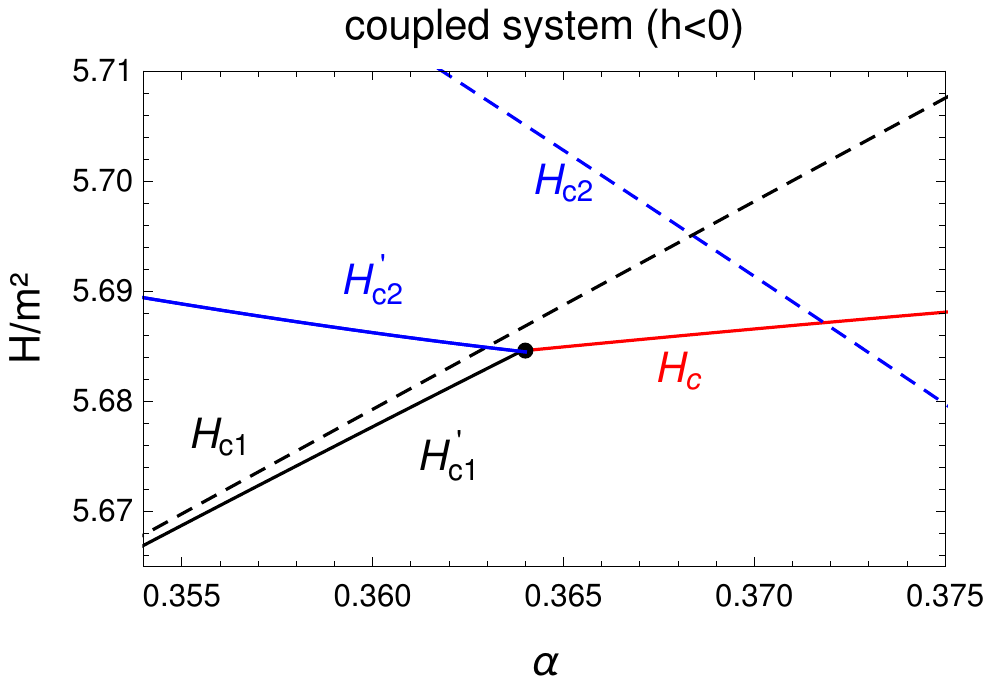}}
\caption{Critical magnetic fields in the type-I/type-II transition region as a function of the parameter $\a$ for a single superconductor, 
$h=0$ (left panel), and a superconductor coupled to a superfluid with negative density coupling, $h<0$ (right panel). All other parameters are taken from
 Fig.\ \ref{fig:phasesl1l2}, i.e., the right panel is a zoom-in to the transition region of the lower left panel of Fig.\ \ref{fig:phasesl1l2}. Solid (dashed) lines are first 
 (second) order phase transitions.
 }
\label{fig:zoom}
\end{center}
\end{figure}

In Fig.\ \ref{fig:zoom}, we present the critical magnetic fields in the region that covers their intersection point(s). In the left panel, we have, for comparison, set the coupling to the 
superfluid to zero, $h=0$, with all other parameters held fixed. As a result, we obtain the expected phase structure of an ordinary superconductor. All three critical magnetic 
fields intersect at one point -- which can be viewed as a check for our numerical calculation of $H_{c1}$ -- and this point corresponds to $\k^2=1/2$. 
For magnetic fields smaller than $H_{c}$ and $H_{c1}$ the superconductor expels the magnetic field completely, and magnetic fields larger than $H_{c}$ and $H_{c2}$ penetrate the system and superconductivity breaks down. In the open "wedge" between $H_{c1}$ and $H_{c2}$, an array of flux tubes 
(with varying flux tube density) is expected to exist, with second-order phase transitions at $H_{c1}$ and $H_{c2}$.

\begin{figure} [h]
\begin{center}
\hbox{\includegraphics[width=0.48\textwidth]{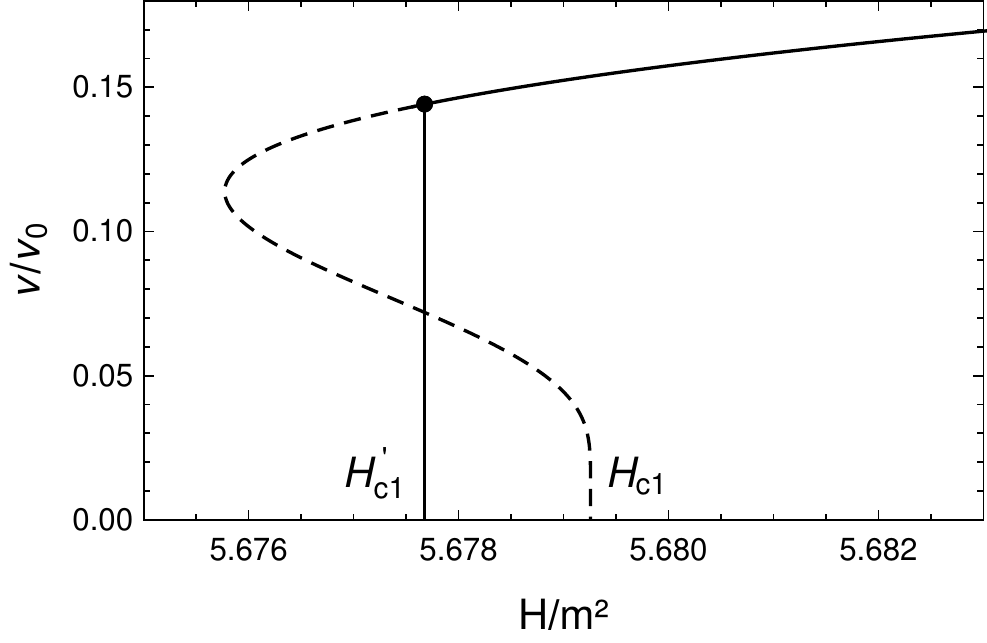}\includegraphics[width=0.52\textwidth]{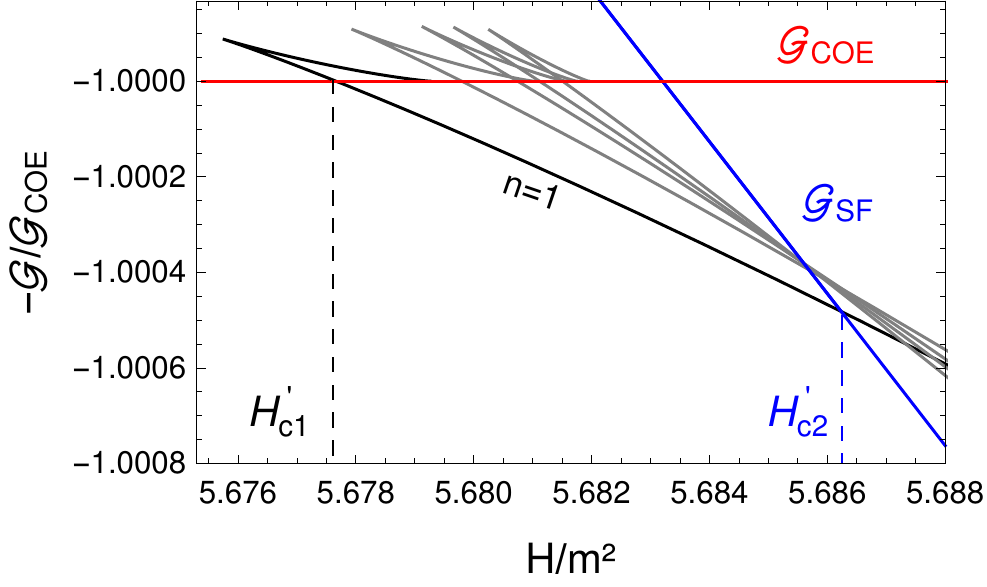}}
\caption{Left panel: flux tube density as a function of $H$ with the parameters of the right panel of Fig.\ \ref{fig:zoom} and $\a=0.360$, in units of $\nu_0=1/(\pi\xi^2)$. 
The dashed line shows the 
unstable and metastable part of the solution and is not realized, i.e., the density jumps at $H=H_{c1}'$ from zero to a finite value indicated by the black dot. For $\nu\to 0$,
the dashed line approaches the mass per unit length of the flux tube, i.e., the "would-be" second-order transition $H_{c1}$. 
Right panel: Gibbs free energies as a function of the external magnetic field $H$ for the Meissner, flux tube, and normal-conducting phases, including higher winding numbers, 
$n=2,4,6,10$, which are energetically disfavored. 
 }
\label{fig:Hnu}
\end{center}
\end{figure}

In the right panel we zoom in to the critical region of the lower left panel of Fig.\ \ref{fig:phasesl1l2}. 
From our analytical results we know the following. $(i)$ The critical magnetic fields $H_c$, $H_{c2}$ intersect at a point given by 
Eq.\ (\ref{HcHc2}), which corresponds for the chosen parameters to $\alpha\simeq 0.37182$. $(ii)$ Just below the curve $H_{c2}$ the flux tube phase is energetically favored over
the normal-conducting phase (not necessarily over the Meissner phase) for all $\alpha < 0.38265$, as we can compute from Eq.\ (\ref{DeltaG}). 
This point is beyond the right end of the scale shown in Fig.\ \ref{fig:zoom}. $(iii)$
The second-order phase transition from the Meissner phase to the flux tube phase turns into a first-order transition at the point given by Eq.\ (\ref{sign}), here 
$\alpha\simeq 0.29236$, which is  beyond the left end of the  scale of the plot. In the single superconductor, these three $\alpha$'s (or $\kappa$'s) coincide. Had we only computed 
$H_c$, $H_{c2}$, and $H_{c1}$, we would have obtained a puzzling collection of potential phase transition lines. However, together with the first-order phase transitions
$H_{c1}'$ and $H_{c2}'$, computed from Eq.\ (\ref{Hcprime}), a consistent picture of the phase structure emerges. Before we comment on this structure, we make the behavior 
at  $H_{c1}'$ more explicit by plotting the flux tube density $\nu$ and the Gibbs free energies in Fig.\ \ref{fig:Hnu}. 
The right panel of this figure includes the results for higher winding numbers. We see that they are energetically disfavored for the parameter set chosen here. 
In Ref.\ \cite{Alford:2007np} it was shown that higher winding numbers become important if the magnetic flux, instead of the external field $H$, is fixed. We did check that our 
numerical results indeed reproduce that observation, but we have not checked systematically whether and for which parameters flux tubes with higher winding numbers are favored 
in an externally given magnetic field $H$. This is an interesting question for future studies. 

\begin{figure} [t]
\begin{center}
\includegraphics[width=\textwidth]{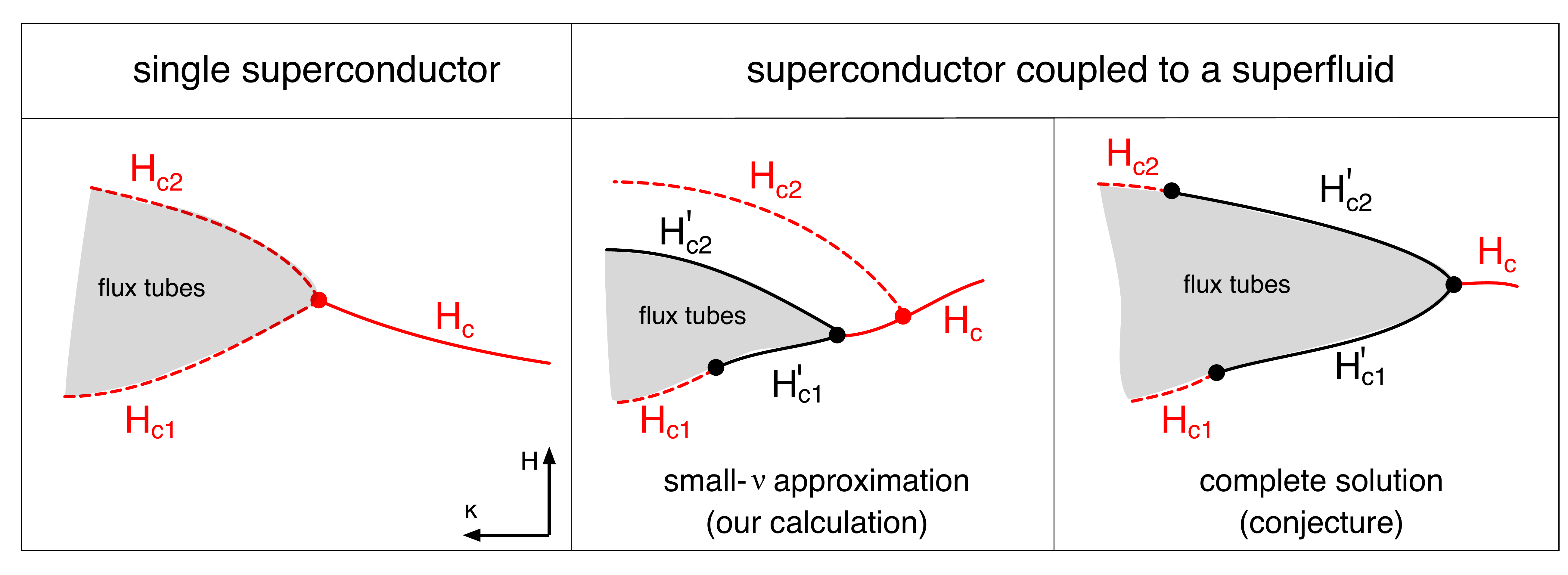}
\caption{Schematic phase structures for a single superconductor and our two-component system in the type-I/type-II transition region. 
Solid (dashed) lines are first (second) order phase 
transitions. Our approximation of small flux tube densities $\nu$ rigorously predicts the critical point at which $H_{c1}$ becomes first order. If we extrapolate our approximation to compute also the upper critical field  -- where $\nu$ is not small --  we arrive at the inconsistent diagram shown in the middle panel: the first-order transition $H_{c2}'$, computed  from our small-$\nu$ approximation, must not be smaller than $H_{c2}$ ($H_{c2}$ is a rigorous result, independent of the approximation). 
The conjectured phase structure in the right panel is the simplest one consistent with our results, including a possible critical point between $H_{c2}$ and 
$H_{c2}'$.  
 }
\label{fig:HHH}
\end{center}
\end{figure}

The most straightforward interpretation of the right panel of Fig.\ \ref{fig:zoom} is to simply ignore the second-order phase transition curves. Then, the topology of the critical region is 
the same as in the left panel, only with first-order instead of second-order transitions at the boundaries of the flux tube phase (with $H_{c1}'$ turning into a second-order phase transition at $\alpha\simeq 0.29236$). However, this cannot be the complete picture. The reason is that after we
have left the flux tube phase through $H_{c2}'$ and keep increasing $H$ we reach $H_{c2}$, and we know that there should be flux tubes just below $H_{c2}$ for 
all $\alpha < 0.38265$. In other words, our result contradicts the observation that $H_{c2}$ is a lower bound for the transition from the flux tube phase to the normal-conducting phase, as explained at the end of Sec.\ \ref{sec:Hc2}. 
This contradiction is resolved when we remember the regime of validity of our approximation for the free energy of the 
flux tube lattice. Our approximation is accurate where $H_{c1}$ turns into $H_{c1}'$ because the distance between the flux tubes is infinitely large at this critical point. 
As we move along $H_{c1}'$ upon increasing $\alpha$, and then along $H_{c2}'$ upon decreasing $\alpha$, our approximation becomes worse and worse. 
Within the present calculation
we can thus not determine the phase structure unambiguously, but it is easy to guess a simple topology of the type-I/type-II transition region that is consistent 
with all our results and takes into account the shortcomings of our approximation. This conjectured phase structure is shown in Fig.\ \ref{fig:HHH}. 

The motivation for the conjecture is as follows. The existence of the first-order line $H'_{c1}$ and its starting point is predicted rigorously in our approach. Let us
move along that line assuming that we go beyond our approximation and know the complete result. As we move towards large $\alpha$, we will 
deviate from the line predicted by our approximation. At some value of $\alpha$, we will intersect the curve $H_c$. In order to resolve the contradiction of our 
phase structure, we expect this intersection to occur "on the other side" of the intersection between $H_{c2}$ and $H_{c}$.
This implies that our approximation underestimates the binding energy of the flux tubes, i.e., we expect the flux tube phase to be more favored in the full result. We have not 
found a simple reason -- other than the inconsistency of the phase structure -- why our approximation distorts the full result in this, and not the other, direction. Now, at the new, correct, 
intersection of $H'_{c1}$ and $H_c$, there must necessarily be a third line attached, namely $H_{c2}'$ (just like in our approximation). The reason is that if we cross $H_{c1}'$ we end up in the flux tube phase and if 
we cross $H_c$ we end up in the normal-conducting phase, and these two phases must be separated by a phase transition line. This critical field $H_{c2}'$ might be 
larger than $H_{c2}$ for all $\alpha$ (below the $\alpha$ of the triple point where $H'_{c1}$, $H'_{c2}$ and $H_c$ intersect) or $H_{c2}'$ might merge with $H_{c2}$, leading to 
an additional critical point. The latter is the scenario shown in the right panel of Fig.\ \ref{fig:HHH}. One might ask whether $H_{c1}'$ and $H_c$ intersect exactly at the point where $H_c$
and the second-order line $H_{c2}$ intersect. In this case, the entire upper critical line would be of second order and given by $H_{c2}$. However, this 
seems to require some fine-tuning of the interaction between the flux tubes since the second-order line $H_{c2}$ does not know anything about this interaction.

\subsection{Flux tube clusters}

\begin{figure} [t]
\begin{center}
\hbox{\includegraphics[width=0.5\textwidth]{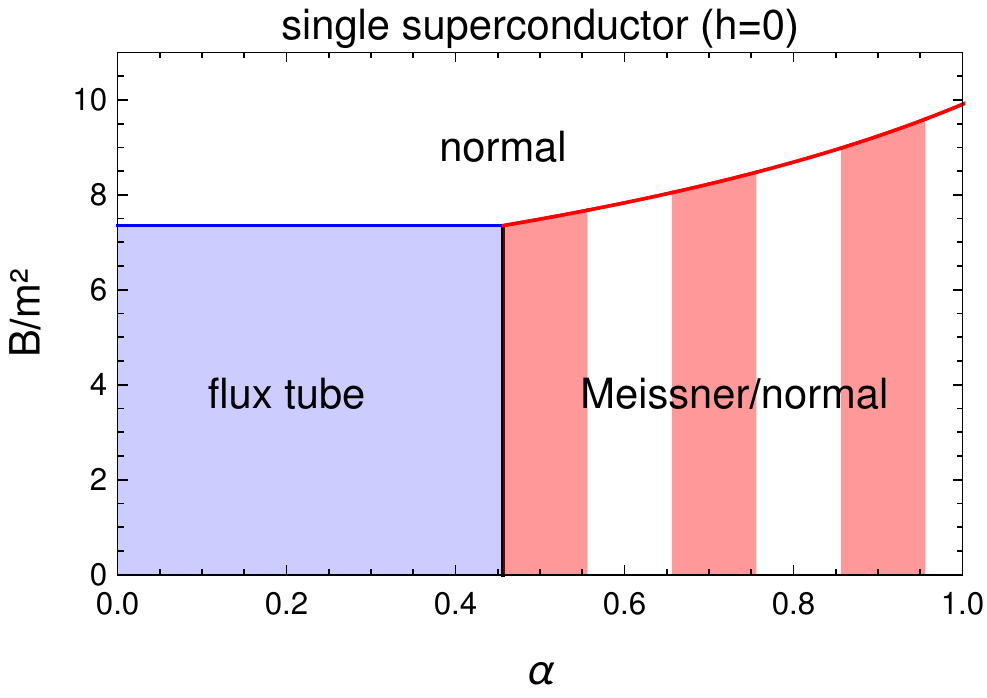}\includegraphics[width=0.5\textwidth]{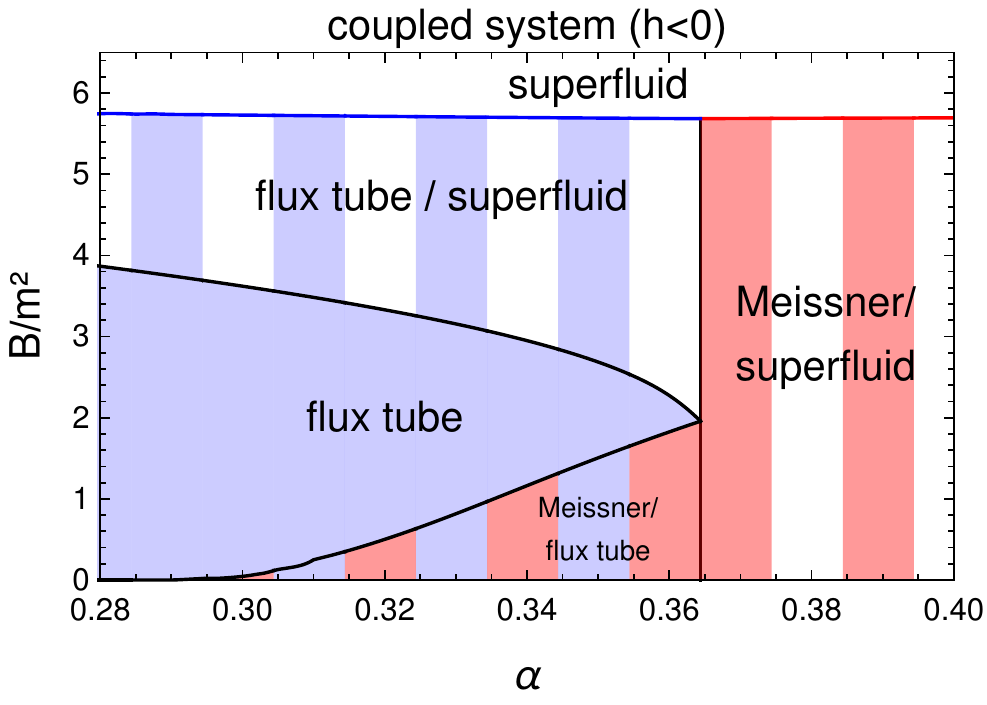}}
\caption{Phases in the $B$-$\alpha$ plane, computed with the parameters
and from the results of Fig.\ \ref{fig:zoom}. In a single superconductor (left panel), the magnetic field penetrates in the form of a flux tube array ("flux tube"), 
through macroscopic regions in a mixed phase ("Meissner/normal") or homogeneously and space filling ("normal"). In a superconductor coupled to a superfluid 
(right panel), it can also penetrate in the form of flux tube clusters, either in 
a mixture with field-free regions ("Meissner/flux tube") or in a mixture with normal-conducting regions ("flux tube/superfluid").
The "Meissner/flux tube" phase is, for the chosen parameters, only possible for $\alpha> 0.29236$ (where the 
phase transition in the $H$-$\alpha$ plane is of first order).
 }
\label{fig:Balfa}
\end{center}
\end{figure}

The first-order phase transitions with $H$ as an external variable translate into mixed phases if we fix the magnetic field $B$ (spatially averaged) 
instead. Again, this can be illustrated by the analogy to the onset of baryonic matter at small temperatures. As a function of $\mu_B$, this onset 
is a first-order transition  with a discontinuity in 
baryon number density $n_B$. If we instead probe this onset with fixed $n_B$ (spatially averaged), we pass through a region of mixed phases, for example nuclei 
in a periodic lattice, until we reach the saturation density. These mixed phases are realized in the outer regions of a neutron star, and it would be an intriguing manifestation  
of this analogy if the mixed flux tube phases discussed here are realized in the core of the star. Each first-order transition in $H$ yields two critical magnetic fields $B$ which 
we compute as follows. At $H_{c1}'$, the lower critical field is $B=0$, and the upper critical field is $\langle B\rangle=\Phi_0 \nu$ [using Eq.\ (\ref{drB})], where $\nu$ is the 
numerically computed flux tube area density as we approach the first-order transition from above; 
at $H_{c2}'$, the lower critical field is $\langle B\rangle=\Phi_0 \nu$, with $\nu$ now being the numerically computed density as we approach  
the first-order transition from below, while the upper critical field is $B=H_{c2}$; at $H_c$, the lower critical field is $B=0$, and the upper one is $B=H_c$. We perform this calculation with the parameters of Fig.\ \ref{fig:zoom}. 
As discussed for the $H$-$\alpha$ phase diagrams above, also for the $B$-$\alpha$ phase structure we do not expect our approximation to yield quantitatively reliable results where the flux tube density is large. Therefore, our results 
reflect the topology of the $B$-$\alpha$ phase diagram correctly, but the precise location of the phase transition lines cannot be determined within our approach. The phase diagrams for the single superconductor and the two-component system are shown in Fig.\ \ref{fig:Balfa}.
In a single superconductor, there is only one possible mixed phase: macroscopic regions in which the magnetic field penetrates, mixed with regions in which the magnetic field 
remains expelled \cite{tinkham2004introduction}. The geometric structure of these regions depends on the details of the system such as the surface tension, and it is beyond the scope of this paper to determine them. 
In the two-component system, two additional mixed phases are possible, both of which contain flux tube clusters. (Unrelated to the first-order phase transitions pointed out here, flux tube clusters have been suggested to exist in neutron stars in the vicinity of superfluid neutron vortices \cite{1995ApJ...447..305S}.) 
Firstly, at $H_{c1}'$, flux tube clusters are immersed in a field-free superconducting region, as predicted for "type-1.5 superconductivity" \cite{PhysRevB.72.180502}. Secondly, at $H_{c2}'$, there is a mixed phase of flux tubes with 
the normal-conducting phase, i.e., superconducting regions that enclose flux tubes and that are themselves surrounded by completely normal-conducting regions. 

\section{Conclusions}
\label{sec:summary}

We have shown that the coupling to a superfluid can have profound effects on the magnetic properties of a superconductor. We have started from a microscopic 
model for two complex scalar fields, coupled to each other via density and gradient coupling terms, with one of the fields being electrically charged. 
By computing the thermal excitations of the system we have derived a Ginzburg-Landau-like effective potential 
for the charged and neutral condensates and the gauge field. This potential has then been evaluated at nonzero temperatures and external magnetic fields, 
computing the two condensates dynamically for all 4 possible phases: condensation of both fields (superconductor + superfluid), condensation of only one field 
(pure superconductor or pure superfluid), or no condensation. We have discussed the structure of the resulting phase diagram in the multi-dimensional parameter space, 
with the main focus on
the transition region between type-I and type-II superconductivity. To this end, we have computed the critical magnetic fields $H_c$, $H_{c2}$
(analytically) and $H_{c1}$ (numerically, based on the profile functions of a magnetic flux tube). In contrast to the standard scenario of a single superconductor, these
three magnetic fields do not intersect in a single point if the superconductor coexists with a superfluid. The phase structure around these intersection points is (at least partially) 
resolved by computing the first-order phase transitions $H_{c2}'$ and $H_{c1}'$. This has been done by employing a simple approximation for the free energy of a flux tube array
that is valid for large flux tube distances and that effectively reduces the calculation to solving the equations of motion for a single flux tube. The 
new critical fields $H_c$, $H_{c2}'$, $H_{c1}'$ {\it do} intersect in a single point, restoring the topology of the transition region, with (segments of) the second-order transition lines
replaced by first-order transitions. In particular, we have identified a new critical point -- and derived an analytical expression for its location -- where the 
second-order flux tube onset $H_{c1}$ turns into a first order transition $H_{c1}'$. The presence of the first-order transitions allows for mixed phases with flux tube clusters, very similar to a type-1.5 superconductor, which consists of two charged fields coupled indirectly through the gauge field. 

There are several possible improvements and extensions of our work. Our approximation for the flux tube array can be improved for instance by determining dynamically 
the values of the condensates far away from the flux tubes instead of using the values of the homogeneous phase. To settle the precise location of the phase transition lines, 
it would be interesting to perform a brute force numerical calculation of the free energy of the flux tube phase, for which our results are a valuable guidance. There are several other interesting aspects of our model which we have mentioned but not worked out in detail. For instance, 
one could perform a more systematic study of the effect of the derivative coupling, which we have included in all our analytical results, but set to zero in the final numerical results
of the phase diagrams. Or one could perform a more detailed study of flux tubes with higher winding numbers, which turned out to be energetically disfavored for the 
parameter regime we have studied, but which are known to potentially play a role in the two-component system. 
One can also study the phase structure at nonzero temperature in more detail and/or improve the large-temperature approximation on which our
Ginzburg-Landau potential was based. Or one can include superfluid vortices, aiming at the phase structure at nonzero magnetic field and externally imposed rotation.  

Our setup and our results are applicable to dense nuclear matter in the core of neutron stars. For instance, one can fit our model parameters, such as the density coupling and gradient coupling, to values predicted for nuclear matter and eventually compute the phase structure as a function of the baryon number density rather than of an abstract 
model parameter. One may also ask whether a potential phase of flux tube clusters would affect the transport properties of the core in a detectable way. 
Moreover, it would be interesting to employ our results in studies of the time evolution of the magnetic field in a neutron star. Here we have computed the ground state in equilibrium for given temperature, magnetic field and chemical potential, but for more phenomenological predictions one needs to know whether and on which time scale this ground state is reached.

\begin{acknowledgments}
We would like to thank  Mark Alford, Nils Andersson, Egor Babaev, Christian Ecker, Carlos Lobo, David M{\"u}ller, and Andreas Windisch for valuable comments and discussions. 
We acknowledge support from the Austrian Science 
Fund (FWF) under project no.\ W1252, and from the {\mbox NewCompStar} network, COST Action MP1304. A.S.\ is supported by the Science \& Technology Facilities Council (STFC) in the form of an Ernest Rutherford Fellowship.
\end{acknowledgments}

\appendix

\section{Derivation of the effective potential}
\label{app:prop}

In this appendix we compute an effective potential in a high-temperature approximation from 
the excitations of the system, taking into account the mixing of the two scalar fields with the photon. Elements of this derivation can be found in discussions 
of the standard abelian Higgs model, see for instance chapter 85 of Ref.\ \cite{Srednicki:2007qs}.

It is convenient to split the complex scalar fields into their real and imaginary parts,
\be
\vf_1=\frac{1}{\sqrt{2}}(\phi_1+i\chi_1) \, , \qquad   \vf_2=\frac{1}{\sqrt{2}}(\phi_2+i\chi_2) \, .
\ee
Then, the Lagrangian (\ref{L}), in the presence of chemical potentials $\mu_1$ and $\mu_2$,  becomes
\bea
{\cal L}  &=&   {\cal L}_1 +{\cal L}_2 + {\cal L}_{\rm int} + {\cal L}_{\rm YM} + {\cal L}_{\rm gf}\, ,
\eea
where we have added a gauge fixing term,
\be
{\cal L}_{\rm gf} = -\frac{(\partial_\mu A^\mu)^2}{2\xi} \, ,
\ee
and where  
\begin{subequations}
\bea
{\cal L}_1 &=& \frac{1}{2}\partial_\mu\phi_1\partial^\mu\phi_1 + \frac{1}{2}\partial_\mu\chi_1\partial^\mu\chi_1 + (qA_\mu-\delta_{0\mu}\mu_1)(\phi_1\partial^\mu\chi_1-\chi_1\partial^\mu\phi_1) 
 \non[2ex]
 &&+\frac{1}{2}(\phi_1^2+\chi_1^2)(\mu_1^2-m_1^2+q^2A_\mu A^\mu-2\mu_1 qA_0) -\frac{\lambda_1}{4}(\phi_1^2+\chi_1^2)^2 \, , \\[2ex]
  {\cal L}_2 &=&\frac{1}{2}\partial_\mu\phi_2\partial^\mu\phi_2 + \frac{1}{2}\partial_\mu\chi_2\partial^\mu\chi_2 -\mu_2(\phi_2\partial_0\chi_2-\chi_2\partial_0\phi_2) 
+\frac{1}{2}(\phi_2^2+\chi_2^2)(\mu_2^2-m_2^2) -\frac{\lambda_2}{4}(\phi_2^2+\chi_2^2)^2 \, , \hspace{0.5cm}\\[2ex]
  {\cal L}_{\rm int} &=& \frac{h}{2}(\phi_1^2+\chi_1^2)(\phi_2^2+\chi_2^2) -\frac{G}{2}(\phi_1\partial_\mu\phi_1+\chi_1\partial_\mu\chi_1)(\phi_2\partial^\mu\phi_2+\chi_2\partial^\mu\chi_2) 
  \, .
 \eea
\end{subequations}
We allow for condensation of both fields by shifting $\phi_1\to \rho_1 + \phi_1$, $\phi_2\to \rho_2 + \phi_2$, i.e., we assume the condensates to be real, and 
from now on $\phi_i$ and $\chi_i$ are fluctuations about the condensates. The dispersion relations of the excitations are computed from the tree-level propagator 
in momentum space. To this end, we introduce the Fourier transformed fields via
\be
\phi_i(X) = \frac{1}{\sqrt{TV}}\sum_K e^{-iK\cdot X}\phi_i(K) \, , \quad \chi_i(X) = \frac{1}{\sqrt{TV}}\sum_K e^{-iK\cdot X}\chi_i(K) \, , \quad A_{\mu}(X) = \frac{1}{\sqrt{TV}}\sum_K e^{-iK\cdot X}A_{\mu}(K) \, ,
\ee
with the space-time four-vector $X=(-i\tau,\vec{r}) $ and the four-momentum $K=(k_0,\vec{k})$, where $k_0=-i\omega_n$ with  
the bosonic Matsubara frequencies $\omega_n = 2\pi n T$, $n\in \mathbb{Z}$. In the imaginary time formalism, we have to replace $A_0\to iA_0$.
The terms of second order in the fluctuations can then be written as  
\be
\int_X {\cal L}^{(2)}= -\frac{1}{2}\sum_K \Xi(-K)^T\frac{S^{-1}(K)}{T^2}\Xi(K) \, , 
\ee
with 
\be
\Xi^T = (\phi_1, \chi_1, \phi_2,  \chi_2,  A_{0}, A_1, A_2, A_3 ) \, .
\ee
The inverse tree-level propagator is an $8\times 8$ matrix, which reads
\bea
S^{-1}(K) = \left(\begin{array}{cc} S_0^{-1}(K) & I(K) \\ [2ex] I^T(-K) & D^{-1}(K) \end{array}\right)\, ,
\eea
with the scalar field sector,
\bea
S_0^{-1}(K) = \left(\begin{array}{cccc} -K^2+\eta_1(\rho_1,\rho_2)+2\lambda_1\rho_1^2 & 2ik_0\mu_1 & \frac{\rho_1\rho_2}{2}(GK^2-4h) & 0 \\[2ex] -2ik_0\mu_1 & -K^2+\eta_1(\rho_1,\rho_2) &0&0 \\[2ex]
\frac{\rho_1\rho_2}{2}(GK^2-4h) & 0 & -K^2+\eta_2(\rho_1,\rho_2)+2\lambda_2\rho_2^2 & 2ik_0\mu_2 \\[2ex] 0&0& -2ik_0\mu_2 & -K^2+\eta_2(\rho_1,\rho_2)\end{array}\right) \, ,
\eea
where $\eta_{1/2}(\rho_1,\rho_2)\equiv -(\mu_{1/2}^2-m_{1/2}^2)+\lambda_{1/2}\rho_{1/2}^2-h\rho_{2/1}^2$, the inverse gauge field propagator,
\bea \label{gaugeprop}
D^{-1}(K) =  
\left(\begin{array}{cccc} -K^2+\sigma k_0^2+4\pi q^2\rho_1^2 & -i\sigma k_0k_1 & -i\sigma k_0k_2 &-i\sigma k_0k_3 \\[2ex]
-i\sigma k_0k_1 & -K^2-\sigma k_1^2+4\pi q^2\rho_1^2 & -\sigma k_1k_2 &-\sigma k_1k_3 \\[2ex]
-i\sigma k_0k_2 & -\sigma k_1k_2 & -K^2-\sigma k_2^2+4\pi q^2\rho_1^2 & -\sigma k_2k_3 \\[2ex]
-i\sigma k_0k_3 & -\sigma k_1k_3 & -\sigma k_2k_3 & -K^2-\sigma k_3^2+4\pi q^2\rho_1^2 \end{array}\right) \, ,
\eea
where $\sigma\equiv 1-1/\xi$, and the off-diagonal blocks that couple the scalar fields to the gauge field,
\be
I(K) =\sqrt{4\pi}q\rho_1 \left(\begin{array}{cccc} 2i\mu_1  & 0&0&0 \\[2ex] -k_0 & ik_1 & ik_2 & ik_3 \\[2ex] 0&0&0&0 \\[2ex] 0&0&0&0 \end{array}\right) \, .
\ee
\begin{figure} [t]
\begin{center}
\includegraphics[width=0.5\textwidth]{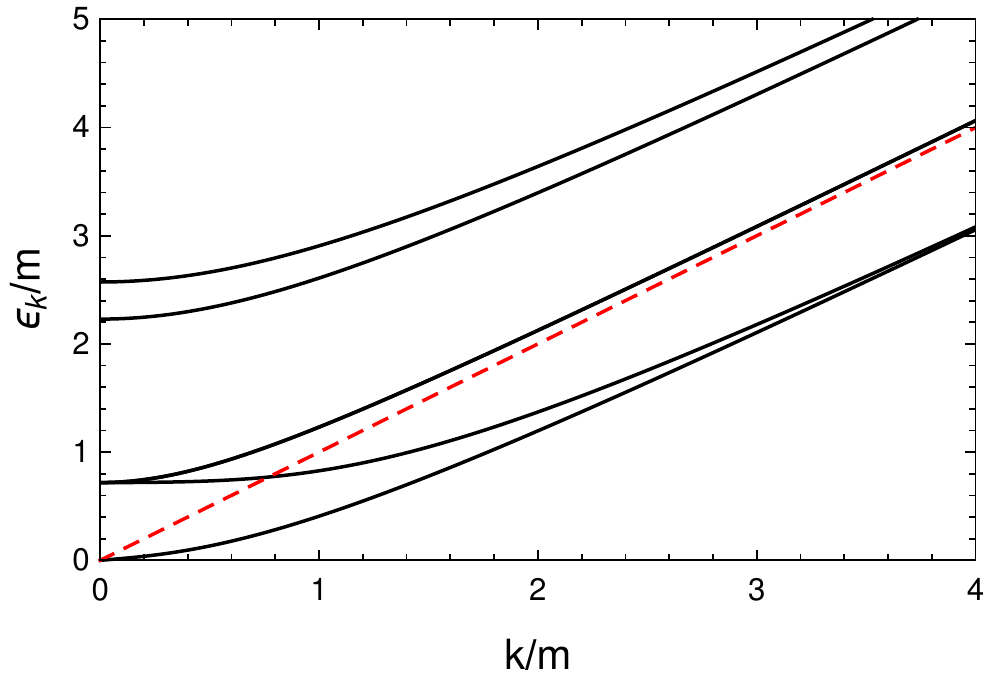}
\caption{Excitation energies for the COE phase, where both charged and neutral fields condense. The dashed (red) line is the diagonal $\epsilon_k=k$ to guide the eye. 
There are 6 modes in total, including one Goldstone mode and three massive gauge modes.  The excitation $\epsilon_k=\sqrt{k^2+4\pi q^2\rho_{01}^2}$, which approaches 
the diagonal from above, is 2-fold degenerate. All other dispersions have very complicated expressions due to the mixing of the gauge field with the scalar fields.
The parameters used for this plot are $m_1 =  m_2\equiv m$, $\mu_1 = 1.2m$, $\mu_2 = 1.1m$, $\lambda_1 = 0.3$, $\lambda_2 = 0.5$, $h = -0.1$, $G = 0$, $q=2e$. While these excitation energies are evaluated at the zero-temperature stationary point, the main purpose of this appendix is to derive an effective 
thermal potential, for which the dispersions for general values of the charged and neutral condensates are needed.}
\label{fig:expand1}
\end{center}
\end{figure}
We are interested in an effective potential for the condensates 
$\rho_1$ and $\rho_2$, and thus we need to keep these condensates general. Nevertheless, it is instructive
to first discuss the dispersions at the zero-temperature stationary point, i.e., we set  $\rho_1=\rho_{01}$ and $\rho_2=\rho_{02}$ with the condensates in the coexistence phase 
$\rho_{01}$ and $\rho_{02}$ from Eq.\ (\ref{COE}). 
Let us first set the cross-coupling between the scalar field to zero, $h=G=0$. The dispersion relations $k_0=\epsilon_k$ are given by the zeros of ${\rm det}\,S^{-1}$. 
Since this is a polynomial of degree 8 in $k_0^2$, we obtain 8 dispersions, 6 of which are physical. The two unphysical ones are of the form $\epsilon_k=k$. These are the usual
unphysical modes of the gauge field, whose contribution to the partition function is canceled by ghost fields. With the given gauge choice, ghosts do not couple to 
any of the fields and merely serve to cancel the unphysical modes. None of the modes depend on the gauge fixing parameter $\xi$, which only appears as a prefactor of the determinant ${\rm det}\,S^{-1}$ and thus does not have to be specified. The 6 physical dispersions are
\begin{subequations}
\bea
\epsilon_k &=& \sqrt{k^2+4\pi q^2\rho_{01}^2} \qquad \mbox{(2-fold)} \, ,  \label{eps1} \\[2ex]
\epsilon_k&=& \sqrt{k^2+3\mu_1^2-m_1^2+2\pi q^2\rho_{01}^2\pm\sqrt{4\mu_1^2k^2+\left(3\mu_1^2-m_1^2-2\pi q^2\rho_{01}^2\right)^2}} \, , \label{eps2} \\[2ex]
\epsilon_k&=& \sqrt{k^2+3\mu_2^2-m_2^2\pm\sqrt{4\mu_2^2k^2+(3\mu_2^2-m_2^2)^2}} \, . \label{eps3}
\eea
\end{subequations}
We have three gauge field modes with mass $\epsilon_{k=0}=\sqrt{4\pi}q\rho_{01}$ 
[the two modes of Eq.\ (\ref{eps1}) and the mode with the 
lower sign in Eq.\ (\ref{eps2})], two more massive modes from the scalar fields, and the Goldstone mode [the mode with the lower sign in Eq.\ (\ref{eps3})]. 
Let us now switch on the couplings $G$ and $h$ between the scalar fields. We find that the unphysical modes remain unaffected and all modes remain independent of the gauge fixing parameter $\xi$.  The mass of the three gauge field modes and the entire dispersion (\ref{eps1}) for two of them is also unchanged. The expressions 
for the remaining dispersions become very complicated. They can easily be computed numerically, and we show the result in Fig.\ \ref{fig:expand1}.

We now compute our effective potential by reinstating the general condensates, i.e., we need to compute the dispersions away from stationary point $\rho_1=\rho_{01}$, 
$\rho_2=\rho_{02}$. We restrict ourselves to the following large-momentum approximation, which is sufficient for the high-temperature approximation we are interested in,
\be \label{epsc1c2}
\epsilon_{k}\simeq k+c_1+\frac{c_2^2}{k} \, ,
\ee
such that 
\be \label{Tln}
T\int \frac{d^3k}{(2\pi)^3}\ln\left(1-e^{-\epsilon_k/T}\right) \simeq  -\frac{\pi^2T^4}{90} +\frac{c_1\zeta(3)T^3}{\pi^2} + \frac{(c_2^2-c_1^2)T^2}{12} \, .
\ee
In general, the dispersions now do depend on the gauge fixing parameter $\xi$. However, in the limit (\ref{epsc1c2}) this dependence drops out, i.e., the coefficients 
$c_1$ and $c_2$ do not depend on $\xi$. Moreover, now the unphysical gauge modes no longer have the simple form $\epsilon_k=k$. Two of the physical gauge modes
keep their simple form (\ref{eps1}), while for the other 4 physical modes the coefficients $c_1$ and $c_2$ are (at least some of them) very lengthy. 
However, adding up the result for all 6 physical modes yields a relatively compact result,
\bea
&&T\sum_{i=1}^6\int \frac{d^3k}{(2\pi)^3}\ln\left(1-e^{-\epsilon_{ki}/T}\right) \simeq  -\frac{\pi^2T^4}{15} -\frac{T^2}{12\left(1-\frac{G^2\rho_1^2\rho_2^2}{4}\right)}
\Bigg\{2(\mu_1^2+\mu_2^2)-(m_1^2+m_2^2)
-\left(2\lambda_1-h+6\pi q^2\right)\rho_1^2\non[2ex]
&&-(2\lambda_2-h)\rho_2^2+Gh\rho_1^2\rho_2^2-\frac{G^2\rho_1^2\rho_2^2}{8}\Big[\mu_1^2+\mu_2^2-(m_1^2+m_2^2)-(\lambda_1-h+12\pi q^2)\rho_1^2-(\lambda_2-h)\rho_2^2\Big]\Bigg\} \non[2ex]
&& \simeq \frac{T^2}{12}\left[(2\lambda_1-h+6\pi q^2)\rho_1^2+(2\lambda_2-h)\rho_2^2-Gh\rho_1^2\rho_2^2\right] + {\rm const.} \, ,
\eea
where, in the second step, we have absorbed all terms that do not depend on $\rho_1$ or $\rho_2$ into "const.", and dropped all higher-order terms in the derivative coupling 
(i.e., we assume $G\mu^2\ll 1$, where $\mu$ stands for all energy scales $\mu_1$, $\mu_2$, $m_1$, $m_2$, $\rho_1$ $\rho_2$). Dropping the constant 
contribution, we add the $T^2$ terms to the potential (\ref{Ux}) and arrive at the potential (\ref{UxT}) in the main text.

\section{Calculation of $H_{c2}$ and Gibbs free energy just below $H_{c2}$}
\label{app:Hc2}

Here we derive Eqs.\ (\ref{HcHc2}) and (\ref{DeltaG}). 
To this end, we need the equations of motion for the scalar fields and the gauge field. We go back to the Lagrangian (\ref{L}), 
take the static limit and replace the parameters $m_i$ and $h$ by their $T$-dependent generalizations $m_{i,T}$ and $h_T$. 
This yields the potential 
\bea \label{freeener}
\FE&=&(\nabla-iq\vec{A})\vf_1\cdot(\nabla+iq\vec{A})\vf_1^*-(\mu_1^2-m_{1,T}^2)|\vf_1|^2+\l_1|\vf_1|^4+\nabla\vf_2\cdot \nabla\vf_2^*-(\mu_2^2-m_{2,T}^2)|\vf_2|^2+\l_2|\vf_2|^4 \,\non[2ex] 
&&-2h_T|\vf_1|^2|\vf_2|^2-\frac{G}{2}\left[\vf_1\vf_2(\nabla+iq\vec{A})\vf_1^*\nabla\vf_2^*+\vf_1\vf_2^*(\nabla+iq\vec{A})\vf_1^*\nabla\vf_2+c.c.\right] +\frac{B^2}{8\pi}\, ,
\eea
and the equations of motion for $\varphi_1^*$, $\varphi_2^*$, and $\vec{A}$ become
\begin{subequations}\label{eom} 
  \bea
  \left[(\nabla-iq\vec{A})^2 + \mu_1^2-m_{1,T}^2-2\l_1|\vf_1|^2+2h_T|\varphi_2|^2 \right]\vf_1&=&G \varphi_1\nabla\cdot{\rm Re}\,(\varphi_2\nabla\varphi_2^*) \, , \label{eom1}\\[2ex]
  \left(\Delta+\mu_2^2-m_{2,T}^2-2\lambda_2|\varphi_2|^2+2h_T|\varphi_1|^2\right)\varphi_2&=&G\varphi_2\nabla\cdot{\rm Re}\,[\varphi_1(\nabla+iq\vec{A})\varphi_1^*] \, , \label{eom2}\\[2ex]
 \nabla\times\vec{B}+8\pi q\,{\rm Im}\,[\varphi_1(\nabla+iq\vec{A})\varphi_1^*] &=& 0 \, .\label{gauge}
  \eea
\end{subequations}
Since the transition from the flux tube phase to the 
normal-conducting phase is assumed to be of second order, the charged condensate becomes infinitesimally small just below $H_{c2}$, and we make the ansatz
$\varphi_1 = \bar{\varphi}_{1}+\delta\varphi_1$ with $\bar{\varphi}_{1} \propto (H_{c2}-H)^{1/2}$, and $\delta\varphi_1$ includes terms of order $(H_{c2}-H)^{3/2}$ and higher, i.e., 
is at least of order $\bar{\varphi}_1^3$.
We also introduce perturbations for the neutral condensate and the gauge field,
$\varphi_2=\bar{\varphi}_2+\delta\varphi_2$, $\vec{A} = (\bar{A}_{y}+\delta A_y) \vec{e}_y$, where $\delta A_y,\delta\varphi_2$ include terms of order $\propto H_{c2}-H$ and higher,
i.e., they are at least of order $\bar{\varphi}_{1}^2$.  As the magnetic field completely penetrates the superconductor at the phase transition, we can choose the 
unperturbed gauge field to be of the form 
$\bar{A}_{y} = x H_{c2}$, and we denote $\delta B = \partial_x\delta A_y$, such that $\vec{B} = (H_{c2}+\delta B)\vec{e}_z$. We assume all functions to be real and 
to depend on $x$ only, not on $y$ and $z$ (solutions with these properties are sufficient for our purpose, the derivation would also work without these
 restrictions but would be somewhat more tedious). We insert this ansatz into the equations 
of motion (\ref{eom}), and keep terms up to order $\bar{\varphi}_{1}^3$.
Then, the linear contributions from Eqs.\ (\ref{eom1}) and (\ref{eom2}) yield two equations for $\bar{\varphi}_1$ and $\bar{\varphi}_2$,
\begin{subequations} \label{lowest}
\bea
{\cal D}_1\bar{\varphi}_{1} &=& 0 \, ,\label{lowest1} \\[2ex]
{\cal D}_2\bar{\varphi}_2 &=& 0 \, ,\label{lowest2}
\eea
\end{subequations}
with
\begin{subequations} 
\bea
{\cal D}_1&\equiv& \partial_x^2-q^2\bar{A}_y^2+\mu_1^2-m_{1,T}^2+2h_T\bar{\varphi}_{2}^2-G\partial_x(\bar{\varphi}_2\partial_x\bar{\varphi}_2) \, , \\[2ex]
{\cal D}_2&\equiv& \partial_x^2+\mu_2^2-m_{2,T}^2-2\lambda_2\bar{\varphi}_{2}^2 \, ,
\eea
\end{subequations}
while the subleading contributions from Eqs.\ (\ref{eom1}) and (\ref{eom2}) and the leading contribution from Eq.\ (\ref{gauge}) yield the following equations for 
the perturbations $\delta\varphi_1$, $\delta\varphi_2$, and $\delta A_y$, 
\begin{subequations}\label{higher} 
\bea
{\cal D}_1\delta \varphi_1 &=& 
\Big[2(q^2\bar{A}_{y}\delta A_y+\lambda_1\bar{\varphi}_{1}^2
-2h_T\bar{\varphi}_2\delta\varphi_2)+G\partial_x^2(\bar{\varphi}_2\delta\varphi_2)\Big]\bar{\varphi}_1 \, , \label{phi31}\\[2ex]
{\cal D}_2\delta \varphi_2&=& \Big[2(2\lambda_2\bar{\varphi}_2\delta\varphi_2-h_T\bar{\varphi}_1^2)+G\partial_x(\bar{\varphi}_1\partial_x\bar{\varphi}_1)\Big]\bar{\varphi}_2 \, ,\label{rho31}\\[2ex]
\partial_x^2\delta A_y &=& -8\pi q^2\bar{A}_y\bar{\varphi}_1^2 \, .\label{gauge21}
\eea
\end{subequations}
Inserting our ansatz into the potential (\ref{freeener}), using partial integration and the equations of motion (\ref{lowest}) and (\ref{higher}), and keeping terms up to 
order $\bar{\varphi}_{1}^4$, we find after some algebra the free energy
\be \label{FGh}
F = \int d^3r \left\{\frac{B^2}{8\pi}-\lambda_1\bar{\varphi}_1^4-\lambda_2\bar{\varphi}_2^4 +\bar{\varphi}_2\delta\varphi_2[2h_T\bar{\varphi}_1^2-G\partial_x(\bar{\varphi}_1\partial_x\bar{\varphi}_1)]\right\} \, .
\ee
We will first compute $H_{c2}$ from Eqs.\ (\ref{lowest}) and afterwards compute the Gibbs free energy just below $H_{c2}$ from Eq.\ (\ref{FGh}). 

We assume the neutral condensate in the SF phase to be homogeneous, and thus Eq.\ (\ref{lowest2}) yields $2\bar{\varphi}_2^2=\rho_{\rm SF}^2$, as expected. 
For the solution of Eq.\ (\ref{lowest1}) we can simply follow the textbook arguments because it has the 
same structure as for a single-component superconductor. It reads 
\be \label{z}
(-\partial_x^2 +q^2H_{c2}^2x^2)\bar{\varphi}_{1}=(\lambda_1\rho_{\rm SC}^2+h_T\rho_{\rm SF}^2)\bar{\varphi}_{1}   \, ,
\ee
and thus is equivalent to the Schr\"{o}dinger equation for the one-dimensional harmonic oscillator, $-\frac{\hbar^2}{2m}\psi''(x) +\frac{m}{2}\omega^2x^2\psi=E\psi$ 
with the identification $E/(\hbar\omega)=(\lambda_1\rho_{\rm SC}^2+h_T\rho_{\rm SF}^2)/(2q H_{c2} )$. Since the eigenvalues are $E_n =(n+\frac{1}{2})\hbar\omega$, 
the largest magnetic field for which the equation allows a physical solution is obtained by setting $n=0$,
\be
H_{c2} = \frac{\lambda_1\rho_{\rm SC}^2}{q}\left(1+\frac{h_T\rho_{\rm SF}^2}{\lambda_1\rho_{\rm SC}^2}\right)
=\frac{1}{q\xi^2}\left(1-\frac{h_T^2}{\lambda_1\lambda_2}\right)\, ,
\ee
in agreement with Eq.\ (13) of Ref.\ \cite{Sinha:2015bva}.  In the second expression we have rewritten the condensates 
$\rho_{\rm SC}$ and $\rho_{\rm SF}$ in terms of the charged condensate in the coexistence phase $\rho_{01}$, see Eq.\ (\ref{COE}),
 and used the definition of the coherence length $\xi$ from Eq.\ (\ref{ellxi}). 
Since the relevant eigenvalue of Eq.\ (\ref{z}) is given by $n=0$, the corresponding eigenfunction is  a Gaussian,
\be 
\bar{\vf}_{1}(x) = C_0 e^{-x^2qH_{c2}/2} \, , \label{eq:SEsol}
\ee
where the exact value of the prefactor $C_0\propto (H_{c2}-H)^{1/2}$ is not relevant for the following. The result shows that, for $H$ just below $H_{c2}$, charged 
condensation with small magnitude of order $(H_{c2}-H)^{1/2}$ occurs in a slab confined in a direction perpendicular to the external magnetic field, here chosen to be the  
$x$-direction, with width $(qH_{c2})^{-1/2}$. Had we allowed for $y$ and $z$ dependencies of the condensate, we could have 
used this linearized approximation to discuss crystalline configurations and determine the preferred lattice structure. Here we continue by checking whether the solution (\ref{eq:SEsol}) is energetically preferred over the normal-conducting phase for $H$ below and close to $H_{c2}$.
To this end, we need to compute the Gibbs free energy, as defined in Eq.\ (\ref{Gibbsdef}), from the free energy (\ref{FGh}). We first solve Eq.\ (\ref{gauge21}) 
with the boundary condition $\delta B(\pm\infty)=H-H_{c2}$ (since $B=H$ in the normal-conducting phase) to find
\be 
\delta B(x) =-(H_{c2}-H)+4\pi q\bar{\varphi}_{1}^2(x) \, .
\ee
Inserting this result into Eq.\ (\ref{FGh}) and using Eq.\ (\ref{eq:SEsol}) yields the Gibbs free energy 
\bea \label{gibbs_Hc2}
{\cal G}_{\rm COE}&=& \mathcal{G}_{\rm SF}+\int d^3r\left\{\left(\frac{1}{2\kappa^2}-1\right)\lambda_1\bar{\varphi}_1^4 +\bar{\varphi}_2\delta\varphi_2\Big[2h_T\bar{\varphi}_1^2-G\partial_x(\bar{\varphi}_1\partial_x\bar{\varphi}_1)\Big]\right\} \, , 
\eea
with $\mathcal{G}_{\rm SF}$ from Eq.\ (\ref{GibbsSF}). 
It remains to compute $\delta\varphi_2$. We use Eq.\ (\ref{rho31}), which can be written as
\be
(\partial_t^2-p^2)\delta\varphi_2(t) = -\frac{h_T p^2 C_0^2}{2\sqrt{2}\lambda_2\rho_{\rm SF}} e^{-t^2}(2+\gamma-2\gamma t^2) \, ,
\ee
with the dimensionless variable $t=\sqrt{q H_{c2}} \,x$ and the dimensionless quantities
\be
p^2 = \frac{2\lambda_2\rho_{\rm SF}^2}{qH_{c2}} \, , \qquad \gamma = \frac{GqH_{c2}}{h_T} \, ,
\ee
where $p$ indicates the magnitude of the neutral condensate and $\gamma$ the magnitude of the gradient coupling $G$ relative to the density coupling $h_T$, both in units given by the critical magnetic field.  With the boundary conditions $\delta\varphi_2(\pm \infty)=0$, this equation has the solution
\be \label{dphi2t}
\delta\varphi_2(t) = \frac{1}{2}\frac{h_T p^2 C_0^2}{2\sqrt{2}\lambda_2\rho_{\rm SF}}\left[\gamma e^{-t^2}+\frac{\sqrt{\pi}}{p}\left(1-\frac{p^2\gamma}{4}\right){\cal Z}(p,t)\right] \, , 
\ee
where we have abbreviated
\be
{\cal Z}(p,t)\equiv e^{p^2/4}\left\{
e^{pt}\left[1-{\rm erf}\left(\frac{p}{2}+t\right)\right]+e^{-pt}\left[1-{\rm erf}\left(\frac{p}{2}-t\right)\right]\right\} \, ,
\ee
with the error function erf. Inserting Eq.\ (\ref{dphi2t}) into Eq.\ (\ref{gibbs_Hc2}) yields 
\be \label{DeltaGfull}
\frac{{\cal G}_{\rm COE}}{V} = \frac{{\cal G}_{\rm SF}}{V} + \lambda_1\langle\bar{\varphi}_1^4\rangle\left(\frac{1}{2\kappa^2}-1+\frac{h_T^2}{\lambda_1\lambda_2}
\left\{\frac{p^2\gamma}{4}\left(1+\frac{\gamma}{4}\right)+\left(1-\frac{p^2\gamma}{4}\right)\left[\left(1+\frac{\gamma}{2}\right){\cal I}_1(p)-\gamma{\cal I}_2(p)\right]\right\}\right) \, ,
\ee
where $\langle\ldots\rangle$ denotes spatial average, and 
\be
{\cal I}_1(p) \equiv \frac{p}{2\sqrt{2}}\int_{-\infty}^\infty dt\,e^{-t^2} {\cal Z}(p,t) \, , \qquad {\cal I}_2(p) \equiv \frac{p}{2\sqrt{2}}\int_{-\infty}^\infty dt\,t^2e^{-t^2} {\cal Z}(p,t) \, .
\ee
We discuss this result for the case without gradient coupling, $\gamma=0$, in the main text.

\section{Interaction between two flux tubes}
\label{app:fl_int}

In this appendix we derive the expression for the interaction energy Eq.\ (\ref{Fint}). We start from the definition (\ref{Fintdef}), i.e., we consider 
two parallel flux tubes $(a)$ and $(b)$ separated by the (dimensionless) distance $R_0$. We divide the total volume $V$ into
two half-spaces $V^{(a)}$ and $V^{(b)}$, which are the simplest versions of two Wigner-Seitz cells: we connect the two flux tubes by a line with length $R_0$, 
and the plane in the center of and 
perpendicular to that line divides $V$ into $V^{(a)}$ and $V^{(b)}$. The interaction free energy is then computed from 
\be \label{FintAB}
F_{\rm int}^\circlearrowleft = 2\int_{V^{(a)}}d^3r \, \left[U_\circlearrowleft^{(a)+(b)} - U^{(a)}_\circlearrowleft - U^{(b)}_\circlearrowleft\right] \, , 
\ee
where, due to the symmetry of the configuration, we have restricted the integration to the half-space $V^{(a)}$, where $U^{(a)}_\circlearrowleft$, $U^{(b)}_\circlearrowleft$ are 
the free energy densities of the two flux tubes in the absence of the other flux tube, and where $U_\circlearrowleft^{(a)+(b)}$ is the total free energy of the flux tubes.
(Recall that by definition $U_\circlearrowleft$ denotes the pure flux tube energy density, with the free energy density of the homogeneous configuration already subtracted.)

We assume $R_0$ to be much larger than the widths of the flux tubes, such that the contribution of flux tube $(b)$ to the free energy is small in $V^{(a)}$. Therefore, we will now compute the free energy density of a "large" contribution that solves the full equations of motion plus a "small" contribution that solves the linearized equations of motion. We shall do so in a  general notation, not referring to the geometry of our two-flux tube setup. Only in Eq.\ (\ref{Fint3}), when we insert the results into the free energy (\ref{FintAB}), we shall come back to this setup and introduce a more explicit notation indicating the contributions of the two different flux tubes. 
Following Ref.\ \cite{Kramer:1971zza}, we define
\be
\vec{Q} \equiv \xi (q\vec{A}-\nabla\psi_1) = -\frac{n(1-a)}{R}\vec{e}_\theta \, ,
\ee
and write   
\begin{subequations}
\bea
\vec{Q} &=& \vec{Q}_0 + \delta\vec{Q} \, , \\[2ex]
f_1 &=& f_{10} + \delta f_1 \, , \\[2ex]
f_2 &=& f_{20} + \delta f_2 \, .
\eea
\end{subequations} 
The equations of motion for a single flux tube to leading order, $\delta\vec{Q} = \delta f_1 = \delta f_2 =0$, are (from now on, in this appendix, all gradients are taken with respect to the dimensionless coordinates)
\begin{subequations}
\bea
0&=&\nabla\times(\nabla\times \vec{Q}_0) + \frac{f_{10}^2}{\kappa^2}\vec{Q}_0  \, , \label{zeroQ}\\[2ex]
0&=&\Delta f_{10}+ f_{10}(1-f_{10}^2-Q_0^2) -\frac{h_T}{\lambda_1} x^2 f_{10}(1-f_{20}^2)- \frac{\Gamma x}{2}f_{10}\nabla\cdot(f_{20}\nabla f_{20}) \, , \label{zerof1}\\[2ex]
0&=&\Delta f_{20}+ \frac{\lambda_2}{\lambda_1}x^2 f_{20}(1-f_{20}^2) - \frac{h_T}{\lambda_1} f_{20}(1-f_{10}^2) -\frac{\Gamma}{2x}f_{20}\nabla\cdot(f_{10}\nabla f_{10})\, , \label{zerof2}
\eea
\end{subequations}
[equivalent to Eqs.\ (\ref{eomA}) in the main text], and the equations of motion of first order in the corrections $\delta\vec{Q}$, $\delta f_1$, $\delta f_2$ become
\begin{subequations}\label{oneQ12}
\bea
0&=&\nabla\times(\nabla\times \delta\vec{Q})+ \frac{f_{10}}{\kappa^2}(f_{10}\delta\vec{Q}+2\delta f_1\vec{Q}_0)    \, , \label{oneQ}\\[2ex]
0&=& -\vec{Q}_0\cdot(2f_{10}\delta\vec{Q}+\delta f_1\vec{Q}_0) \non[2ex]
&&+\Delta \delta f_{1}+ \delta f_1(1-3f_{10}^2)-\frac{h_T}{\lambda_1} x^2 [\delta f_1(1-f_{20}^2) -2f_{10}f_{20}\delta f_2]
- \frac{{\Gamma }x}{2}[\delta f_1\nabla\cdot (f_{20}\nabla f_{20})+f_{10}\Delta(f_{20} \delta f_2)] \, ,\label{onef1} \\[2ex]
0&=&\Delta \delta f_{2}+ \frac{\lambda_2}{\lambda_1}x^2\delta f_2(1-3f_{20}^2)-\frac{h_T}{\lambda_1} [\delta f_2(1-f_{10}^2) -2f_{10}f_{20}\delta f_1]
- \frac{{\Gamma }}{2x}[\delta f_2\nabla\cdot (f_{10}\nabla f_{10})+f_{20}\Delta(f_{10} \delta f_1)]\, . \hspace{0.5cm} \label{onef2}
\eea
\end{subequations}
We denote the free energy density, up to second order and after using the equations of motions, by $U_0 + \delta U$, where 
\bea
U_0 &=& \frac{\rho_{01}^2}{2} \left\{\kappa^2(\nabla\times \vec{Q}_0)^2 +(\nabla f_{10})^2+f_{10}^2Q_0^2+\frac{(1-f_{10}^2)^2}{2} +x^2
\left[(\nabla f_{20})^2+ \frac{\lambda_2}{\lambda_1}x^2\frac{(1-f_{20}^2)^2}{2}\right]\right. \non[2ex]
&& \left.-\frac{h_T}{\lambda_1} x^2 (1-f_{10}^2)(1-f_{20}^2)-{\Gamma}x f_{10}f_{20}\nabla f_{10}\cdot\nabla f_{20}\right\}  
\eea
is the free energy density of a single flux tube from Eq.\ (\ref{Efl}), and the first-order and second-order corrections can be written as a total derivative, 
\bea
\delta U &=& \rho_{01}^2 \nabla\cdot \Bigg\{\kappa^2 \delta\vec{Q}\times\left[\nabla\times\left(\vec{Q} _0+\frac{\delta\vec{Q}}{2}\right)\right] 
+\delta f_1\nabla\left(f_{10}+\frac{\delta f_1}{2}\right)
+x^2\delta f_2\nabla\left(f_{20}+\frac{\delta f_2}{2}\right) \non[2ex]
&& \hspace{1cm}-\frac{{\Gamma}x}{2} \left[\delta f_1\left(f_{10}+\frac{\delta f_1}{2}\right)f_{20}\nabla f_{20}+\delta f_2\left(f_{20}+\frac{\delta f_2}{2}\right)f_{10}\nabla f_{10}+\frac{1}{2}\nabla(f_{10}f_{20}\delta f_1 \delta f_2)\right]
\Bigg\}   \, . 
\eea
Notice that any explicit dependence on the density coupling $h_T$ has disappeared, while the derivative coupling $\Gamma$ does appear explicitly.

We can now go back to the interaction free energy (\ref{FintAB}) and identify the full free energy $U_\circlearrowleft^{(a)+(b)}$ in the half-space $V^{(a)}$with $U_0+\delta U$. 
In $V^{(a)}$, $U_\circlearrowleft^{(a)}$ is given by setting $\delta \vec{Q}=\delta f_1 = \delta f_2=0$ in $U_0 + \delta U$
(which simply leaves $U_0$), and
$U^{(b)}_\circlearrowleft$ is obtained by setting $\vec{Q}_0=0$, $f_{10} = f_{20}=1$ in $U_0 + \delta U$ (which leaves various terms from $\delta U$). Consequently, we find
\bea \label{Fint3}
F_{\rm int}^\circlearrowleft
&\simeq&2 \rho_{01}^2 \int_{\partial V^{(a)}} d\vec{S}\cdot\Bigg\{\kappa^2 \delta\vec{Q}^{(b)}\times\left(\nabla\times\vec{Q} _0^{(a)}\right) 
+\delta f_1^{(b)}\nabla f_{10}^{(a)}+x^2\delta f_2^{(b)}\nabla f_{20}^{(a)} -\frac{\Gamma x}{2}\left[\delta f_1^{(b)}\left(f_{10}^{(a)}
+\frac{\delta f_1^{(b)}}{2}\right)f_{20}^{(a)}\nabla f_{20}^{(a)} \right.\non[2ex]
&&\left.+\delta f_2^{(b)}\left(f_{20}^{(a)}+\frac{\delta f_2^{(b)}}{2}\right)f_{10}^{(a)}\nabla f_{10}^{(a)}+\frac{1}{2}\nabla(f_{10}^{(a)}f_{20}^{(a)}\delta f_1^{(b)} \delta f_2^{(b)})-\frac{1}{2}\nabla(\delta f_1^{(b)}\delta f_2^{(b)}) \right] \Bigg\} \, , 
\eea
where we have rewritten the volume integral as a surface integral and where we have made the contributions from the two flux tubes $(a)$ and $(b)$ explicit. 
Since the derivatives of all fields vanish at infinity, the integration surface is reduced to the 
plane that separates the two 
Wigner-Seitz cells. We now use the geometry of the setup to simplify this expression: we align the $z$-axis with flux tube $(a)$, such that this flux tube sits 
in the origin of the $x$-$y$ plane, with the $x$-axis connecting the two flux tubes. Therefore, $\vec{Q}^{(a)}$, $f_{10}^{(a)}$, $f_{20}^{(a)}$ are functions only of $R$, while 
$\delta \vec{Q}^{(b)}$, $\delta f_{1}^{(b)}$, $\delta f_{2}^{(b)}$ also depend on the azimuthal angle $\theta$. However, since we only need the functions and their gradients at the boundary between the two Wigner-Seitz cells and since this boundary is by assumption far away not only from flux tube $(b)$ but also from flux tube $(a)$, we can write ($i=1,2$)
\begin{subequations} \label{surface}
\bea
\vec{Q}_0^{(a)} &\simeq& \delta \vec{Q}^{(a)} \equiv - \delta Q \,\vec{e}_\theta = \delta Q (\sin\theta \, \vec{e}_x - \cos\theta \, \vec{e}_y) \, , \qquad 
\delta \vec{Q}^{(b)} = -\delta Q (\sin\theta \, \vec{e}_x + \cos\theta \, \vec{e}_y) \, , \\[2ex]
f_{i0}^{(a)} &\simeq& 1- \delta f_i^{(a)} \, , \qquad \delta f_{i}^{(b)}  = \delta f_i^{(a)}  \equiv \delta f_{i} \, , \\[2ex]
\nabla f_{i0}^{(a)} &\simeq& -\nabla \delta f_i^{(a)} = -\delta f_i'\,\vec{e}_R = -\delta f_i'(\cos\theta \, \vec{e}_x + \sin\theta \, \vec{e}_y) \, , \qquad 
\nabla \delta f_{i}^{(b)} = \delta f_i' (-\cos\theta \, \vec{e}_x + \sin\theta \, \vec{e}_y) \, .
\eea
\end{subequations}
Note in particular that, at the relevant surface, $\delta f_{i}^{(b)}  = \delta f_i^{(a)}$, but $d\vec{S}\cdot \nabla \delta f_i^{(a)} = -d\vec{S}\cdot \nabla \delta f_i^{(b)}$.
Now, $\delta Q$ and $\delta f_i$ are functions only of $R$. Inserting Eqs.\ (\ref{surface}) into Eq.\ (\ref{Fint3}) yields 
\bea
\frac{F_{\rm int}^\circlearrowleft}{L} &=& 2 \rho_{01}^2R_0\int_{R_0/2}^\infty \frac{dR}{\sqrt{R^2-(R_0/2)^2}} \left\{-\kappa^2 \delta Q\left(\frac{\delta Q}{R}+\delta Q'\right) + \delta f_1 \delta f_1' + x^2\delta f_2\delta f_2' \right.\non[2ex]
&& \hspace{4.3cm}\left.-\frac{\Gamma x}{4}[2(1-\delta f_1-\delta f_2)+\delta f_1\delta f_2](\delta f_1\delta f_2)'\right\} \, .
\eea
We can employ this result by inserting the modified Bessel functions from Eq.\ (\ref{asympsol}),
\begin{subequations} 
\bea
\delta Q &\simeq&  -nC K_1(R/\kappa) \, , \\[2ex]
\delta f_1 &\simeq& -D_+ \gamma_+ K_0(\sqrt{\nu_+}R) -D_- \gamma_- K_0(\sqrt{\nu_-}R) \, , \\[2ex]
\delta f_2 &\simeq& -D_+  K_0(\sqrt{\nu_+}R) -D_-  K_0(\sqrt{\nu_-}R) \, . 
\eea
\end{subequations} 
We may also extrapolate this result down to 
smaller distances by reinstating the full numerical functions through $\delta Q \to Q = -n(1-a)/R$ and $\delta f_i \to 1-f_i$, which yields the result (\ref{Fint}) in the main text.

\section{Asymptotic approximation of flux tube interaction with gradient coupling}
\label{app:asymp}

In the main text, we discuss the large-distance behavior of the flux tube interaction without gradient coupling. In the presence of a gradient coupling,
the interaction is more complicated, but, as we show in this appendix,  an equally compact expression can be derived if we are only interested in the leading order contributions, i.e., 
the exponential behavior. 

We start by inserting the  asymptotic solutions (\ref{asympsol}) into the expression for the interaction free energy (\ref{Fint}). The result is an integral over a sum of many terms, 
each of which is a product of 2, 3, or 4 modified Bessel functions of the second kind. In each product, one factor is $K_1$ and the remaining ones are $K_0$. 
The integral over the terms with 2 Bessel functions that have the same argument can be expressed again as a Bessel function with the help of Eq.\ (\ref{Kint}).
For the integral over all other products we use the expansion,  
\be \label{Besselexp}
K_n(z) = \sqrt{\frac{\pi}{2z}} e^{-z}\left[1+\frac{4n^2-1}{8z}+{\cal O}\left(\frac{1}{z^2}\right)\right]  \, ,
\ee
and only keep terms with the smallest exponential suppression. These terms are found as follows. With Eq.\  (\ref{Besselexp}) we approximate
\be \label{K0K1}
e^{-\alpha R}\simeq \frac{\alpha R}{\pi}K_0(\alpha R/2)K_1(\alpha R/2) \, . 
\ee
Then, we approximate each product of Bessel functions $K_0K_1$, $K_0K_0K_1$, $K_0K_0K_0K_1$ by the leading order term, and re-express the exponential 
as a product $K_0K_1$ with the help of Eq.\ (\ref{K0K1}). If we have started with a product $K_0K_1$ with different arguments, we arrive
at an expression which we can integrate using Eq.\ (\ref{Kint}). If we have started with a product of 3 or 4 Bessel function, we do not exactly reproduce the 
integrand of Eq.\ (\ref{Kint}) because there is an additional factor $R^{-1/2}$ (for 3 Bessel functions) or $R^{-1}$ (for 4 Bessel functions). The resulting integral can be 
expressed in terms of the so-called Meijer G-function, which we expand again since we are anyway only interested in the asymptotic behavior. As a result, we 
obtain 
\begin{subequations}  
\bea
\int_{R_0/2}^\infty dR\,\frac{K_0(\alpha_1 R)K_1(\alpha_2  R)}{\sqrt{R^2-(R_0/2)^2}} &\sim& e^{-\frac{\alpha_1+\alpha_2}{2}R_0} \, ,\\[2ex]
\int_{R_0/2}^\infty dR\,\frac{K_0(\alpha_1 R)K_0(\alpha_2 R)K_1(\alpha_3  R)}{\sqrt{R^2-(R_0/2)^2}} &\sim& e^{-\frac{\alpha_1+\alpha_2+\alpha_3}{2}R_0} \, ,\\[2ex]
\int_{R_0/2}^\infty dR\,\frac{K_0(\alpha_1 R)K_0(\alpha_2 R)K_0(\alpha_3 R)K_1(\alpha_4  R)}{\sqrt{R^2-(R_0/2)^2}} &\sim& e^{-\frac{\alpha_1+\alpha_2+\alpha_3+\alpha_4}{2}R_0} \, .
\eea
\end{subequations}
For each of the terms in the interaction energy we need to replace $\alpha_i$ by either $\sqrt{\nu_+}$ or $\sqrt{\nu_-}$. From Eq.\ (\ref{nugam}) we see that
$\sqrt{\nu_+}>\sqrt{\nu_-}$. Therefore, the largest contribution we obtain is $\exp(-\sqrt{\nu_-}R_0)$, and this contribution is only created by the product of 2 Bessel functions
with the same argument $\sqrt{\nu_-}$ because 2 Bessel functions with different arguments give rise to $\exp[-(\sqrt{\nu_+}+\sqrt{\nu_-})R_0/2]$, which is suppressed  more strongly, 
3 Bessel functions give rise to suppressions of at least $\exp[-3\sqrt{\nu_-}R_0/2]$ etc. The largest contributions are thus given by the terms where we can apply the integral 
(\ref{Kint}), and we obtain 
\be 
\frac{F_{\rm int}^{\circlearrowleft}(R_0)}{L} \simeq2\pi\rho_{01}^2[\kappa^2 n^2 C^2 K_0(R_0/\kappa) - D_+^2(\gamma_-^2+x^2-\Gamma x\gamma_-) K_0(R_0\sqrt{\nu_-})] \, .
\ee
Therefore, if $\gamma_-^2+x^2-\Gamma x\gamma_->0$, one can use the same arguments as in the main text for the discussion of the attractiveness of the flux tube 
interaction at large distances, only with a more complicated eigenvalue $\nu_-$, which now depends on the gradient coupling $\Gamma$.

\bibliography{refs1}

\begin{thebibliography}{52}
\expandafter\ifx\csname natexlab\endcsname\relax\def\natexlab#1{#1}\fi
\expandafter\ifx\csname bibnamefont\endcsname\relax
  \def\bibnamefont#1{#1}\fi
\expandafter\ifx\csname bibfnamefont\endcsname\relax
  \def\bibfnamefont#1{#1}\fi
\expandafter\ifx\csname citenamefont\endcsname\relax
  \def\citenamefont#1{#1}\fi
\expandafter\ifx\csname url\endcsname\relax
  \def\url#1{\texttt{#1}}\fi
\expandafter\ifx\csname urlprefix\endcsname\relax\def\urlprefix{URL }\fi
\providecommand{\bibinfo}[2]{#2}
\providecommand{\eprint}[2][]{\url{#2}}

\bibitem[{\citenamefont{Bogoliubov}(1958)}]{Bogolyubov1958}
\bibinfo{author}{\bibfnamefont{N.}~\bibnamefont{Bogoliubov}},
  \bibinfo{journal}{Doklady Akad. Nauk SSSR} \textbf{\bibinfo{volume}{119}},
  \bibinfo{pages}{52} (\bibinfo{year}{1958}).

\bibitem[{\citenamefont{Migdal}(1959)}]{MIGDAL1959655}
\bibinfo{author}{\bibfnamefont{A.}~\bibnamefont{Migdal}},
  \bibinfo{journal}{Nuclear Physics} \textbf{\bibinfo{volume}{13}},
  \bibinfo{pages}{655 } (\bibinfo{year}{1959}).

\bibitem[{\citenamefont{Page et~al.}(2014)\citenamefont{Page, Lattimer,
  Prakash, and Steiner}}]{Page:2013hxa}
\bibinfo{author}{\bibfnamefont{D.}~\bibnamefont{Page}},
  \bibinfo{author}{\bibfnamefont{J.~M.} \bibnamefont{Lattimer}},
  \bibinfo{author}{\bibfnamefont{M.}~\bibnamefont{Prakash}}, \bibnamefont{and}
  \bibinfo{author}{\bibfnamefont{A.~W.} \bibnamefont{Steiner}}, in
  \emph{\bibinfo{booktitle}{Novel Superfluids: Volume 2}}, edited by
  \bibinfo{editor}{\bibfnamefont{K.~H.} \bibnamefont{Bennemann}}
  \bibnamefont{and} \bibinfo{editor}{\bibfnamefont{J.~B.}
  \bibnamefont{Ketterson}} (\bibinfo{publisher}{Oxford University Press, New
  York}, \bibinfo{year}{2014}), p. \bibinfo{pages}{505}, \eprint{1302.6626}.

\bibitem[{\citenamefont{Sedrakian and Clark}(2006)}]{Sedrakian:2006xm}
\bibinfo{author}{\bibfnamefont{A.}~\bibnamefont{Sedrakian}} \bibnamefont{and}
  \bibinfo{author}{\bibfnamefont{J.~W.} \bibnamefont{Clark}},
  \bibinfo{journal}{Ser. Adv. Quant. Many Body Theor.}
  \textbf{\bibinfo{volume}{8}}, \bibinfo{pages}{135} (\bibinfo{year}{2006}),
  \eprint{nucl-th/0607028}.

\bibitem[{\citenamefont{Haber and Schmitt}(2017)}]{Haber:2016ljn}
\bibinfo{author}{\bibfnamefont{A.}~\bibnamefont{Haber}} \bibnamefont{and}
  \bibinfo{author}{\bibfnamefont{A.}~\bibnamefont{Schmitt}},
  \bibinfo{journal}{EPJ Web Conf.} \textbf{\bibinfo{volume}{137}},
  \bibinfo{pages}{09003} (\bibinfo{year}{2017}), \eprint{1612.01865}.

\bibitem[{\citenamefont{Chamel and Haensel}(2006)}]{Chamel:2006rc}
\bibinfo{author}{\bibfnamefont{N.}~\bibnamefont{Chamel}} \bibnamefont{and}
  \bibinfo{author}{\bibfnamefont{P.}~\bibnamefont{Haensel}},
  \bibinfo{journal}{Phys. Rev.} \textbf{\bibinfo{volume}{C73}},
  \bibinfo{pages}{045802} (\bibinfo{year}{2006}), \eprint{nucl-th/0603018}.

\bibitem[{\citenamefont{Glampedakis et~al.}(2011)\citenamefont{Glampedakis,
  Andersson, and Samuelsson}}]{Glampedakis:2010sk}
\bibinfo{author}{\bibfnamefont{K.}~\bibnamefont{Glampedakis}},
  \bibinfo{author}{\bibfnamefont{N.}~\bibnamefont{Andersson}},
  \bibnamefont{and}
  \bibinfo{author}{\bibfnamefont{L.}~\bibnamefont{Samuelsson}},
  \bibinfo{journal}{Mon. Not. Roy. Astron. Soc.}
  \textbf{\bibinfo{volume}{410}}, \bibinfo{pages}{805} (\bibinfo{year}{2011}),
  \eprint{1001.4046}.

\bibitem[{\citenamefont{Wambach et~al.}(1993)\citenamefont{Wambach, Ainsworth,
  and Pines}}]{wambach1993quasiparticle}
\bibinfo{author}{\bibfnamefont{J.}~\bibnamefont{Wambach}},
  \bibinfo{author}{\bibfnamefont{T.}~\bibnamefont{Ainsworth}},
  \bibnamefont{and} \bibinfo{author}{\bibfnamefont{D.}~\bibnamefont{Pines}},
  \bibinfo{journal}{Nuclear Physics A} \textbf{\bibinfo{volume}{555}},
  \bibinfo{pages}{128} (\bibinfo{year}{1993}).

\bibitem[{\citenamefont{Gusakov et~al.}(2009)\citenamefont{Gusakov, Kantor, and
  Haensel}}]{Gusakov:2009kc}
\bibinfo{author}{\bibfnamefont{M.~E.} \bibnamefont{Gusakov}},
  \bibinfo{author}{\bibfnamefont{E.~M.} \bibnamefont{Kantor}},
  \bibnamefont{and} \bibinfo{author}{\bibfnamefont{P.}~\bibnamefont{Haensel}},
  \bibinfo{journal}{Phys. Rev.} \textbf{\bibinfo{volume}{C79}},
  \bibinfo{pages}{055806} (\bibinfo{year}{2009}), \eprint{0904.3467}.

\bibitem[{\citenamefont{Alford et~al.}(1999)\citenamefont{Alford, Rajagopal,
  and Wilczek}}]{Alford:1998mk}
\bibinfo{author}{\bibfnamefont{M.~G.} \bibnamefont{Alford}},
  \bibinfo{author}{\bibfnamefont{K.}~\bibnamefont{Rajagopal}},
  \bibnamefont{and} \bibinfo{author}{\bibfnamefont{F.}~\bibnamefont{Wilczek}},
  \bibinfo{journal}{Nucl. Phys.} \textbf{\bibinfo{volume}{B537}},
  \bibinfo{pages}{443} (\bibinfo{year}{1999}), \eprint{hep-ph/9804403}.

\bibitem[{\citenamefont{Iida}(2005)}]{Iida:2004if}
\bibinfo{author}{\bibfnamefont{K.}~\bibnamefont{Iida}}, \bibinfo{journal}{Phys.
  Rev.} \textbf{\bibinfo{volume}{D71}}, \bibinfo{pages}{054011}
  (\bibinfo{year}{2005}), \eprint{hep-ph/0412426}.

\bibitem[{\citenamefont{Giannakis and Ren}(2003)}]{Giannakis:2003am}
\bibinfo{author}{\bibfnamefont{I.}~\bibnamefont{Giannakis}} \bibnamefont{and}
  \bibinfo{author}{\bibfnamefont{H.-c.} \bibnamefont{Ren}},
  \bibinfo{journal}{Nucl. Phys.} \textbf{\bibinfo{volume}{B669}},
  \bibinfo{pages}{462} (\bibinfo{year}{2003}), \eprint{hep-ph/0305235}.

\bibitem[{\citenamefont{Eto et~al.}(2014)\citenamefont{Eto, Hirono, Nitta, and
  Yasui}}]{Eto:2013hoa}
\bibinfo{author}{\bibfnamefont{M.}~\bibnamefont{Eto}},
  \bibinfo{author}{\bibfnamefont{Y.}~\bibnamefont{Hirono}},
  \bibinfo{author}{\bibfnamefont{M.}~\bibnamefont{Nitta}}, \bibnamefont{and}
  \bibinfo{author}{\bibfnamefont{S.}~\bibnamefont{Yasui}},
  \bibinfo{journal}{PTEP} \textbf{\bibinfo{volume}{2014}},
  \bibinfo{pages}{012D01} (\bibinfo{year}{2014}), \eprint{1308.1535}.

\bibitem[{\citenamefont{Glampedakis et~al.}(2012)\citenamefont{Glampedakis,
  Jones, and Samuelsson}}]{Glampedakis:2012qp}
\bibinfo{author}{\bibfnamefont{K.}~\bibnamefont{Glampedakis}},
  \bibinfo{author}{\bibfnamefont{D.~I.} \bibnamefont{Jones}}, \bibnamefont{and}
  \bibinfo{author}{\bibfnamefont{L.}~\bibnamefont{Samuelsson}},
  \bibinfo{journal}{Phys. Rev. Lett.} \textbf{\bibinfo{volume}{109}},
  \bibinfo{pages}{081103} (\bibinfo{year}{2012}), \eprint{1204.3781}.

\bibitem[{\citenamefont{Alford and Sedrakian}(2010)}]{Alford:2010qf}
\bibinfo{author}{\bibfnamefont{M.~G.} \bibnamefont{Alford}} \bibnamefont{and}
  \bibinfo{author}{\bibfnamefont{A.}~\bibnamefont{Sedrakian}},
  \bibinfo{journal}{J. Phys.} \textbf{\bibinfo{volume}{G37}},
  \bibinfo{pages}{075202} (\bibinfo{year}{2010}), \eprint{1001.3346}.

\bibitem[{\citenamefont{Bedaque and Sch{\"a}fer}(2002)}]{Bedaque:2001je}
\bibinfo{author}{\bibfnamefont{P.~F.} \bibnamefont{Bedaque}} \bibnamefont{and}
  \bibinfo{author}{\bibfnamefont{T.}~\bibnamefont{Sch{\"a}fer}},
  \bibinfo{journal}{Nucl. Phys.} \textbf{\bibinfo{volume}{A697}},
  \bibinfo{pages}{802} (\bibinfo{year}{2002}), \eprint{hep-ph/0105150}.

\bibitem[{\citenamefont{Alford et~al.}(2008{\natexlab{a}})\citenamefont{Alford,
  Schmitt, Rajagopal, and Sch{\"a}fer}}]{Alford:2007xm}
\bibinfo{author}{\bibfnamefont{M.~G.} \bibnamefont{Alford}},
  \bibinfo{author}{\bibfnamefont{A.}~\bibnamefont{Schmitt}},
  \bibinfo{author}{\bibfnamefont{K.}~\bibnamefont{Rajagopal}},
  \bibnamefont{and}
  \bibinfo{author}{\bibfnamefont{T.}~\bibnamefont{Sch{\"a}fer}},
  \bibinfo{journal}{Rev.Mod.Phys.} \textbf{\bibinfo{volume}{80}},
  \bibinfo{pages}{1455} (\bibinfo{year}{2008}{\natexlab{a}}),
  \eprint{0709.4635}.

\bibitem[{\citenamefont{{Ferrier-Barbut}
  et~al.}(2014)\citenamefont{{Ferrier-Barbut}, {Delehaye}, {Laurent}, {Grier},
  {Pierce}, {Rem}, {Chevy}, and {Salomon}}}]{2014Sci...345.1035F}
\bibinfo{author}{\bibfnamefont{I.}~\bibnamefont{{Ferrier-Barbut}}},
  \bibinfo{author}{\bibfnamefont{M.}~\bibnamefont{{Delehaye}}},
  \bibinfo{author}{\bibfnamefont{S.}~\bibnamefont{{Laurent}}},
  \bibinfo{author}{\bibfnamefont{A.~T.} \bibnamefont{{Grier}}},
  \bibinfo{author}{\bibfnamefont{M.}~\bibnamefont{{Pierce}}},
  \bibinfo{author}{\bibfnamefont{B.~S.} \bibnamefont{{Rem}}},
  \bibinfo{author}{\bibfnamefont{F.}~\bibnamefont{{Chevy}}}, \bibnamefont{and}
  \bibinfo{author}{\bibfnamefont{C.}~\bibnamefont{{Salomon}}},
  \bibinfo{journal}{Science} \textbf{\bibinfo{volume}{345}},
  \bibinfo{pages}{1035} (\bibinfo{year}{2014}), \eprint{1404.2548}.

\bibitem[{\citenamefont{{Delehaye} et~al.}(2015)\citenamefont{{Delehaye},
  {Laurent}, {Ferrier-Barbut}, {Jin}, {Chevy}, and
  {Salomon}}}]{2015PhRvL.115z5303D}
\bibinfo{author}{\bibfnamefont{M.}~\bibnamefont{{Delehaye}}},
  \bibinfo{author}{\bibfnamefont{S.}~\bibnamefont{{Laurent}}},
  \bibinfo{author}{\bibfnamefont{I.}~\bibnamefont{{Ferrier-Barbut}}},
  \bibinfo{author}{\bibfnamefont{S.}~\bibnamefont{{Jin}}},
  \bibinfo{author}{\bibfnamefont{F.}~\bibnamefont{{Chevy}}}, \bibnamefont{and}
  \bibinfo{author}{\bibfnamefont{C.}~\bibnamefont{{Salomon}}},
  \bibinfo{journal}{Physical Review Letters} \textbf{\bibinfo{volume}{115}},
  \bibinfo{eid}{265303} (\bibinfo{year}{2015}), \eprint{1510.06709}.

\bibitem[{\citenamefont{{Lin} et~al.}(2009)\citenamefont{{Lin}, {Compton},
  {Jim{\'e}nez-Garcia}, {Porto}, and {Spielman}}}]{2009Natur.462..628L}
\bibinfo{author}{\bibfnamefont{Y.-J.} \bibnamefont{{Lin}}},
  \bibinfo{author}{\bibfnamefont{R.~L.} \bibnamefont{{Compton}}},
  \bibinfo{author}{\bibfnamefont{K.}~\bibnamefont{{Jim{\'e}nez-Garcia}}},
  \bibinfo{author}{\bibfnamefont{J.~V.} \bibnamefont{{Porto}}},
  \bibnamefont{and} \bibinfo{author}{\bibfnamefont{I.~B.}
  \bibnamefont{{Spielman}}}, \bibinfo{journal}{Nature}
  \textbf{\bibinfo{volume}{462}}, \bibinfo{pages}{628} (\bibinfo{year}{2009}),
  \eprint{1007.0294}.

\bibitem[{\citenamefont{{Dalibard} et~al.}(2011)\citenamefont{{Dalibard},
  {Gerbier}, {Juzeli{\= u}nas}, and {{\"O}hberg}}}]{2011RvMP...83.1523D}
\bibinfo{author}{\bibfnamefont{J.}~\bibnamefont{{Dalibard}}},
  \bibinfo{author}{\bibfnamefont{F.}~\bibnamefont{{Gerbier}}},
  \bibinfo{author}{\bibfnamefont{G.}~\bibnamefont{{Juzeli{\= u}nas}}},
  \bibnamefont{and}
  \bibinfo{author}{\bibfnamefont{P.}~\bibnamefont{{{\"O}hberg}}},
  \bibinfo{journal}{Reviews of Modern Physics} \textbf{\bibinfo{volume}{83}},
  \bibinfo{pages}{1523} (\bibinfo{year}{2011}), \eprint{1008.5378}.

\bibitem[{\citenamefont{{Goldman} et~al.}(2014)\citenamefont{{Goldman},
  {Juzeli{\= u}nas}, {{\"O}hberg}, and {Spielman}}}]{2014RPPh...77l6401G}
\bibinfo{author}{\bibfnamefont{N.}~\bibnamefont{{Goldman}}},
  \bibinfo{author}{\bibfnamefont{G.}~\bibnamefont{{Juzeli{\= u}nas}}},
  \bibinfo{author}{\bibfnamefont{P.}~\bibnamefont{{{\"O}hberg}}},
  \bibnamefont{and} \bibinfo{author}{\bibfnamefont{I.~B.}
  \bibnamefont{{Spielman}}}, \bibinfo{journal}{Reports on Progress in Physics}
  \textbf{\bibinfo{volume}{77}}, \bibinfo{eid}{126401} (\bibinfo{year}{2014}),
  \eprint{1308.6533}.

\bibitem[{\citenamefont{Carlstrom et~al.}(2011)\citenamefont{Carlstrom, Babaev,
  and Speight}}]{Carlstrom:2010wn}
\bibinfo{author}{\bibfnamefont{J.}~\bibnamefont{Carlstrom}},
  \bibinfo{author}{\bibfnamefont{E.}~\bibnamefont{Babaev}}, \bibnamefont{and}
  \bibinfo{author}{\bibfnamefont{M.}~\bibnamefont{Speight}},
  \bibinfo{journal}{Phys. Rev.} \textbf{\bibinfo{volume}{B83}},
  \bibinfo{pages}{174509} (\bibinfo{year}{2011}), \eprint{1009.2196}.

\bibitem[{\citenamefont{Brandt and Das}(2011)}]{Brandt2011}
\bibinfo{author}{\bibfnamefont{E.~H.} \bibnamefont{Brandt}} \bibnamefont{and}
  \bibinfo{author}{\bibfnamefont{M.~P.} \bibnamefont{Das}},
  \bibinfo{journal}{Journal of Superconductivity and Novel Magnetism}
  \textbf{\bibinfo{volume}{24}}, \bibinfo{pages}{57} (\bibinfo{year}{2011}).

\bibitem[{\citenamefont{Babaev and Silaev}(2013)}]{2012arXiv1206.6786B}
\bibinfo{author}{\bibfnamefont{E.}~\bibnamefont{Babaev}} \bibnamefont{and}
  \bibinfo{author}{\bibfnamefont{M.}~\bibnamefont{Silaev}},
  \bibinfo{journal}{Journal of Superconductivity and Novel Magnetism}
  \textbf{\bibinfo{volume}{26}}, \bibinfo{pages}{2045} (\bibinfo{year}{2013}),
  \eprint{1206.6786}.

\bibitem[{\citenamefont{Wu et~al.}(2016)\citenamefont{Wu, Wu, and
  Zhang}}]{Wu:2015sqk}
\bibinfo{author}{\bibfnamefont{M.-S.} \bibnamefont{Wu}},
  \bibinfo{author}{\bibfnamefont{S.-Y.} \bibnamefont{Wu}}, \bibnamefont{and}
  \bibinfo{author}{\bibfnamefont{H.-Q.} \bibnamefont{Zhang}},
  \bibinfo{journal}{JHEP} \textbf{\bibinfo{volume}{05}}, \bibinfo{pages}{011}
  (\bibinfo{year}{2016}), \eprint{1511.01325}.

\bibitem[{\citenamefont{Babaev et~al.}(2004)\citenamefont{Babaev, Sudbo, and
  Ashcroft}}]{babaev2004superconductor}
\bibinfo{author}{\bibfnamefont{E.}~\bibnamefont{Babaev}},
  \bibinfo{author}{\bibfnamefont{A.}~\bibnamefont{Sudbo}}, \bibnamefont{and}
  \bibinfo{author}{\bibfnamefont{N.}~\bibnamefont{Ashcroft}},
  \bibinfo{journal}{Nature} \textbf{\bibinfo{volume}{431}},
  \bibinfo{pages}{666} (\bibinfo{year}{2004}).

\bibitem[{\citenamefont{{Tuoriniemi} et~al.}(2002)\citenamefont{{Tuoriniemi},
  {Martikainen}, {Pentti}, {Sebedash}, {Boldarev}, and
  {Pickett}}}]{2002JLTP..129..531T}
\bibinfo{author}{\bibfnamefont{J.}~\bibnamefont{{Tuoriniemi}}},
  \bibinfo{author}{\bibfnamefont{J.}~\bibnamefont{{Martikainen}}},
  \bibinfo{author}{\bibfnamefont{E.}~\bibnamefont{{Pentti}}},
  \bibinfo{author}{\bibfnamefont{A.}~\bibnamefont{{Sebedash}}},
  \bibinfo{author}{\bibfnamefont{S.}~\bibnamefont{{Boldarev}}},
  \bibnamefont{and}
  \bibinfo{author}{\bibfnamefont{G.}~\bibnamefont{{Pickett}}},
  \bibinfo{journal}{Journal of Low Temperature Physics}
  \textbf{\bibinfo{volume}{129}}, \bibinfo{pages}{531} (\bibinfo{year}{2002}).

\bibitem[{\citenamefont{Rysti et~al.}(2012)\citenamefont{Rysti, Tuoriniemi, and
  Salmela}}]{PhysRevB.85.134529}
\bibinfo{author}{\bibfnamefont{J.}~\bibnamefont{Rysti}},
  \bibinfo{author}{\bibfnamefont{J.}~\bibnamefont{Tuoriniemi}},
  \bibnamefont{and} \bibinfo{author}{\bibfnamefont{A.}~\bibnamefont{Salmela}},
  \bibinfo{journal}{Phys. Rev. B} \textbf{\bibinfo{volume}{85}},
  \bibinfo{pages}{134529} (\bibinfo{year}{2012}).

\bibitem[{\citenamefont{Haber et~al.}(2016)\citenamefont{Haber, Schmitt, and
  Stetina}}]{Haber:2015exa}
\bibinfo{author}{\bibfnamefont{A.}~\bibnamefont{Haber}},
  \bibinfo{author}{\bibfnamefont{A.}~\bibnamefont{Schmitt}}, \bibnamefont{and}
  \bibinfo{author}{\bibfnamefont{S.}~\bibnamefont{Stetina}},
  \bibinfo{journal}{Phys. Rev.} \textbf{\bibinfo{volume}{D93}},
  \bibinfo{pages}{025011} (\bibinfo{year}{2016}), \eprint{1510.01982}.

\bibitem[{\citenamefont{Alpar et~al.}(1984)\citenamefont{Alpar, Langer, and
  Sauls}}]{Alpar:1984zz}
\bibinfo{author}{\bibfnamefont{M.~A.} \bibnamefont{Alpar}},
  \bibinfo{author}{\bibfnamefont{S.~A.} \bibnamefont{Langer}},
  \bibnamefont{and} \bibinfo{author}{\bibfnamefont{J.~A.} \bibnamefont{Sauls}},
  \bibinfo{journal}{Astrophys. J.} \textbf{\bibinfo{volume}{282}},
  \bibinfo{pages}{533} (\bibinfo{year}{1984}).

\bibitem[{\citenamefont{Alford and Good}(2008)}]{Alford:2007np}
\bibinfo{author}{\bibfnamefont{M.~G.} \bibnamefont{Alford}} \bibnamefont{and}
  \bibinfo{author}{\bibfnamefont{G.}~\bibnamefont{Good}},
  \bibinfo{journal}{Phys. Rev.} \textbf{\bibinfo{volume}{B78}},
  \bibinfo{pages}{024510} (\bibinfo{year}{2008}), \eprint{0712.1810}.

\bibitem[{\citenamefont{{Kobyakov} and {Pethick}}(2017)}]{2015arXiv150400570K}
\bibinfo{author}{\bibfnamefont{D.~N.} \bibnamefont{{Kobyakov}}}
  \bibnamefont{and} \bibinfo{author}{\bibfnamefont{C.~J.}
  \bibnamefont{{Pethick}}}, \bibinfo{journal}{Astrophys.\ J.}
  \textbf{\bibinfo{volume}{836}}, \bibinfo{eid}{203} (\bibinfo{year}{2017}),
  \eprint{1504.00570}.

\bibitem[{\citenamefont{Sinha and Sedrakian}(2015)}]{Sinha:2015bva}
\bibinfo{author}{\bibfnamefont{M.}~\bibnamefont{Sinha}} \bibnamefont{and}
  \bibinfo{author}{\bibfnamefont{A.}~\bibnamefont{Sedrakian}},
  \bibinfo{journal}{Phys. Rev.} \textbf{\bibinfo{volume}{C91}},
  \bibinfo{pages}{035805} (\bibinfo{year}{2015}), \eprint{1502.02979}.

\bibitem[{\citenamefont{Kramer}(1971)}]{Kramer:1971zza}
\bibinfo{author}{\bibfnamefont{L.}~\bibnamefont{Kramer}},
  \bibinfo{journal}{Phys. Rev.} \textbf{\bibinfo{volume}{B3}},
  \bibinfo{pages}{3821} (\bibinfo{year}{1971}).

\bibitem[{\citenamefont{Speight}(1997)}]{Speight:1996px}
\bibinfo{author}{\bibfnamefont{J.~M.} \bibnamefont{Speight}},
  \bibinfo{journal}{Phys. Rev.} \textbf{\bibinfo{volume}{D55}},
  \bibinfo{pages}{3830} (\bibinfo{year}{1997}), \eprint{hep-th/9603155}.

\bibitem[{\citenamefont{Buckley
  et~al.}(2004{\natexlab{a}})\citenamefont{Buckley, Metlitski, and
  Zhitnitsky}}]{Buckley:2003zf}
\bibinfo{author}{\bibfnamefont{K.~B.~W.} \bibnamefont{Buckley}},
  \bibinfo{author}{\bibfnamefont{M.~A.} \bibnamefont{Metlitski}},
  \bibnamefont{and} \bibinfo{author}{\bibfnamefont{A.~R.}
  \bibnamefont{Zhitnitsky}}, \bibinfo{journal}{Phys. Rev. Lett.}
  \textbf{\bibinfo{volume}{92}}, \bibinfo{pages}{151102}
  (\bibinfo{year}{2004}{\natexlab{a}}), \eprint{astro-ph/0308148}.

\bibitem[{\citenamefont{Buckley
  et~al.}(2004{\natexlab{b}})\citenamefont{Buckley, Metlitski, and
  Zhitnitsky}}]{Buckley:2004ca}
\bibinfo{author}{\bibfnamefont{K.~B.~W.} \bibnamefont{Buckley}},
  \bibinfo{author}{\bibfnamefont{M.~A.} \bibnamefont{Metlitski}},
  \bibnamefont{and} \bibinfo{author}{\bibfnamefont{A.~R.}
  \bibnamefont{Zhitnitsky}}, \bibinfo{journal}{Phys. Rev.}
  \textbf{\bibinfo{volume}{C69}}, \bibinfo{pages}{055803}
  (\bibinfo{year}{2004}{\natexlab{b}}), \eprint{hep-ph/0403230}.

\bibitem[{\citenamefont{Alford et~al.}(2005)\citenamefont{Alford, Good, and
  Reddy}}]{Alford:2005ku}
\bibinfo{author}{\bibfnamefont{M.}~\bibnamefont{Alford}},
  \bibinfo{author}{\bibfnamefont{G.}~\bibnamefont{Good}}, \bibnamefont{and}
  \bibinfo{author}{\bibfnamefont{S.}~\bibnamefont{Reddy}},
  \bibinfo{journal}{Phys. Rev.} \textbf{\bibinfo{volume}{C72}},
  \bibinfo{pages}{055801} (\bibinfo{year}{2005}), \eprint{nucl-th/0505025}.

\bibitem[{\citenamefont{Bettencourt and Rivers}(1995)}]{Bettencourt:1994kf}
\bibinfo{author}{\bibfnamefont{L.~M.~A.} \bibnamefont{Bettencourt}}
  \bibnamefont{and} \bibinfo{author}{\bibfnamefont{R.~J.}
  \bibnamefont{Rivers}}, \bibinfo{journal}{Phys. Rev.}
  \textbf{\bibinfo{volume}{D51}}, \bibinfo{pages}{1842} (\bibinfo{year}{1995}),
  \eprint{hep-ph/9405222}.

\bibitem[{\citenamefont{MacKenzie et~al.}(2003)\citenamefont{MacKenzie, Vachon,
  and Wichoski}}]{MacKenzie:2003jp}
\bibinfo{author}{\bibfnamefont{R.}~\bibnamefont{MacKenzie}},
  \bibinfo{author}{\bibfnamefont{M.~A.} \bibnamefont{Vachon}},
  \bibnamefont{and} \bibinfo{author}{\bibfnamefont{U.~F.}
  \bibnamefont{Wichoski}}, \bibinfo{journal}{Phys. Rev.}
  \textbf{\bibinfo{volume}{D67}}, \bibinfo{pages}{105024}
  (\bibinfo{year}{2003}), \eprint{hep-th/0301188}.

\bibitem[{\citenamefont{Babaev and Speight}(2005)}]{PhysRevB.72.180502}
\bibinfo{author}{\bibfnamefont{E.}~\bibnamefont{Babaev}} \bibnamefont{and}
  \bibinfo{author}{\bibfnamefont{M.}~\bibnamefont{Speight}},
  \bibinfo{journal}{Phys. Rev. B} \textbf{\bibinfo{volume}{72}},
  \bibinfo{pages}{180502} (\bibinfo{year}{2005}).

\bibitem[{\citenamefont{Babaev et~al.}(2010)\citenamefont{Babaev, Carlstr\"om,
  and Speight}}]{PhysRevLett.105.067003}
\bibinfo{author}{\bibfnamefont{E.}~\bibnamefont{Babaev}},
  \bibinfo{author}{\bibfnamefont{J.}~\bibnamefont{Carlstr\"om}},
  \bibnamefont{and} \bibinfo{author}{\bibfnamefont{M.}~\bibnamefont{Speight}},
  \bibinfo{journal}{Phys. Rev. Lett.} \textbf{\bibinfo{volume}{105}},
  \bibinfo{pages}{067003} (\bibinfo{year}{2010}).

\bibitem[{\citenamefont{Alford et~al.}(2014)\citenamefont{Alford, Mallavarapu,
  Schmitt, and Stetina}}]{Alford:2013koa}
\bibinfo{author}{\bibfnamefont{M.~G.} \bibnamefont{Alford}},
  \bibinfo{author}{\bibfnamefont{S.~K.} \bibnamefont{Mallavarapu}},
  \bibinfo{author}{\bibfnamefont{A.}~\bibnamefont{Schmitt}}, \bibnamefont{and}
  \bibinfo{author}{\bibfnamefont{S.}~\bibnamefont{Stetina}},
  \bibinfo{journal}{Phys. Rev.} \textbf{\bibinfo{volume}{D89}},
  \bibinfo{pages}{085005} (\bibinfo{year}{2014}), \eprint{1310.5953}.

\bibitem[{\citenamefont{Fejos and Hatsuda}(2016)}]{Fejos:2016wza}
\bibinfo{author}{\bibfnamefont{G.}~\bibnamefont{Fejos}} \bibnamefont{and}
  \bibinfo{author}{\bibfnamefont{T.}~\bibnamefont{Hatsuda}},
  \bibinfo{journal}{Phys. Rev.} \textbf{\bibinfo{volume}{D93}},
  \bibinfo{pages}{121701} (\bibinfo{year}{2016}), \eprint{1604.05849}.

\bibitem[{\citenamefont{Kapusta}(1981)}]{Kapusta:1981aa}
\bibinfo{author}{\bibfnamefont{J.~I.} \bibnamefont{Kapusta}},
  \bibinfo{journal}{Phys. Rev.} \textbf{\bibinfo{volume}{D24}},
  \bibinfo{pages}{426} (\bibinfo{year}{1981}).

\bibitem[{\citenamefont{Alford et~al.}(2008{\natexlab{b}})\citenamefont{Alford,
  Braby, and Schmitt}}]{Alford:2007qa}
\bibinfo{author}{\bibfnamefont{M.~G.} \bibnamefont{Alford}},
  \bibinfo{author}{\bibfnamefont{M.}~\bibnamefont{Braby}}, \bibnamefont{and}
  \bibinfo{author}{\bibfnamefont{A.}~\bibnamefont{Schmitt}},
  \bibinfo{journal}{J. Phys.} \textbf{\bibinfo{volume}{G35}},
  \bibinfo{pages}{025002} (\bibinfo{year}{2008}{\natexlab{b}}),
  \eprint{arXiv:0707.2389 [nucl-th]}.

\bibitem[{\citenamefont{Tinkham}(2004)}]{tinkham2004introduction}
\bibinfo{author}{\bibfnamefont{M.}~\bibnamefont{Tinkham}},
  \emph{\bibinfo{title}{Introduction to Superconductivity}}
  (\bibinfo{publisher}{Dover Publications, New York}, \bibinfo{year}{2004}),
  ISBN \bibinfo{isbn}{9780486435039}.

\bibitem[{\citenamefont{Forgacs and Luk‡{\'a}cs}(2016)}]{Forgacs:2016ndn}
\bibinfo{author}{\bibfnamefont{P.}~\bibnamefont{Forgacs}} \bibnamefont{and}
  \bibinfo{author}{\bibfnamefont{{\'A}.}~\bibnamefont{Luk‡{\'a}cs}},
  \bibinfo{journal}{Phys. Lett.} \textbf{\bibinfo{volume}{B762}},
  \bibinfo{pages}{271} (\bibinfo{year}{2016}), \eprint{1603.03291}.

\bibitem[{\citenamefont{Forga‡cs and Luk‡{\'a}cs}(2016)}]{Forgacs:2016iva}
\bibinfo{author}{\bibfnamefont{P.}~\bibnamefont{Forga‡cs}} \bibnamefont{and}
  \bibinfo{author}{\bibfnamefont{{\'A}.}~\bibnamefont{Luk‡{\'a}cs}},
  \bibinfo{journal}{Phys. Rev.} \textbf{\bibinfo{volume}{D94}},
  \bibinfo{pages}{125018} (\bibinfo{year}{2016}), \eprint{1608.00021}.

\bibitem[{\citenamefont{{Sedrakian} and
  {Sedrakian}}(1995)}]{1995ApJ...447..305S}
\bibinfo{author}{\bibfnamefont{A.~D.} \bibnamefont{{Sedrakian}}}
  \bibnamefont{and} \bibinfo{author}{\bibfnamefont{D.~M.}
  \bibnamefont{{Sedrakian}}}, \bibinfo{journal}{Astrophys.\ J.}
  \textbf{\bibinfo{volume}{447}}, \bibinfo{pages}{305} (\bibinfo{year}{1995}).

\bibitem[{\citenamefont{Srednicki}(2007)}]{Srednicki:2007qs}
\bibinfo{author}{\bibfnamefont{M.}~\bibnamefont{Srednicki}},
  \emph{\bibinfo{title}{{Quantum field theory}}} (\bibinfo{publisher}{Cambridge
  University Press, Cambridge, England}, \bibinfo{year}{2007}), ISBN
  \bibinfo{isbn}{9780521864497}.

\end{thebibliography}

\end{document}